\documentclass{article}
\usepackage{deluxetable}
\newcommand{\affil}[1]{$^{\rm #1}$}
\usepackage[authoryear]{natbib}
\bibpunct{ (}{)}{;}{a}{}{,}
\usepackage{graphicx}
\date{} 
\newcommand{\kms}{\mbox{km\,s$^{-1}$}}
\def\LCDM {$\Lambda$CDM~}
\def\kms {\ifmmode{{\rm ~km~s}^{-1}}\else{~km~s$^{-1}$}\fi}
\def\lsun {\ifmmode{{\rm ~L}_\odot}\else{~L$_\odot$}\fi}
\def\deg {^{\circ} }
\def\sqdeg {\,deg$^2$}
\def\arcsec {\,arcsec}
\def\arcmin {\,arcmin}
\def\ujybm {\,$\mu$Jy/beam}

\newcommand{\HI}{H\,{\sc i}}

\newcommand{\Msun}{~M$_{\odot}$}

\newbox\grsign \setbox\grsign=\hbox{$>$} \newdimen\grdimen \grdimen=\ht\grsign
\newbox\simlessbox \newbox\simgreatbox
\setbox\simgreatbox=\hbox{\raise.5ex\hbox{$>$}\llap
 {\lower.5ex\hbox{$\sim$}}}\ht1=\grdimen\dp1=0pt
\setbox\simlessbox=\hbox{\raise.5ex\hbox{$<$}\llap
 {\lower.5ex\hbox{$\sim$}}}\ht2=\grdimen\dp2=0pt

\def \etal {\rm ~{\it \etal},~}

\def\apj {{\it Ap.~J.}}
\def\apjl {{\it Ap.~J.\ (Letters)}}
\def\apjs {{\it Ap.~J.\ Suppl.}}
\def\aj {{\it A.~J.}}
\def\aa {{\it Astr.~Ap.}}
\def\aap {{\it Astr.~Ap.}}
\def\aas {{\it Astr.~Ap.\ Suppl.}}

\def\araa {{\it Ann.\ Rev.\ Astr.\ Ap.}}

\def\jcap {{\it J. Cosm. Astroparticle Phys.}}
\def\mnras {{\it MNRAS}}
\def\memras {{\it Mem. RAS}}
\def\na {{\it New Astronomy}}
\def\nat {{\it Nature}}
\def\pasa {{\it PASA}}
\def\pasp {{\it PASP}}
\def\prd {{\it Phys. Rev. D}}
\def\procspie {{\it Proc. SPIE}}

\def\ngal {70 }
\def\nagn {20 }
\def\nsf {50 }

\title{\large\bf\flushleft {EMU: Evolutionary Map of the Universe}}
\author{\parbox{\textwidth}{\flushleft
\vspace{-0.5cm}
{\it
Ray P.\ Norris\affil{1},
A.\,M.\ Hopkins\affil{2, 36},
J.\ Afonso\affil{3},
S.\ Brown\affil{1},
J.\,J.\ Condon\affil{4},
L.\ Dunne\affil{5},
I.\ Feain\affil{1},
R. Hollow\affil{1},
M.\ Jarvis\affil{6,38},
M.\ Johnston-Hollitt\affil{7},
E.\ Lenc\affil{1},
E.\ Middelberg\affil{8},
P.\ Padovani\affil{9},
I.\ Prandoni\affil{10},
L.\ Rudnick\affil{11},
N.\ Seymour\affil{12},
G.\ Umana\affil{13},
H.\ Andernach\affil{14},
D.\ M. Alexander\affil{21},
P. N. Appleton\affil{15},
D.Bacon\affil{16},
J. Banfield\affil{1},
W. Becker\affil{17},
M. J. I. Brown\affil{18},
P. Ciliegi\affil{19},
C. Jackson\affil{1},
S. Eales\affil{20},
A. C. Edge\affil{21},
B.M. Gaensler\affil{22, 36},
G. Giovannini\affil{10},
C. A. Hales\affil{1,22},
P. Hancock\affil{22, 36},
M.Y.Huynh\affil{23},
E. Ibar\affil{24},
R.\,J.\ Ivison\affil{24, 25},
R. Kennicutt\affil{26},
Amy E. Kimball\affil{4},
A. M. Koekemoer\affil{27},
B. S. Koribalski\affil{1},
\'A. R. L\'opez-S\'anchez\affil{2, 37},
M. Y. Mao\affil{1,2,28},
T. Murphy\affil{22, 36},
H. Messias\affil{29},
K.\ A.\ Pimbblet\affil{18},
A. Raccanelli\affil{16},
K. E. Randall\affil{1,22},
T. H. Reiprich\affil{30},
I. G. Roseboom\affil{31}
H. R\"ottgering\affil{32},
D.J. Saikia\affil{33},
R.G.Sharp\affil{34},
O.B.Slee\affil{1},
Ian Smail\affil{21},
M. A. Thompson\affil{6},
J. S. Urquhart\affil{1},
J. V. Wall\affil{35},
G.-B. Zhao\affil{16}
 \\
\vspace{0.4cm}
}}}
\begin{document}
\twocolumn[
\begin{changemargin}{.8cm}{.5cm}
\begin{minipage}{.9\textwidth}
\vspace{-1cm}
\maketitle
%
%

{\bf Abstract: EMU is a wide-field radio continuum survey planned for the new Australian Square Kilometre Array Pathfinder (ASKAP) telescope. The primary goal of EMU is to make a deep
 (rms $\sim$ 10\ujybm) radio continuum survey of the entire Southern Sky at 1.3 GHz, extending as far North as $+30\deg$
declination, with a resolution of 10\arcsec. EMU is expected to detect and catalogue about \ngal million galaxies, including typical star-forming galaxies up to z$\sim$1, powerful starbursts to even greater redshifts, and AGNs to the edge of the visible Universe. It will undoubtedly discover new classes of object. This paper defines the science goals and parameters of the survey, and describes the development of techniques necessary to maximise the science return from EMU.
}


\medskip
\medskip
\end{minipage}
\end{changemargin}

]

\section{Introduction}
\subsection{Background}
Deep continuum surveys of the radio sky have a distinguished history both for discovering new classes of object and for providing radio counterparts to astronomical objects studied at other wavelengths. The earliest large surveys, such as the 3C catalogue  \citep{Edge59} and the Molonglo Reference Catalogue  \citep{Large81}, gave us the first insight into the physics of radio galaxies and radio-loud quasars, but were insufficiently sensitive to detect any but the nearest radio-quiet or star-forming galaxies. Later radio surveys reached flux densities where normal star-forming galaxies were detected, but were still largely dominated by radio-loud active galactic nuclei (AGN). Only very long integrations in narrow deep fields made it possible to start probing star-forming galaxies beyond the local Universe. This paper describes a planned survey, EMU (Evolutionary Map of the Universe), which will reach a similar sensitivity ($\sim$ 10 \ujybm) to those deep surveys, but over the entire visible sky. At that sensitivity, EMU will be able to trace the evolution of galaxies over most of the lifetime of the Universe.
 
 Fig. 1 shows the major 20-cm continuum radio surveys. The largest existing radio survey, shown in the top right, is the wide but shallow NRAO VLA Sky Survey (NVSS),  whose release paper \citep{Condon98} is one of the most cited papers in astronomy. The most sensitive existing radio survey is the deep but narrow Lockman Hole observation  \citep{Owen08} in the lower left. All current surveys are bounded by a diagonal line that roughly marks the limit of available telescope time of current-generation radio telescopes. The region to the left of this line is currently unexplored, and this area of observational phase space presumably contains as many potential new discoveries as the region to the right.

\begin{figure}[h]
\begin{center}
\includegraphics[scale=0.4, angle=0]{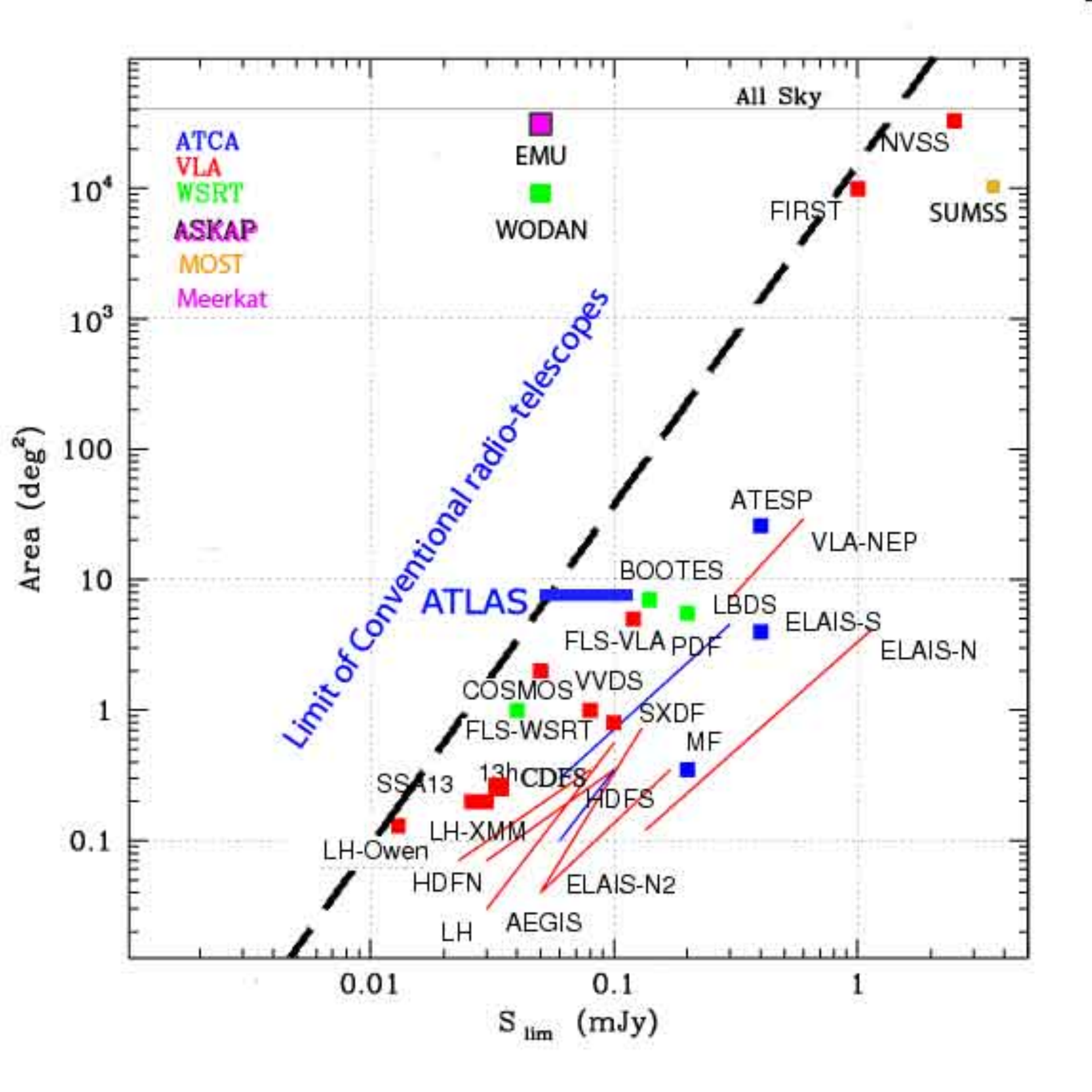}
\caption{Comparison of EMU with existing deep 20 cm radio surveys. Horizontal axis is 5-$\sigma$ sensitivity, and vertical axis shows the sky coverage. The diagonal dashed line shows the approximate envelope of existing surveys, which is largely determined by the availability of telescope time. The squares in the top-left represent the EMU survey, discussed in this paper, and the complementary WODAN \citep{Rottgering10b} survey which has been proposed for the upgraded Westerbork telescope to cover the sky North of $+30\deg$. Surveys represented by diagonal lines are those which range from a wide shallow area to a smaller deep area. The horizontal line for ATLAS extends in sensitivity from the intermediate published data releases  \citep{Norris06, Middelberg08a, Hales11} to the final data release  \citep{Grant11}.  }
\label{fig1}
\end{center}
\end{figure}

The Square Kilometre Array (SKA) is a proposed major internationally-funded radio telescope  \citep{Dewdney09} whose construction is expected to be completed in 2022. It will be many times more sensitive than any existing radio telescope, and will answer fundamental questions about the Universe \citep{Carilli04}. It is likely to consist of between 1000 and 1500 15-meter dishes in a central area of diameter 
5 km, surrounded by an equal number of dishes in a region stretching up to thousands of km.

The Australian SKA Pathfinder (ASKAP) is a new radio telescope being built both to test and develop aspects of potential SKA technology, and to develop SKA science. ASKAP is being built on the Australian candidate SKA site in Western Australia, at the Murchison Radio-astronomy Observatory, with a planned completion date of late 2012. In addition to developing SKA science and technology, ASKAP is a major telescope in its own right, likely to generate significant new astronomical discoveries.

\subsection{ASKAP}
\label{askap}
ASKAP  \citep{Johnston07, Johnston08, Deboer09} will consist of 36 12-metre antennas spread over a region 6 km in diameter. Although the array of antennas is no larger than many existing radio telescopes, the feed array at the focus of each antenna is revolutionary, with a phased-array feed \citep[PAF:][]{Bunton10} of 96 dual-polarisation pixels, designed to work in a frequency band of 700--1800 MHz, with an instantaneous bandwidth of 300 MHz. This will replace the single-pixel feeds that are almost universal in current-generation synthesis radio telescopes. As a result, ASKAP will have a  field of view up to 30 \sqdeg\, enabling it to survey the sky up to thirty times faster than existing synthesis arrays, and allowing surveys of a scope that cannot be contemplated with current-generation telescopes. To ensure good calibration, the antennas are a novel 3-axis design, with the feed and reflector rotating  to mimic  the effect of an equatorial mount,  ensuring a constant position angle of the PAF and sidelobes on the sky. The pointing accuracy of each antenna is significantly better than 30 arcsec.

The ASKAP array configuration  \citep{Gupta08} balances the need for high sensitivity to extended structures (particularly for neutral hydrogen surveys) with the need for high resolution for continuum projects such as EMU. To achieve this, 30 antennas follow a roughly Gaussian distribution with a scale of $\sim$\,700 m, corresponding to a point spread function of $\sim$ 30\arcsec~using natural weighting, with a further six antennas extending to a maximum baseline of 6 km, corresponding to a point spread function of $\sim$ 10\arcsec~ using uniform weighting. The positions of the antennas are optimised for \emph{uv} coverage (i.e. coverage in the Fourier plane) between declination $-50\deg$ and $+10\deg$, but give excellent \emph{uv} coverage between declination $-90\deg$ and $+30\deg$.

The PAF  is still under development, but the performance of prototypes gives us confidence that the EMU survey is feasible as planned. The PAF will consist of 96 dual-polarisation receivers, each with a  system temperature  $\sim$ 50 K, which are combined in a beam-former to form up to 36 beams. Each of these beams has the same primary beam response as a single-pixel feed ($\sim 1.2 \deg$ full-width half-maximum at 1.4 GHz), distributed in a uniform grid across an envelope of 30 \sqdeg. The optimum weighting and number of beams is still being studied, but the current expectation is that 36 beams will be used for EMU, with the sensitivity over the 30\sqdeg\ FOV (field of view) expected to be uniform to $\sim$ 20\%. This will be improved to $<$10\% uniformity by dithering, with no significant loss of sensitivity, so that the images from the 36 beams can be jointly imaged and deconvolved as a single image covering the FOV.
Consequently, it is expected that the telescope will dwell on one position in the sky for 12 hours, reaching an rms sensitivity of $\sim$ 10 \ujybm\  over a $\sim$ 30\sqdeg\ FOV.  The strategy for achieving this is still under development, and is discussed in \S \ref{strategy}.

Although high spatial resolution is essential for EMU, the short spacings of ASKAP also deliver excellent sensitivity to low-surface brightness emission, which is essential for a number of science drivers such as studies of radio emission from nearby clusters (\S \ref{clusters}). The $\sim$ 10\ujybm\ rms continuum sensitivity in 12 hours is approximately constant for beam sizes from 10 to 30 arcsec, then increases to $\sim$ 20\ujybm\ for a 90 arcsec beam and $\sim$ 40\ujybm\ for a 3 arcmin beam.

Science data processing  \citep{Cornwell11} will take place in an automated pipeline processor in real time. To keep up with the large data rate ($\sim$ 2.5 GB/s, or 100 PB/year), all science data processing steps, from the output of the correlator to science-qualified images, spectra, and catalogues, are performed in automated pipelines running on a highly distributed parallel processing computer. These steps include flagging bad data, calibration, imaging, source-finding, and archiving.

A typical ASKAP field will contain about 50\,Jy of flux in compact or slightly resolved sources. ASKAP
can observe the entire visible 20 cm continuum sky to an rms sensitivity of $\sim$1 mJy/beam in one day, so that initial observations will produce a global sky model (an accurate description of all sources stronger than $\sim$ 1 mJy) which significantly simplifies subsequent processing, as strong sources will be subtracted from the visibility data before processing. This sky model also means that antenna complex gains can be self-calibrated in one minute without any need to switch to calibrator sources. It is expected that individual receiver gains will be sufficiently stable that the dominant causes of antenna complex gain variation (i.e. ionosphere and troposphere) will be common to all pixels, so that a gain solution in one beam of an antenna can be transferred to other beams of that antenna.

In continuum mode, ASKAP will observe a 300 MHz band, split into 1 MHz channels, with full Stokes parameters measured in each channel. The data will be processed in a  multi-frequency synthesis mode, in which data from each channel are correctly gridded in the \emph{uv} plane. As well as producing images and source catalogues, the processing pipeline will also measure spectral index, spectral curvature, and all polarisation products across the band.

Completion of the Boolardy Engineering Test Array (BETA), which is a 6-antenna subset of ASKAP, is expected in late 2011. BETA will be equipped with 6 PAFs, and all the necessary beamformers, correlators, and processing hardware to produce images over the full 30\sqdeg\  field. The primary goal of BETA is to enable engineering tests and commissioning activities while the remaining ASKAP hardware is being constructed.  If engineering commissioning proceeds as expected, science observations on BETA will commence in 2012, on a small number of test fields on which good radio-astronomical and ancillary data already exist. These test BETA observations will be used to debug and fine-tune not only ASKAP, but also the processes for handling the data.

The full ASKAP array is expected to be commissioned in early 2013, and the science surveys are expected to start in late 2013. There is no proprietary period on ASKAP data, with all data being placed in the public domain after quality control, so data are expected to start flowing to the astronomical community by the end of 2013.

Expressions of Interest for ASKAP survey projects were sought in November 2008, and full proposals were solicited in mid-2009 \citep{Ball09}. Of the 38 initial expressions of interest, ten proposals were eventually selected, with two, EMU (Evolutionary Map of the Universe) and WALLABY \cite[Wide-field ASKAP L-band Legacy All-sky Blink SurveY: ][]{Koribalski11}, being selected as highest priority. EMU  is an all-sky continuum survey whilst WALLABY is an all-sky survey for neutral hydrogen. ASKAP design is now being driven by the requirement to maximise the science return from these ten projects, with a particular focus on maximising the science from EMU and WALLABY.

It is planned that EMU, WALLABY, and some other projects will observe commensally, i.e., they will agree on an observing schedule, and will observe the sky in both continuum and HI modes at the same time, splitting the two data streams into two separate processing pipelines. More information on all the ASKAP projects, including links to their individual websites, can be found on http://askap.org.

\begin{figure}[h]
\begin{center}
\includegraphics[scale=0.3, angle=0]{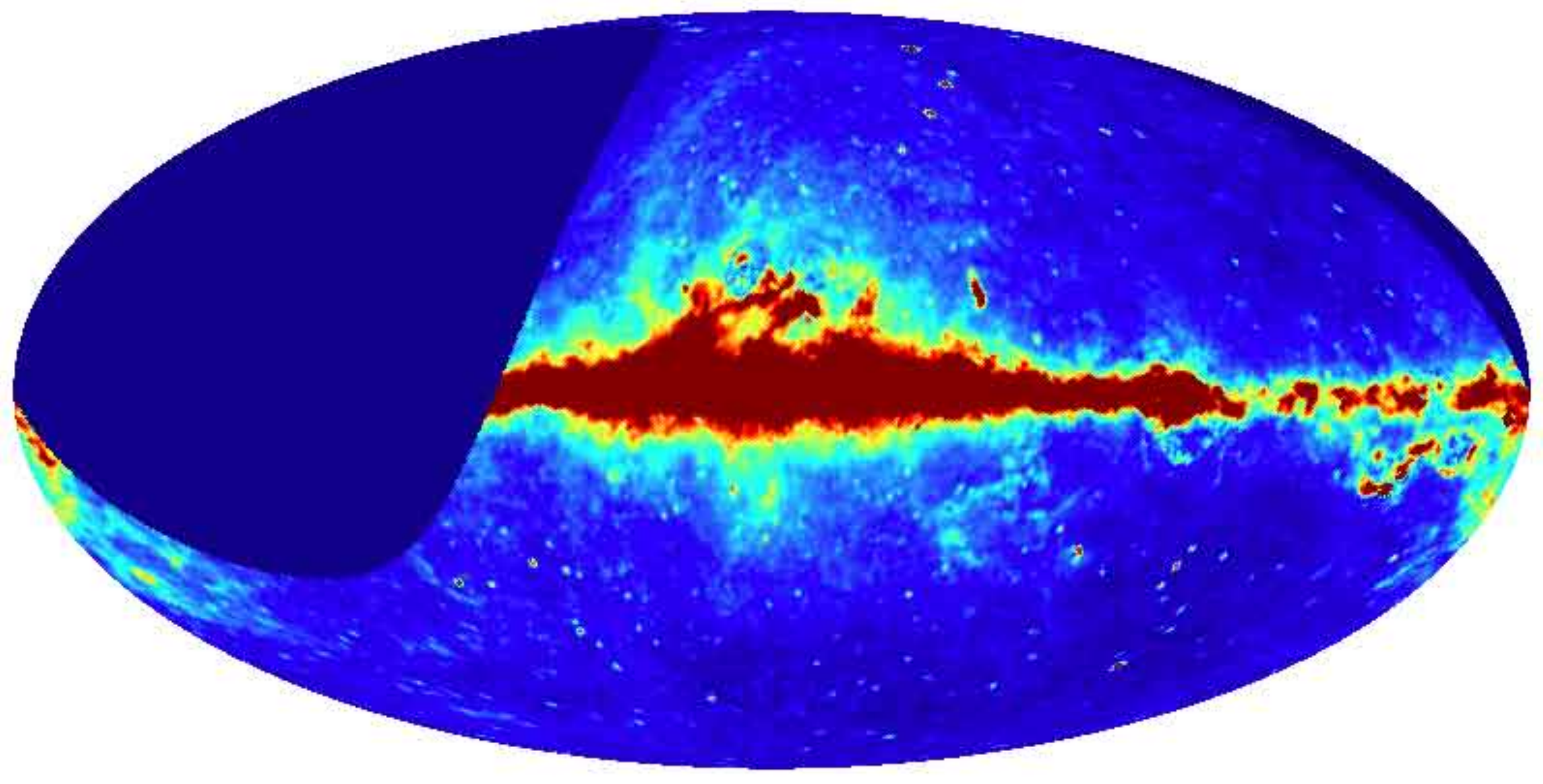}
\caption{A representation of the EMU sky coverage in Galactic coordinates overlaid on 23 GHz WMAP data  \citep{Gold11}. The dark area in the top left is the part of the  sky {\bf NOT} covered by EMU.
 }
\label{sky}
\end{center}
\end{figure}

\subsection{EMU}
The primary goal of EMU is to make a deep (10 \ujybm\ rms) radio continuum survey of the entire Southern Sky, extending as far North as $+30\deg$. EMU will cover roughly the same fraction (75\%) of the sky as the benchmark NVSS survey \citep{Condon98}, but will be 45 times more sensitive, and will have an angular resolution (10 arcsec) 4.5 times better. Because of the excellent short-spacing \emph{uv} coverage of ASKAP, EMU will also have higher sensitivity to extended structures. The sky coverage of EMU is shown in Fig.\ref{sky}, and the EMU specifications are summarised in Table \ref{specs}. Like most radio surveys, EMU will adopt a 5-$\sigma$ cutoff, leading to a source detection threshold of 50 \ujybm. EMU is expected to generate a catalogue of about \ngal million galaxies, and all
radio data from the EMU survey will be placed in the public domain as soon as the data quality has been assured.

\begin{table}[h]
\begin{center}
\caption{EMU Specifications}
\label{specs}
\begin{tabular}{ll}
\hline
Instantaneous FOV & 30\sqdeg\  \\
Area of survey & entire sky south \\ & of $+30\deg$ dec. \\
Synthesised beamwidth & 10 arcsec~FWHM \\
Frequency range & 1130-1430 MHz \\
Rms sensitivity & 10 \ujybm \\
Total integration time & $\sim 1.5$ years$^1$ \\
Number of sources & $\sim \ngal$ million \\
\hline
\end{tabular}
\medskip\\
\end{center}
$^1$ The primary specification is the sensitivity, rather than the integration time. If for any reason ASKAP is less sensitive than expected, EMU will increase the integration time rather than lose sensitivity. Conversely, an increase in sensitivity of ASKAP may reduce the total integration time. \\
\end{table}

Currently, only a total of about 5\sqdeg\  of the sky has been surveyed at 20 cm to the planned 10\ujybm\ rms of EMU, in fields such as the {\it Hubble}, {\it Chandra}, COSMOS and Phoenix deep fields  \citep{Huynh05, Miller08, Schinnerer07, Hopkins03, Biggs06, Morrison2010}, with a further 7\sqdeg\  expected in the immediate future as part of the ATLAS survey \citep{Norris06, Middelberg08a, Hales11, Grant11}.

Surveys at this depth extend beyond the traditional domains of radio astronomy, where sources are predominantly radio-loud galaxies and quasars, into the regime of star-forming galaxies. At this depth, even the most common active galactic nuclei (AGN) are radio-quiet AGNs, which make up most of the X-ray extragalactic sources.
As a result, the role of radio astronomy is changing. Whereas most traditional radio-astronomical surveys had most impact on the niche area of radio-loud AGNs, current radio-astronomical surveys are dominated by the same galaxies as are studied by optical and IR surveys, making radio-astronomical surveys such as EMU an increasingly important component of multi-wavelength studies of galactic evolution.

\begin{figure}[h]
\begin{center}
\includegraphics[width=8cm, angle=0]{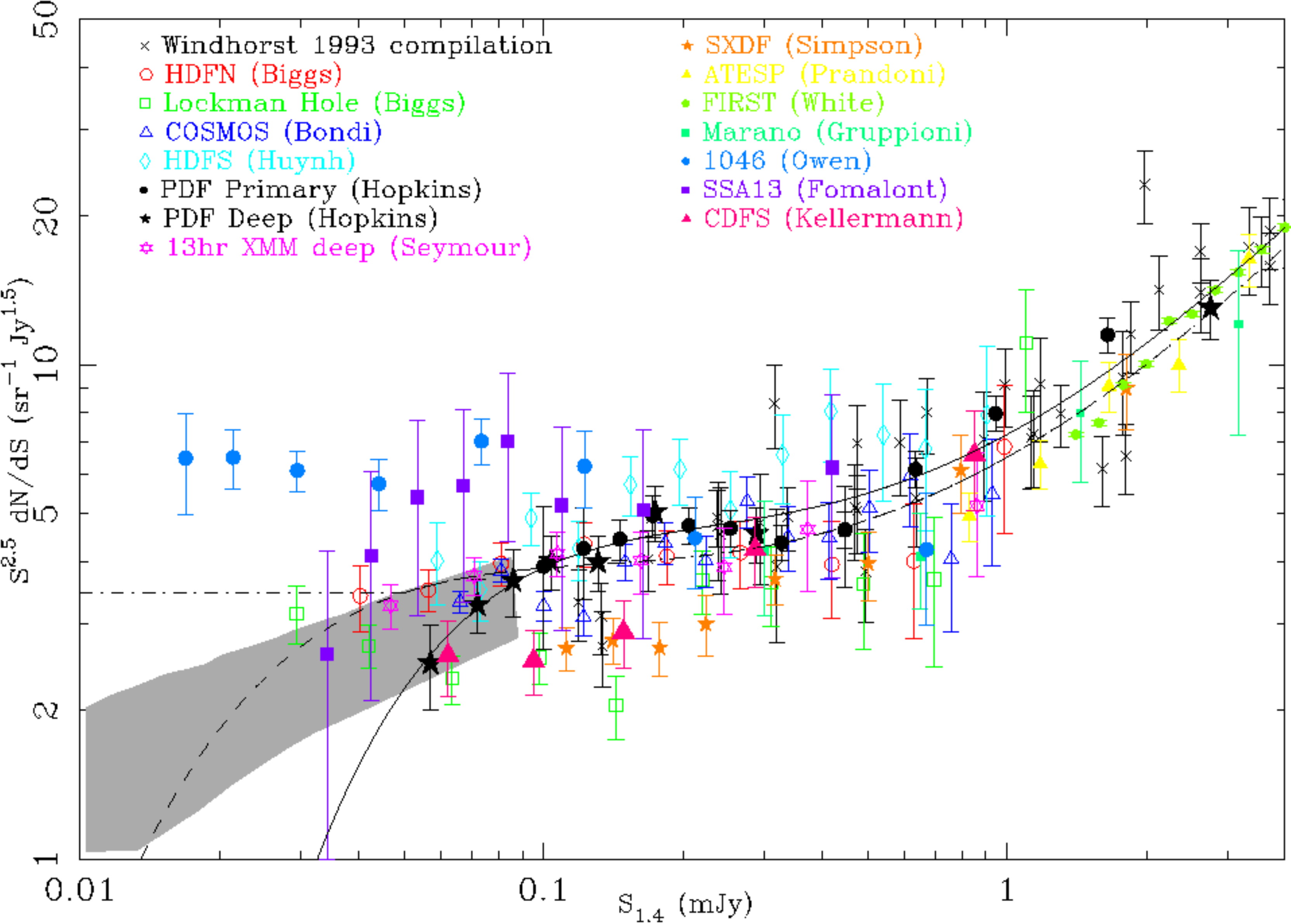}
\caption{Distribution of differential radio source counts at 1.4 GHz, based on and updated from the distribution shown in \cite{Hopkins03}.  The solid curve is the polynomial fit from \cite{Hopkins03},
the dashed curve is an updated polynomial fit and is the one used to estimate
the EMU source numbers. The horizontal dot-dashed line  represents a non-evolving population in a Euclidean universe.
The shaded region shows the prediction based on fluctuations due to weak confusing sources ( a ``P(D) analysis'') from
\citet{Condon74, Mitchell85}.
 }
\label{srccnt}
\end{center}
\end{figure}

Because only a small area of sky has been surveyed to the depth of EMU, it is difficult to estimate precisely how many galaxies it will detect. Most surveys to this sensitivity cover only a small area of sky, so that source counts at this level are significantly affected by sample variance, completeness, and bias issues. Our estimate for the number per \sqdeg\ above a flux density of 50 \ujybm\ is based on an extrapolation from source counts at higher flux densities \citep[2263 sources/\sqdeg; ][]{Jackson05} , the compilation shown in Fig. \ref{srccnt}  (2278 sources/\sqdeg),   and the COSMOS survey  \citep[2261 sources/\sqdeg; ][]{Scoville07, Schinnerer07}. These three figures are in good agreement and predict a total of $\sim$ \ngal million sources in EMU, which is therefore the number adopted throughout this paper.

Estimating the fraction of these radio sources which are AGN is difficult. Below 1 mJy, star-forming galaxies start to become a major component of the 1.4 GHz source counts, dominating below $\sim$ 0.15 mJy  \citep{Seymour08, Ibar09}, but, even at these levels, there is still a significant proportion of low-luminosity AGNs   
\citep{Jarvis04, Afonso05, Afonso06, Norris06, Simpson06, Smolcic08, Seymour08, Mignano08, Padovani09}.

\begin{figure}[h]
\begin{center}
\includegraphics[width=7cm, angle=0]{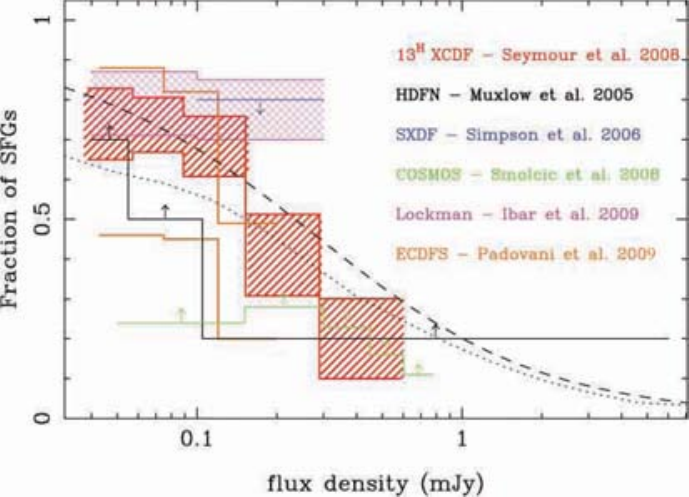}
\caption{Differential fraction of star-forming galaxies as a function of 1.4 GHz
flux density, from a selection of recent deep surveys. Shaded boxes,
and the two lines for Padovani et al.,  show the range of uncertainty
in the survey results. Arrows indicate constraints from other surveys.
These results show that the fraction of star-forming galaxies
increases rapidly below 1 mJy and, at the $50\,\mu$Jy survey limit of
EMU, about 75\% of sources will be star-forming galaxies.
}
\label{SFfraction}
\end{center}
\end{figure}

\cite{Seymour08} have presented the  most comprehensive attempt so far to divide radio sources into AGN and SF galaxies, and their result, together with other recent estimates, is shown in Fig. \ref{SFfraction}. From these we estimate that about 75\% of EMU sources will be star-forming galaxies.

\begin{figure}[h]
\begin{center}
\includegraphics[width=8cm, angle=0]{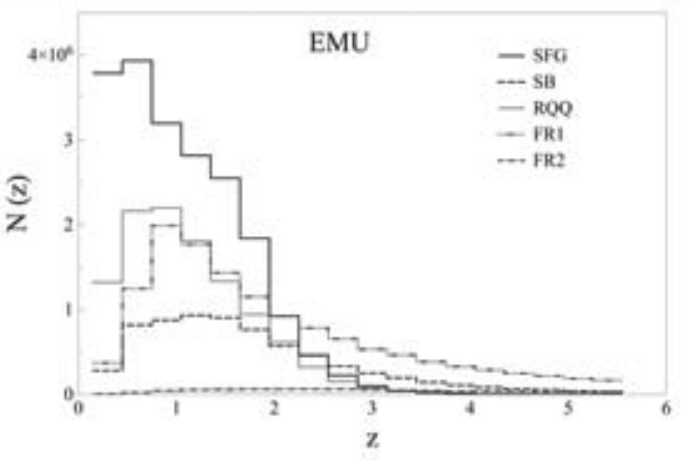}
\caption{Expected redshift distribution of EMU sources, based on the SKADS simulations  \citep{Wilman08, Wilman10}. The five lines show the distributions for star-forming galaxies (SFG), starburst galaxies(SB), radio-quiet quasars (RQQ), and radio-loud galaxies of Fanaroff-Riley types I and II \citep[FRI \& FR2; ][]{FRI}. Vertical scale shows the total number of sources expected to be detected by EMU.
}
\label{nz}
\end{center}
\end{figure}

To estimate the redshift distribution of  AGN and SF galaxies, we use the SKADS simulation \citep{Wilman08, Wilman10}, shown in Fig. \ref{nz}.  About \nsf million of the EMU sources are expected to be star-forming galaxies (see \S \ref{SFAGN}) at redshifts up to z $\sim$ 3, with a mean redshift of z $\sim$ 1.08.  The remainder are AGNs with a mean z $\sim$ 1.88, and extend up to z $\sim$ 6. However, if any FRII \citep{FRI} galaxies exist beyond that redshift (e.g. L $\sim 3.3\times10^{25} W Hz^{-1}$ at z = 10), EMU will detect them.

Confusion of radio sources, discussed more thoroughly in \S \ref{s_confusion}, is well-understood at this level, since previous surveys have already imaged small areas of sky to this depth and beyond.

EMU differs from many previous surveys in that a goal of the project is to cross-identify the detected radio sources with major surveys at other wavelengths, and produce public-domain VO-accessible catalogues as ``value-added'' data products. This is facilitated by the growth in the number of large southern hemisphere telescopes and associated planned major surveys spanning all wavelengths, discussed below in \S \ref{crossid}.

\subsection{Science}

Broadly, the key science goals for EMU are:
\begin{itemize}
\item To trace the evolution of star-forming galaxies from $z=2$ to the present day, using a wavelength unbiased by dust or molecular emission,
\item To trace the evolution of massive black holes throughout the history of the Universe, and understand their relationship to star formation,
\item To use the distribution of radio sources to explore the large-scale structure and cosmological parameters of the Universe, and to test fundamental physics,
\item To determine how radio sources populate dark matter halos, as a step towards understanding the underlying astrophysics of clusters and halos,
\item To create the most sensitive wide-field atlas of Galactic continuum emission yet made in the Southern Hemisphere, addressing areas such as star formation, supernovae, and Galactic structure,
\item To explore an uncharted region of observational parameter space, with a high likelihood of finding new classes of object.
\end{itemize}

Table 1 gives an overview of EMU specifications. In addition to the well-defined scientific goals outlined above, and the obvious legacy value, the large EMU dataset will include extremely rare objects, only possible by covering large areas.

A challenge for EMU will be the lack of spectroscopic redshifts, since no existing or planned redshift survey can cover more than a tiny fraction of EMU's \ngal million sources. 
As discussed in \S \ref{redshifts}, $\sim$ 30\% of EMU sources will have multi-wavelength optical/IR photometric data at the time of data release, increasing to $\sim$ 70\% in 2020. We expect these to provide accurate photometric redshifts for the majority of star-forming galaxies in EMU, and a minority of AGN (for which photometric redshifts tend to be unreliable).  In addition, many of the EMU sources will have  ``statistical redshifts'', which are valuable for some statistical tests. For example, most polarised sources are AGNs (mean $z \sim 1.88$), while most unpolarised sources are star-forming galaxies ( mean $z \sim 1.08$). More precise statistical redshifts can be derived where optical/IR photometry is available, as discussed in \S \ref{redshifts}.

A further goal of EMU is to test and develop strategies for the SKA. Many aspects of ASKAP, such as the automated observing, calibration, and data reduction processes, and the phased-array feeds, are potential technologies for the SKA, and it will be important to test whether these approaches deliver the planned results.

\subsection{Relationship to other surveys}
\label{others}
The following radio surveys are particularly complementary to the scientific goals of EMU.
\begin{itemize}
\item The WODAN survey  \citep{Rottgering10b} has been proposed for the Westerbork telescope which is currently being upgraded with a phased array feed \citep{Oosterloo10}. WODAN will cover the northern 25\% of the sky (i.e. North of declination $+30\deg$) that is inaccessible to ASKAP, to an rms sensitivity of 10 \ujybm\ and a spatial resolution of 15 \arcsec. Together, EMU and WODAN will provide full-sky 1.3\,GHz imaging at $\sim$ 10--15 \arcsec\ resolution to an rms noise level of 10 \ujybm, providing an unprecedented sensitive all-sky radio survey as a legacy for astronomers at all wavelengths. The WODAN survey will overlap with EMU by a few degrees of declination to provide a comparison and cross-validation, to ensure consistent calibration, and to check on completeness and potential sources of bias between the surveys.
\item The LOFAR continuum survey  \citep{Rottgering10a} will cover the northern half of the sky (i.e. North of declination $0\deg$) with the new LOFAR telescope operating at low frequencies (15--200 MHz).
LOFAR will be especially complementary to WODAN and EMU in surveying the sky at high sensitivity and resolution but at a much lower frequency.
\item The MIGHTEE survey \citep{Heyden10} on the Meerkat telescope \citep{Jonas10}  will probe to much fainter flux densities (0.1-1$\mu$Jy rms) over smaller areas ($\sim 35$ square degrees) at higher angular resolution, providing the completeness as a function of flux density for the EMU and WODAN Surveys. The higher sensitivity and resolution will enable exploration of the AGN and star-forming galaxy populations to higher redshifts and lower luminosities.

\item POSSUM  \citep{Gaensler10} is an all-sky ASKAP survey of linear polarisation. It is expected that POSSUM will be commensal with EMU, and that the two surveys will overlap considerably in their analysis pipelines and source catalogues. POSSUM will provide a catalogue of polarised fluxes and Faraday rotation measures for approximately 3 million compact extragalactic sources. These data will be used to determine the large-scale magnetic field geometry of the Milky Way, to study the turbulent properties of the interstellar medium, and to constrain the evolution of intergalactic magnetic fields as a function of cosmic time. POSSUM will also be a valuable counterpart to EMU, in that it will provide polarisation properties or upper limits to polarisation for all sources detected by EMU.
\item FLASH  \citep{Ball09} is an ASKAP survey whose goal is to detect extragalactic neutral hydrogen absorption. To do so it will observe at frequencies outside the 1130-1430 MHz band of EMU, thus yielding valuable spectral index information for those sources common to both surveys.
\item DINGO  \citep{Ball09} is an ASKAP survey whose goal is to detect faint extragalactic neutral hydrogen emission, and to do so it will spend many days on one ASKAP pointing. As a by-product, it will thus provide sensitive continuum images over smaller areas (several tens of\sqdeg\ ), allowing EMU to explore fainter flux densities in an optimal tiered survey structure, and also to quantify the effects of confusion at this level. However, the continuum images from DINGO will be severely confusion-limited at flux densities below a few \ujybm. It may be possible to transcend this limit  by subtracting known sources from the image, such as those star-forming galaxies which are seen in infrared images and whose radio flux can be predicted using the IR-radio correlation. However, this challenge is currently external to the core EMU project.
\item VAST \citep{Vast10} is an ASKAP survey that will observe partly commensally with EMU, with the goal of detecting transients and variable sources. EMU has no planned transient capability, since all information on variability of EMU sources will be available from VAST. This separation enables each of EMU and VAST to focus on its specific science goals, although significant coordination between the projects will clearly be essential.
\item WALLABY \citep{Koribalski11} is an HI survey which will deliver high-sensitivity spectral line data over
the same area of sky as EMU, and will observe commensally with EMU. Observations will
give a velocity coverage of --2,000 to
+77,000\kms\ ($z = 0 - 0.26$) and
velocity resolution of 4\kms. The angular resolution for WALLABY will be 30 \arcsec,
a factor three lower than EMU, as computing resources to make the large
spectral line data cubes are restricted to baselines shorter than $\sim$2~km.
Nearly all the $\sim 5 \times 10^5$ sources detected by WALLABY will also be detected by EMU, and WALLABY will provide an HI redshift for each of these, adding significantly to the redshift information for low-redshift EMU sources.
\end{itemize}

This paper defines the EMU survey, setting out its science goals in  \S\,2, and identifying the challenges to achieve these goals. \S\,3  describes how these challenges are being addressed in the EMU Design Study, and  \S\,4 describes  the survey operational plan, primary data products, and the data release plans and policy.

\section{EMU Science Goals}
\label{science}

\subsection{Star-forming galaxies and AGNs}
\label{SFAGN}

The fraction of star-forming galaxies as a function of flux density is shown in Figure \ref{SFfraction}. Of the $\sim$ \ngal million sources detected by EMU to a  $5\,\sigma$ limit of $50\,\mu$Jy, about \nagn million galaxies are expected to be dominated by Active Galactic Nuclei (AGN), and \nsf million to be dominated by star formation (SF). However, there is considerable overlap between the two classes, with composite AGN/SF galaxies becoming more common at low flux densities  \citep[e.g.][]{Chapman03,Norris06,Seymour08}.

It is unclear what fraction of putative SF galaxies have a significant AGN component. However, the excellent agreement in star-formation rate between radio and other star-forming indicators \citep[e.g.][]{Cram98, Bell03} suggests that an AGN is not a major contributor to the radio emission in such galaxies.

Detected AGNs and star-forming galaxies span a significant fraction of the age of the Universe, almost reaching the era of re-ionisation for radio AGNs and the most extreme starbursts.
Particularly at high redshift, both AGN and star formation processes are likely to be important in a large fraction of galaxies, but neither the fraction of the luminosity (bolometric and radio) generated by each process, nor how they are influenced by feedback, is currently known.
EMU will explore the evolution of these populations and its dependence on galaxy mass, environment, SF history, and interaction/merger history. It will quantify these effects in detail, by providing a deep homogeneously selected sample of both AGN and SF galaxies over the majority of cosmic history, unbiased by dust obscuration, and so provide a comprehensive overview of galaxy evolution.

The EMU analysis pipeline, which will encompass automated multi-wavelength cross-identification of sources between the EMU catalogue and other complementary surveys, will also include a variety of measures appropriate for distinguishing between, and quantifying the proportions of, AGN and star-forming activity. These will include:
\begin{itemize}
\item Radio morphology \citep[e.g.][]{Biggs08, Biggs10},
\item Radio spectral index \citep[e.g.][]{Ibar09, Ibar10},
\item Radio-far-infrared ratio,
\item Radio-near-infrared ratio,
\item Radio polarisation, from the POSSUM project  \citep{Gaensler10},
\item Radio variability, from the VAST project  \citep{Vast10},
\item optical and IR colours, to be used in SED analysis and comparison with templates.
\end{itemize}
These diagnostics and the algorithm which encodes them will be trained prior to the EMU data using the ATLAS data  \citep{Mao11b}.

\subsection{Evolution of star-forming \\ galaxies}
Tracing when and where stars formed across cosmic time is one of the key questions in galaxy evolution today. 
There is now evidence that star formation depends both on galaxy stellar mass  \citep[see Figure \ref{SFR}]{Feulner05, Juneau05, Papovich06, Mobasher09} and environment  \citep{Lewis02, Gomez03}, and these dependencies also evolve with time  \citep{Elbaz07}. Massive galaxies appear to form their stars early and quickly, progressively becoming less active after $z\sim 2$, while lower mass galaxies become dominant at lower redshifts. As a result, the cosmic SFR density is dominated by progressively less massive galaxies in less dense environments at lower redshifts. This process is known as cosmic ``downsizing''  \citep{Cowie96}, although it has been challenged by results from  \cite{Zheng07} and  \cite{Dunne09}.

EMU, with \nsf million SF galaxies and a large collection of complementary data, will accurately determine the true dependence of star formation on mass, environment, and redshift, in all types of galaxy from normal star-forming galaxies up to the most  UltraLuminous InfraRed Galaxies (ULIRGs). Radio observations have a considerable advantage over most other diagnostics of star formation as radio luminosities are directly related to star formation rates  \citep{Condon92, Haarsma00, Bell03} through a calibration reliable at least up to $z\sim 2$  \citep{Garn09}, and no corrections for absorption by gas or dust are required. However, to measure star formation rates from radio luminosities requires that any AGNs are removed from radio samples, and this can largely be achieved using a variety of indicators, as discussed in \S \ref{SFAGN}. For tracing evolution, EMU sources will necessarily be limited to those for which redshifts can be measured or estimated, as discussed in \S \ref{redshifts}. Fortunately, photometric redshifts prove to be very
accurate for SF galaxies \citep[see e.g. ][]{Rowan08} and a large fraction of star-forming galaxies detected by EMU will eventually have reliable photometric redshifts.

 \begin{figure}[h]
\begin{center}
\includegraphics[width=7cm, angle=0]{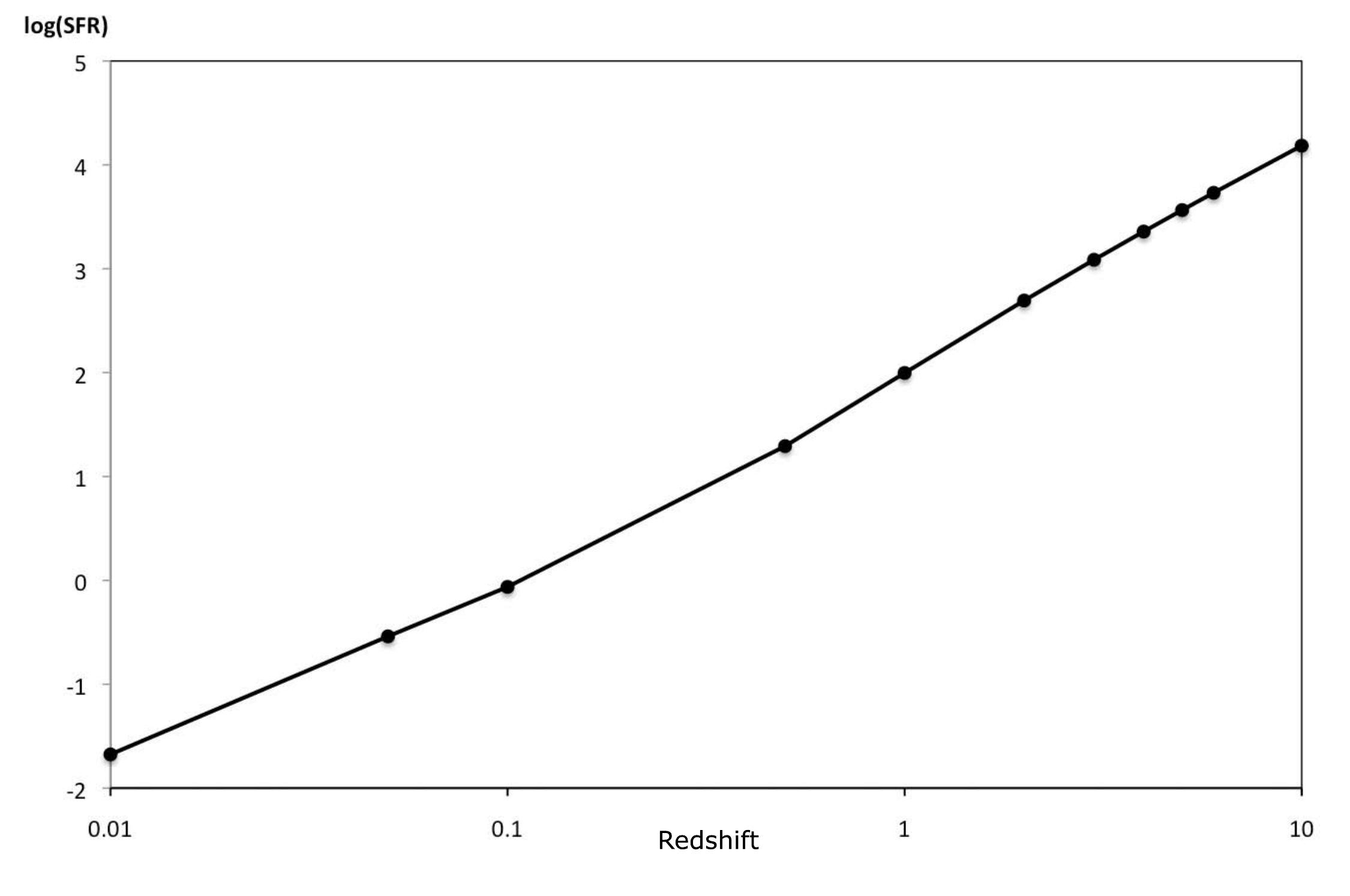}
\caption{Star formation rate (in \Msun/yr) of individual star-forming galaxies detectable by EMU at 5\,$\sigma$ as a function of redshift. The calculation of SFR follows \cite{Bell03} modified following \cite{Seymour08}, and the k-correction assumes a spectral index  of -0.7.
}
\label{SFR-z}
\end{center}
\end{figure}

EMU's surface brightness sensitivity of 0.06\,K is well below the median face-on surface brightness of a typical spiral galaxy of $\sim 1\,$K at 1.4\,GHz. So, unlike previous large radio surveys, EMU will have the brightness sensitivity needed to detect nearly all resolved normal spiral galaxies. Luminous starburst galaxies, with star formation rates (SFRs) around $100\,$M$_{\odot}\,$yr$^{-1}$, will be detectable to $z\sim 2$ (see Fig. \ref{SFR-z}), while ordinary disk galaxies like the Milky Way, with SFRs of only a few M$_{\odot}\,$yr$^{-1}$, will be visible to $z\sim 0.3$, and many of these will have spectroscopic redshifts measured by WALLABY. Star-formation rates of classes of galaxy at higher redshifts can be estimated using stacking (see \S \ref{stacking}), provided that AGNs have been eliminated from the stacked sample.

EMU will therefore trace the co-moving SFR density  \citep[e.g.,][]{Lilly96, Madau96, Hopkins04, Hopkins06} up to high redshift with greater accuracy than previously possible, and thereby determine when most of the stars in the Universe formed. The wide area of this survey will overcome the problems of low number statistics and sample variance sometimes associated with deep but small-area radio surveys \citep{Hopkins03}.

\label{bpt}
 \begin{figure}[h]
\begin{center}
\includegraphics[width=7cm, angle=0]{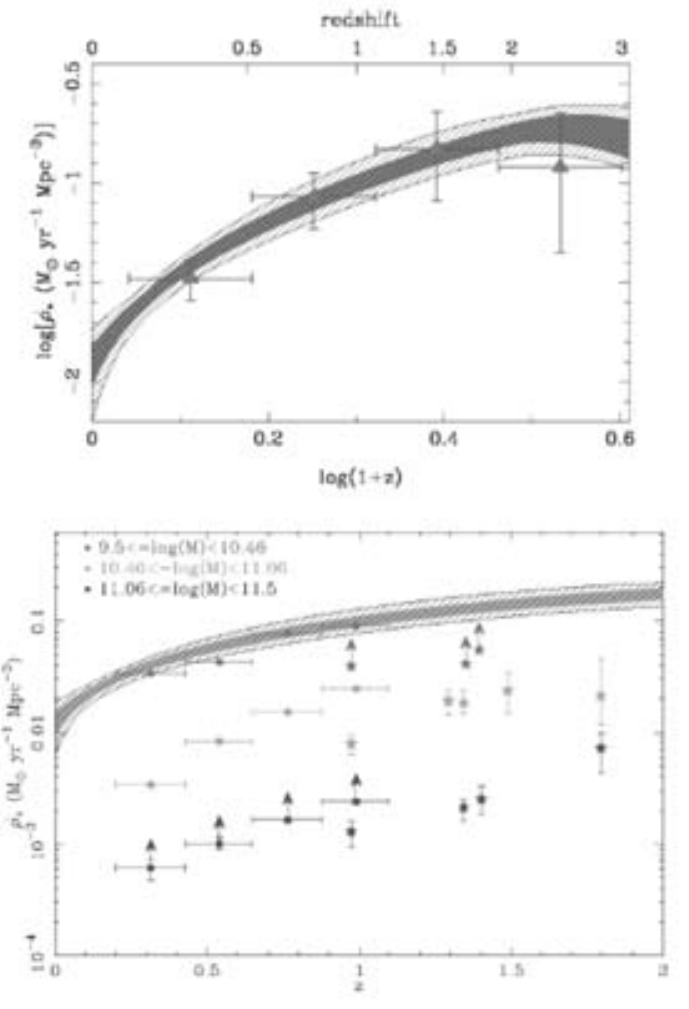}
\caption{Evolution with redshift of the star formation rate density (SFRD) in galaxies, $\dot{\rho}_*$. The grey and shaded regions in both panels are the 1-$\sigma $ and 3-$\sigma $ confidence region fits to the compilation of SFRD data from  \cite{Hopkins06}. Top: Measured SFRD taken from  \cite{Mauch07} and  \cite{Seymour08}. Bottom: The SFRD shown as a function of galaxy mass, adapted from Fig. 7 of  \cite{Mobasher09}.
}
\label{SFR}
\end{center}
\end{figure}

\subsection{Evolution of AGNs}

AGNs play a major role in the framework of galaxy formation. During their short active lifetime, they release an enormous amount of energy in the form of ionising radiation or relativistic jets, which can have a significant effect on the host galaxy and its surroundings.

The peak of QSO activity took place at z$\sim$2  \citep[e.g.][]{Schmidt68, Shaver96, Croom04, Hasinger05}, at epochs when SF activity was also extreme. Intriguingly, embedded AGNs have been found in 20-30\% of z$\sim$2 massive star-forming galaxies  \citep{Daddi07}. It has even been suggested that the AGN jets could be responsible for the formation of galaxies at high redshift  \citep{Elbaz09,Klamer04}.

At lower redshifts, AGNs appear to downsize, in a similar way to that of  SF galaxies, so that the peak of the AGN activity appears to shift significantly to lower redshifts for lower power AGNs.
This suggests \citep{Cowie96} that a feedback mechanism couples AGNs to galaxy evolution  \citep{Hasinger05, Bongiorno07, Padovani09}, via supernovae, starbursts or AGNs  \citep{Croton06, Bower06}. AGN jets either push back and heat the infalling gas, reducing the cooling flows building up the galaxy, or shock-heat and collapse the gas clouds, inducing star formation. The AGN energetic feedback appears to be an important ingredient for reproducing the galaxy stellar mass function  \citep{Croton06, Best06}, and the remarkable black hole vs. bulge mass (or velocity dispersion) correlation  \citep{Gebhardt00,Ferrarese00, Springel05}.

Recent work work on low-z 3CR galaxies by \cite{Ogle10}
suggest that radio jets can lead to
significant shock-heating of the host molecular disk, leading
to a significant enhancement of warm molecular hydrogen emission.  Despite their large molecular 
masses, these systems
seem to be very inefficient at making stars, suggesting that shock-heating by radio jets can play a negative-feedback role in star formation 
in these systems.

Low power radio-loud AGNs (P$<10^{25} WHz^{-1}$) appear quite different from the high-power radio-loud AGNs. Many do not show the luminous narrow lines 
expected in the framework of AGN unification \citep[see e.g. ][]{Hine79, Laing94, Jackson97}. They also lack the expected
infrared emission from a dusty torus \citep[see e.g. ][]{Whysong04,
Ogle06} and do not show accretion related X-ray emission
\citep{Hardcastle06, Evans06}. The sum of this observational
evidence suggests that low power radio sources correspond to a
distinct phase of AGN, accreting through a radiatively inefficient ``radio mode'', rather than the radiatively efficient accretion ``quasar mode'' typical of optical or X-ray selected AGNs  \citep{Croton06, Bower06}.
The physical reasons behind these two different accretion modes are still unclear \citep[e.g.][]{Hardcastle07, Herbert10}.

The key questions related to AGN evolution that EMU will address include: (1) The relationship between AGN and SF activity; (2) The evolution of low and intermediate power AGNs, exploring the so-called ``AGN cosmic downsizing'' scenario,  (3) The relative contribution of different accretion regimes (radio vs. quasar modes) in low and intermediate luminosity radio AGNs, its evolution with redshift, and the role played by the environment; (4) The relative contribution of radiative versus jet-driven (kinetic) feedback to the global AGN feedback in models of galaxy formation; and (5) the evolution of radio galaxies as a function of environment, spectral index, redshift, radio luminosity, and source size.

\subsection{High-redshift AGNs and the \\ Epoch of Re-ionisation}
\label{hizagn}
There have been many attempts to find very high-redshift radio galaxies  \citep{Breuck01, Jarvis01a, Jarvis01b, Best03, Cruz06, Cruz07} but all suffer from the difficulty of finding these extremely rare galaxies. EMU will identify numerous radio galaxies at z$>$4 and isolate the most distant objects by cross-matching with large near- and mid-infrared surveys, as described in \S \ref{crossid}. EMU will also provide spectral indices and (with POSSUM) polarisation information which can also help identify high-redshift AGNs.

High-redshift AGNs from EMU will constrain the existence and demographics of the most massive galaxies over the history of the Universe \citep[e.g.][]{Jarvis01b,Jarvis01c, Wall05}. These sources provide constraints on the co-evolution of galaxy bulges and central supermassive black holes  \citep{Magorrian98, Mclure06}, trace (proto-)clusters at early times \citep[e.g.][]{Miley06}, and may be crucial in determining the impact that powerful radio activity has on the host galaxy \citep[e.g.][]{Croton06, Bower06} and its larger scale environment  \citep{Rawlings04, Gopal01, Elbaz09}.

Infrared-Faint Radio Sources (IFRS) are probably a particular class of radio source that are characterised by a very high radio-infrared ratio $ (S_{20cm}/S_{3.6\mu m} > 500)$ and a low infrared flux density. First identified by  \cite{Norris06}, there is strong circumstantial evidence that they are high-redshift radio galaxies, based on their SED  \citep{Garn08, Huynh10}, their steep spectral index \citep{Middelberg11}, their VLBI cores \citep{Norris07, Middelberg08b}, and their extreme faintness in the infrared 
\citep[$S_{3.6\mu m}<1.2\mu$Jy; ][]{Norris11a}. Observations suggest $\sim$ 5 IFRS's occur per \sqdeg\  at a flux limit of $\sim 100\mu$Jy  \citep{Norris06,Middelberg08a}, implying that EMU will detect at least 
$1.5 \times 10^5$ , but the lack of corresponding deep infrared data will prevent their identification from other unremarkable non-detections. However, many will be located in smaller deep infrared fields such as those observed with HERMES  \citep{Oliver10}, which should yield several thousand IFRS's, enabling us to compile solid statistical data on their distribution, spectral index, and polarisation.
Thousands more will be selected as candidate high-redshift AGNs through their steep spectrum and polarisation (see \S \ref{redshifts}), and it expected that they will turn out to be an important class of high-redshift AGN.

A major objective of current extragalactic astronomy is to understand the Epoch of Reionisation (EoR), when ultraviolet photons from the first stars and quasars ionised the primordial neutral hydrogen. This process can in principle be studied by measuring the neutral hydrogen fraction in the early Universe using the 21 cm forest in front of a bright distant source  \citep{Carilli02}, analogous to the Ly-$\alpha$ forest seen against bright quasars at lower redshifts  \citep[e.g.][]{Peroux05}. Such observations require a population of z$>$6 radio-loud background sources, but  the highest redshift radio galaxy currently known lies at only z=5.2  \citep{Breugel99}. EMU provides an excellent opportunity to identify such sources, producing a large sample of distant radio sources for investigating the formation and evolution of the most massive galaxies at the highest redshifts.

\subsection{CSS and GPS sources}
Compact Steep Spectrum (CSS) sources are typically defined to be radio sources with a spectral index
$ < -0.5$, and with a size less
than about 20 kpc, and hence of sub-galactic dimensions. Gigahertz Peaked Spectrum (GPS)  sources are radio sources whose spectrum reaches a maximum in the frequency range 1--10 GHz. Typically they are significantly smaller (hundreds of pc) than CSS sources, and confined to the circumnuclear regions of the host galaxy.

Approximately 10\% and 30\% of bright centimetre wavelength sources are GPS and CSS sources respectively. If similar statistics hold for AGNs at low flux densities, EMU will discover $\sim$ 2 million GPS sources and $\sim$ 5 million CSS sources. Most studies of CSS and GPS objects have been confined to sources selected from strong source surveys,  and current samples only probe down to $\sim$ 10mJy at 1.4 GHz \citep{tschager03,snellen98,snellen99, Randall11}. EMU will offer a complete, faint sample of these intriguing objects, providing a probe into the evolution of young radio AGNs, and showing how these objects fit into models of galaxy evolution.

There is a consensus  \citep{Odea98,cfanti09a,rfanti09b,morganti09, Snellen09} that GPS and CSS sources  represent the start of the evolutionary path for large-scale radio sources.  
It is generally accepted that most GPS sources evolve into CSS sources, which gradually transform into the largest radio sources known, Fanaroff-Riley Type I and II galaxies \citep{FRI}, depending on their initial luminosity. These sources offer an ideal resource to investigate early galaxy evolution and formation, as well as AGN feedback, as they are young AGNs but also have star formation occurring due to interactions and mergers \citep{labiano08,morganti08}. Measurements of component advance speeds for a few compact sources yield ages of about 10$^3$ yr, while spectral studies indicate ages less than about 10$^5$ yr \citep{Odea98, Owsianik99, Murgia99}.  

These sources are thought to be fuelled by the infall of gas to the
supermassive black holes, triggered by interactions with companions and mergers.
CSS and GPS sources show evidence, from both their structural and
polarisation asymmetries, that the jets are interacting with an
asymmetric external environment  \citep{Saikia03, Cotton03}.
CSS and GPS sources  show evidence of H{\sc i} absorption more often than other radio-loud
AGNs, with the GPS objects showing the highest incidence
of absorption \citep[][and references therein]{Gupta06}.
The EMU sample of CSS and GPS sources over a large redshift
range will enable the variation of the H{\sc i}
properties with cosmic epoch to be determined.

Several CSS and GPS objects
show evidence of diffuse extended emission which may
be a remnant of an earlier cycle of activity.
An important goal of AGN physics is to understand the
episodic nature of nuclear or jet activity \citep{Saikia09}.
EMU with its high
surface brightness sensitivity will be an ideal survey to
probe the existence of such diffuse emission for a large
number of sources, and constrain the time scales of episodic activity for these compact objects.

\begin{figure}[h]
\begin{center}
\includegraphics[scale=1, angle=0]{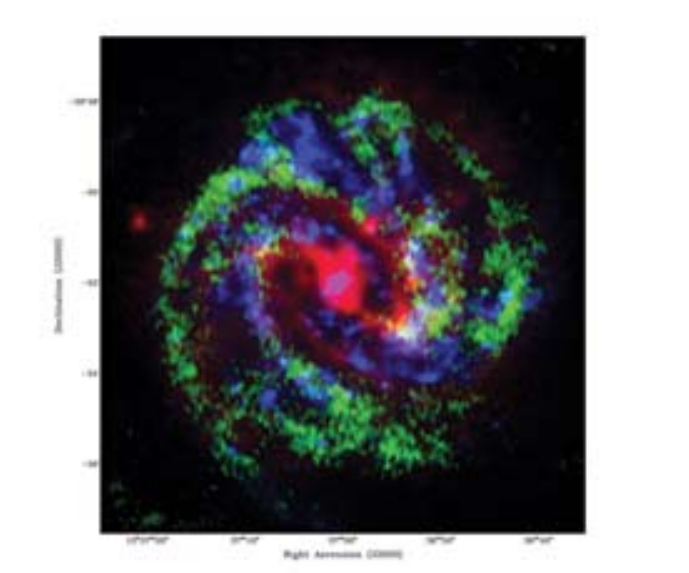}
\caption{This multi-wavelength composite image of the inner part of the
  spiral galaxy M\,83 highlights the synergy between the two key ASKAP large
  survey science projects: WALLABY and EMU. The ATCA 20-cm radio continuum
  emission is shown in red, the ATCA \HI\ distribution in green, and the
  GALEX FUV emission in blue \citep{Koribalski08, Galex07}.}
\label{wallaby}
\end{center}
\end{figure}

\subsection{Low redshift galaxies, and synergies with WALLABY}
WALLABY is expected to detect HI emission from $\sim 5 \times 10^5$ galaxies to a depth of $z = 0.05$, with massive galaxies detected
out to $z = 0.25$. The majority of these galaxies will be
spiral, with typical \HI\ masses of a few times 10$^9$\Msun. Their
large gas reservoir fuels star formation, implying that all spiral galaxies
detected by WALLABY will have 20-cm radio continuum emission detectable by
EMU. As a result, WALLABY is expected to contribute $\sim 5 \times 10^5$ redshifts to EMU.

In return, EMU will be able to measure star formation rates of the galaxies studied by WALLABY, enabling detailed studies of the factors that influence star formation rates in the local Universe. A goal of WALLABY is to measure local and global star formation (SF) rates
for all gas-rich spirals and compare their SF and \HI\ distributions (see
Fig.~\ref{wallaby}).

Although most WALLABY data will have a 30-arcsec resolution,  high-resolution (10 \arcsec) \HI\ 'postage stamps' will be obtained
of particularly interesting nearby galaxies, allowing a more detailed analysis and
comparison with data at other wavelengths.

\subsection{Galaxy clusters}
\label{clusters}
\begin{figure}[h]
\begin{center}
\includegraphics[scale=1, angle=0]{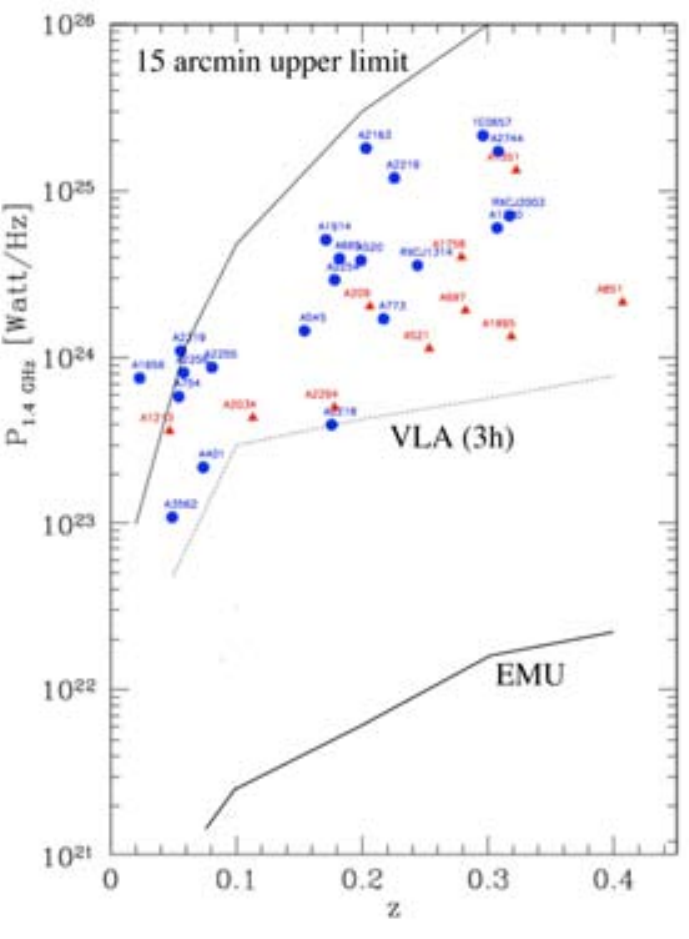}
\caption{1.4 GHz radio power of detected cluster halos as a function of redshift showing the detection limits of previous cluster observations, adapted from  \cite{Giovannini09}, and the calculated detection limit of EMU, assuming a halo with a  diameter of 1 Mpc. The upper line shows the limit corresponding to a scale size of 15 arcmin, which is approximately the largest size object that can be imaged with the VLA unless single-dish data are added to the interferometry data.
}
\label{cluster}
\end{center}
\end{figure}

Several different types of diffuse radio emission are associated with clusters of galaxies \citep{Kempner04}, including haloes around the centres of clusters, relics (representing shocks from cluster-cluster collisions) at the periphery, and tailed radio galaxies which are an important barometer of the intra-cluster medium. These three classes of radio source are important as tracers of clusters, and are also diagnostics of the physics of clusters, particularly when combined with X-Ray data.
However, the number of detected cluster radio sources is limited by current telescope sensitivities (see Fig \ref{cluster}). EMU will not only give us large samples of such sources, but will push beyond the present limits to detect diffuse sources with a range of powers over a larger redshift range, greatly improving our understanding of these sources.

Most clusters have been found either through X-ray surveys (Rosati et al. 1998; Romer et al. 2001; Pierre et al. 2003) or by searching for over-densities in optical colour-position space (Gladders \& Yee 2005; Wilson et al. 2008; Kodama et al. 2007). As a result,  tens of thousands of clusters are currently known, but only a few at z$ > $ 1 (Wilson et al. 2008, Kodama et al. 2007), with the highest redshift at z = 2.07 (Gobat et al. 2011).

\subsubsection{Haloes}

Radio haloes are found in clusters and groups of galaxies, indicating synchrotron emission powered by diffuse and faint (0.1 - 1 $\mu$G) magnetic fields and relativistic particles. So far about 35 radio halos are known with z $<$ 0.4 and only 2
at z $\sim$ 0.5, generally discovered by making deep radio surveys of hot
and bright X-ray clusters \citep{Venturi08, Giovannini09}.
A strong correlation is present between the total halo radio power and the
cluster X-ray luminosity \citep{Giovannini09}. Since the cluster mass
and X-ray luminosity are correlated it follows that halo radio power 
correlates with the cluster mass \citep{Feretti00, Govoni01}.
\citet{Brunetti09, Cassano10, Schuecker01} have suggested that radio haloes in the centres of clusters are distributed bimodally, with haloes generally found only in those clusters which have recently undergone a merger, resulting in a disturbed appearance at X-ray wavelengths.

The ATLBS survey  \citep{Subrahmanyan10}, which has surveyed 8.4\sqdeg\  to an rms sensitivity of 80 \ujybm\  on a scale size of 50 \arcsec~ at 1.4 GHz, has detected tens of diffuse sources, of which about 20 have been tentatively identified as cluster and group haloes  \citep{Saripalli11}. EMU will have even better sensitivity to low-surface-brightness structures than ATLBS, so if the ATLBS numbers are confirmed, then  EMU will discover $\sim 6 \times 10^4$  cluster and group haloes, which significantly increases the number of known clusters. An important science goal will then be to compare the X-ray properties
(luminosity, temperature and, for the brighter clusters, morphologies)
of these radio-selected clusters to those of the X-ray selected
population from the eROSITA all-sky X-ray survey \citep{Predehl10}.

\subsubsection{Relics}
On the periphery of clusters, elongated radio ``relics'' are found, which probably represent the signatures of shock structures generated in cluster mergers 
\citep{Rottgering97, Brown11a, Weeren10}.
They provide important diagnostics for the dynamics of accretion and mergers by which clusters form  \citep{Barrena09}. Large populations of these structures will appear in EMU, and are likely to lead to new cluster identifications especially beyond $z\sim 0.5$. For example, only 44 radio relics are currently known  \citep{Hoeft11}, and few have been discovered in current radio surveys because of the relatively poor sensitivity of most surveys to low-surface-brightness structures.
One probable relic has been discovered in the seven square degrees of ATLAS \citep{Middelberg08a, Mao10}. Since EMU will have greater sensitivity to such low-surface-brightness structures than ATLAS, this suggests that  EMU should detect $>$ 4000 relics, although this number is clearly very uncertain. As a means of finding clusters, it is less effective than other radio and X-ray techniques, but will be invaluable for studying shock structures accompanying cluster mergers. 
Furthermore, relics show evidence  of ordered large scale 
magnetic fields in the periphery of galaxy clusters, in regions with 
a very low density of galaxies and thermal gas.

\subsubsection{Tailed radio galaxies}
Head-tail, wide-angle tail (WAT), and narrow-angle-tail galaxies (collectively named ``tailed radio galaxies'')  are believed to represent radio-loud AGNs in which the jets are distorted by the intra-cluster medium \citep{Mao09,Mao10}. They can also contribute to the diffuse emission  \citep{Rudnick09}, especially after the jets from the nucleus of the host galaxy have turned off; EMU's high resolution and sensitivity, especially when combined with polarisation information from the commensal POSSUM survey  \citep{Gaensler10}, will allow us to distinguish these from the (largely unpolarised) cluster-wide halo emission. Even more importantly, such tailed galaxies can be detected out to high redshifts\citep{Wing11, Mao10}, providing a powerful diagnostic for finding clusters.
From the WAT's discovered in the ATLAS fields, \cite{Mao11c} and \cite{Deghan11} have estimated that EMU will detect at least 26000, and possibly as many as $2 \times 10^5$ WATs, depending on their luminosity function and density evolution.

\subsubsection{AGN feed back in galaxy clusters and mini-halo sources}

The relatively cool and dense gas at the centres of many galaxy
clusters and groups emits copious X-ray radiation by thermal bremsstrahlung
and line emission. In the absence of external sources of heating,
this high emission should lead to very rapid cooling (t$_{cool} <$
1 Gyr) and very high rates of mass deposition onto the central
cluster galaxy (up to $\sim$1000 \Msun /yr), in turn causing
very high star formation rates and strong X-ray line emission \citep[e.g.][]{Fabian77, Cowie77,
Peterson06, McNamara07}. The lack of
such obvious observational signatures \citep[e.g. ][]{Peterson01, Peterson03}
implies that some central source of
heating must be present. The most plausible source of heating
is feedback from the central AGN.
EMU will enable a statistically significant study of the correlation between the radio emission from AGNs at the cluster centre with the thermal and non thermal
cluster properties, and explore how it evolves with redshift. 

Moreover a few cooling-core clusters exhibit signs of
diffuse synchrotron emission that extends far from the dominant
radio galaxy at the cluster center, forming what is referred to as a
mini-halo. These diffuse radio sources are extended on a moderate
scale (typically $\sim$ 500 kpc) and, in common with large-scale
halos, have a steep spectrum and a very low surface brightness.
Because of a combination of small angular size and the
strong radio emission of the central radio galaxy, the detection
of a mini-halo requires data of a much higher dynamic range
and resolution than those in available surveys.
The study of radio
emission from the center of cooling-core clusters is of large
importance not only in understanding the feedback mechanism
involved in the energy transfer between the AGN and the ambient
medium \citep[e.g.][]{McNamara07} but also in the
formation process of the non-thermal mini-halos. The energy 
released by the central AGN may also play a role in the formation
of these extended structures \citep[e.g.][]{Fujita07}.

\subsubsection{The impact of EMU on cluster research}

The key impacts of EMU are likely to be:
\begin{itemize}
\item To increase the number of known clusters beyond the few tens of thousands currently known. EMU will detect at least $3 \times 10^4$ new clusters\citep{Mao11c, Norris11c, Deghan11}, which will roughly double the number of known clusters.  Depending on the redshift distribution and luminosity function of cluster radio sources, EMU may detect as many as $\sim 2 \times 10^5 $ cluster sources \citep{Mao11c, Norris11c}. eRosita is also expected to detect $\sim 10^5 $ clusters at X-ray wavelengths \citep{Predehl10, Pillepich11}, and comparison of these two complementary surveys will be transformational.

\item To detect clusters at high redshifts. In principle, radio sources can be used to detect clusters even beyond z=1, where current constraints on large-scale structure are weaker, and traditional detection techniques like X-ray surveys and use of the Red Cluster Sequence (Gladders \& Yee 2005) become less effective. A few clusters have already been detected  up to high redshift using radio detections (Blanton et al. 2003, Wing et al. 2011), but extrapolation beyond z=1 is uncertain because the luminosity function and evolution of cluster radio sources are unknown. Furthermore, inverse Compton cooling of electrons by the cosmic microwave background is expected to quench their synchrotron radio emission at $z \gg 1$, although confidence in this expectation is reduced by the failure to detect a similar effect in the far-IR-radio correlation at high redshift (Mao et al. 2011a).

\item To explore radio properties of clusters detected at other wavelengths. It will be important to compare  the properties of radio-selected clusters from the unbiased EMU survey to those of the X-ray selected
population from surveys such as the eROSITA all-sky X-ray survey (Predehl et al. 2010), and the  Sunyaev-ZelÕdovich surveys made with the South Pole Telescope (Williamson et al. 2011), Atacama Cosmology Telescope (Marriage et al. 2010) and Planck (Planck Collaboration, 2011) surveys. EMU will enable not only  statistical approaches \citep[e.g. halo occupation distribution modelling: ][]{Peacock00} but also  directly observed overdensities using tracers such as WATs  \citep{Mao10}.

 \item To detect new forms of cluster radio emission. For example, high-resolution numerical simulations  \citep{Battaglia09} predict an additional network of weak shocks throughout the cluster volume which can only be seen with sufficient resolution and sensitivity, but which will be detectable with the short spacings of ASKAP.
 
 \end{itemize}
 
Additional cluster-related goals to be addressed by EMU include:
\begin{itemize}
\item Do the radio properties of AGNs depend more on the properties of their host galaxy or of their local environment, such as gas temperature or cooling time \citep[e.g. ][]{Mittal09}?
\item Are galaxies of a given mass more likely to host a radio source if they are a central or satellite galaxy within a halo?
\item Is star formation truncated in halos above a given mass by AGN feedback or virial shock heating?
\item How does the observed correlation and indicated feedback cycle between cluster central radio AGNs and the cooling intracluster medium evolve with redshift \citep[e.g. ][]{Santos10}?
\end{itemize}

\subsection{Cosmic filaments, and the Warm-Hot Intergalactic Medium \\
 (WHIM)}

Approximately half of the Universe's baryons are currently ``missing''  \citep{Cen99} and are likely to be contained in the elusive Warm-Hot Intergalactic Medium (WHIM), where temperatures of $10^{5} - 10^{7}$ K make them extremely difficult to detect \citep{Fraser11}. By subtracting the contribution of compact radio sources, studies of low-surface-brightness emission with EMU can illuminate the otherwise invisible baryons associated with large scale structure. Perhaps the most important diffuse radio sources are those illuminating shock structures in the WHIM, which occur during infall onto and along large-scale-structure filaments  \citep{Miniati01, Dolag08}.

Because of the sensitivity and short spacings of ASKAP, EMU will be able to
 detect faint radio emission from cosmological
filaments, increasing our knowledge of the physical properties of large-scale structures. The shortest spacing of ASKAP, 20m (resulting in sampling of the $uv$ plane down to 8m in a joint deconvolution), ensures that EMU will be sensitive to structures as large as 1$\deg$, whilst even larger scale sizes can be imaged by adding in single-dish data.

In a few cases, the synchrotron emission from filaments may be detected directly. One putative example is the bridge of radio emission extending over the
1 Mpc between the central halo of the Coma cluster and the
peripheral relic source 1253+275 \citep{Kim89, Kronberg07}.  Another is the radio emission from                  
ZwCl 2341.1+0000                                                     
which is a linear structure some  6 Mpc in length \citep{Bagchi02, Giovannini10}.      

Even though EMU will be able to detect only the brightest shock structures in the WHIM directly, statistical characterisation of an ensemble of fainter shock emission is possible  \citep{Keshet04, Brown10}.
Faint synchrotron emission due to WHIM shocks is highly correlated with large-scale structure (LSS) on $\sim$1-4~Mpc scales  \citep{Pfrommer07, Ryu08, Skillman08, Skillman10}, and can thus be detected below the EMU noise limits through LSS cross-correlation  \citep{Brown10} using the statistical redshifts described in \S \ref{redshifts}, or in some cases high-quality photometric or spectroscopic redshifts.
The sky coverage of EMU coupled with its high sensitivity to low surface-brightness emission makes it ideal for detecting the synchrotron cosmic-web at cosmologically important redshifts $0.1 < z < 1.0$  
\citep{Brown11b}, corresponding to $\sim$0.1-0.5~Mpc on arcmin scales.

\subsection{Cosmology and Fundamental Physics}
\label{cosmology}


\begin{figure}[h]
\begin{center}
\includegraphics[width=7cm, angle=0]{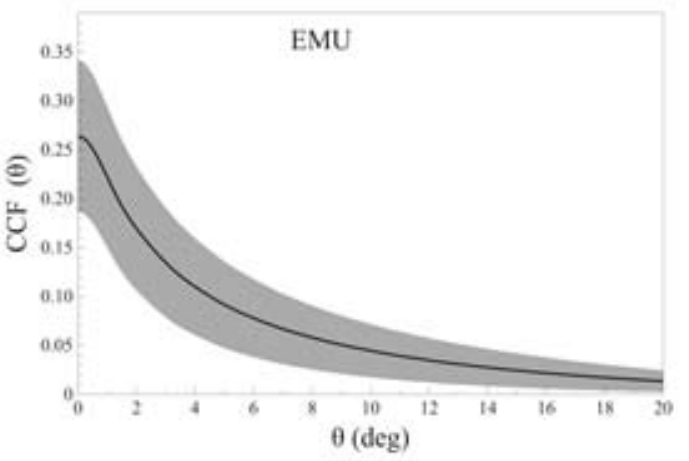}
\caption{Predicted cross-correlation of the CMB (measured by WMAP etc.) with EMU sources, assuming a standard \LCDM cosmology. The solid lines is the theoretical prediction, and the shaded region shows cosmic variance errors. 
}
\label{CCF}
\end{center}
\end{figure}

\begin{figure}[h]
\begin{center}
\includegraphics[width=7cm, angle=0]{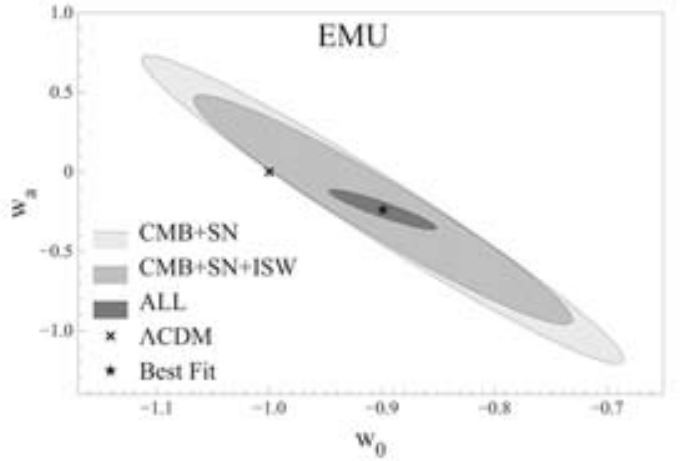}
\caption{Predicted constraints from EMU on dark energy parameters, assuming no redshifts are available. The outer grey ellipse shows the current constraints from Type IA supernovae and CMB fluctuations. The intermediate ellipse shows the constraints from EMU using the ISW effect, providing only a modest improvement over existing measurements. The inner (dark) ellipse shows the constraints from EMU using all effects, including cosmic magnification, and shows a significant improvement over existing measurements. In particular, EMU will resolve the difference between the current best measurement (star) and the value (cross) predicted for a standard \LCDM cosmology with non-evolving dark energy.
}
\label{DE}
\end{center}
\end{figure}

\begin{figure}[h]
\begin{center}
\includegraphics[width=7cm, angle=0]{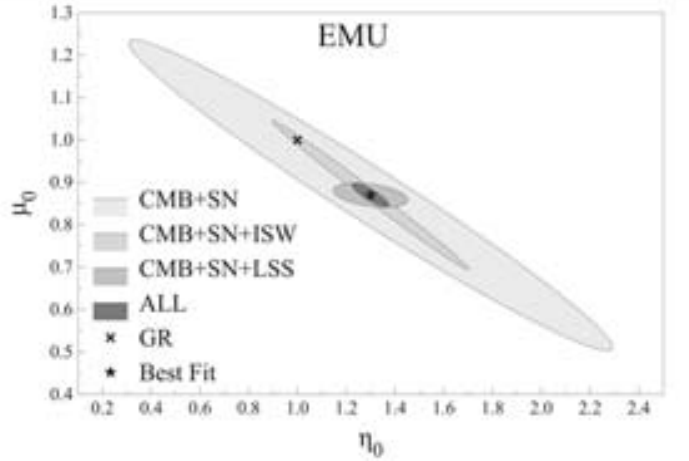}
\caption{Forecast of constraints for modified gravity parameters, with symbols the same as for Fig. \ref{DE}. The largest ellipse shows constraints from current measurements, and the smallest ellipse shows the constraints from EMU using all effects, but assuming no redshifts are available. EMU will resolve the difference between the current best measurement (star) and the value (cross) for standard General Relativity, either confirming or ruling out some alternative models.
}
\label{MG}
\end{center}
\end{figure}


\subsubsection{Background}

The existence of dark energy is one of the most profound problems in cosmology. The evidence for its presence is indirect: it is implied by the supernovae Type Ia {\it Hubble} relation  \citep{Riess98, Perlmutter99}, by the combined inference of flat geometry ($\Omega_{tot}=1$) from Cosmic Microwave Background (CMB) measurements and low mass density ($\Omega_{mass} \sim 0.3$) from large-scale structure measurements \citep{Cole05,Eisenstein05}, and from galaxy
cluster measurements  \citep{Vikhlinin09, Mantz10, Allen08}.

A further puzzle is the nature of gravity. While General Relativity (GR) is consistent with all current observational measurements  \citep[e.g.][]{Uzan03}, it fails to connect gravity with the other fundamental forces, models for which lead to hypotheses such as ``brane-worlds''  \citep{Brax03} in which  
gravity is modified by leakage into dimensions other than the brane  
dimensions. Such models predict deviations from GR which have not yet been fully tested.

EMU will be able to perform independent tests of models of dark energy and deviations from GR using a combination of (a) the auto-correlation of radio source positions, (b) the cross-correlation of radio sources with the CMB (the late-time Integrated Sachs Wolfe (ISW) effect), and (c) cross-correlation of radio sources with foreground objects (cosmic magnification).

We assume throughout most of this section that redshifts of EMU sources are {\bf not} available, and at the end consider the impact if statistical redshifts become available. Throughout this section, we assume a 5\,$\sigma$ detection limit, which is expected to be reliable for the EMU survey, while \cite{Raccanelli11} use a conservative 10\,$\sigma$ limit, which gives rise to slightly different constraints.

\subsubsection{Autocorrelations}
The auto-correlation function (ACF) of source counts, also known as the  two-point correlation of radio sources, measures the degree of clustering of the radio  sources
\citep[see e.g.][]{Blake02, Raccanelli11}.  Here we consider the ACF for EMU sources that we will measure in two dimensions on the sky, ignoring any potential availability of redshifts, and making no distinction between source  
types, which have different biases and number densities as a function  
of redshift.

Despite these limitations, the ACF is of value for two purposes. First,  
cosmological constraints from this probe can be combined with those  
from other probes to substantially improve the net cosmological  
constraint, as shown by the combined constraints in Figures \ref{DE} and \ref{MG}.
Second, the behaviour of the ACF on large scales constrains
non-Gaussianity (i.e. the extent to which density fluctuations in the early universe were distributed with non-Gaussian noise)
hence  providing a window on early-Universe physics. 
Non-Gaussianity indicates
a positive skewness of the matter density probability distribution, which would lead to an increased bias for large-scale halos
 \citep[see e.g.][]{Dalal08}.
EMU will be able to detect non-Gaussianity parameterized  
by the $f_{NL}$ parameter at the level $f_{NL} \ge 8,$   
\citep{Raccanelli11}. Such a detection would be difficult to reconcile with  
a simple, single scalar field inflation model for the early  
Universe.

\subsubsection{The Integrated Sachs-Wolfe Effect}
The ISW effect can be understood as follows. Travelling from the last scattering surface to us, CMB photons pass through gravitational potential wells of large structures such as superclusters that lie along our line of sight. In an Einstein-de Sitter universe, the blueshift of a photon falling into a well is cancelled by the redshift as it climbs out. However, in a universe with dark energy or modified GR, the local gravitational potential varies with time, stretching the potential well while photons are traversing it, so that the blueshift on entry is not fully cancelled by the redshift on exit. Thus the net photon energy is increased, producing CMB temperature anisotropies which make the CMB appear slightly hotter in the direction of superclusters.

The effect is very weak, but can be detected by cross-correlating CMB maps with tracers of large-scale structure  \citep{Crittenden96} such as radio sources. The detection of the effect has been repeatedly confirmed
\citep[e.g.][and references therein]{Giannantonio08a}; for instance, SDSS galaxies in the Sloan Digital Sky Survey \citep[SDSS][]{York00} have been cross-correlated with the CMB  \citep{Granett08} to achieve a 4-$\sigma$ result. In the radio, cross-correlation by  \cite{Raccanelli08} of the NVSS radio galaxies with the CMB anisotropies also gave a tentative detection of the ISW effect. The NVSS detection is partly limited by shot noise, and so the far greater number of galaxies in EMU should achieve a robust measurement of the effect.

EMU will also be able to constrain the redshift
evolution of the
equation of state of dark energy, which would allow
us to distinguish between different models of
dark energy, such as a cosmological constant  \citep{Tegmark04},
quintessence  \citep{Ziatev99}, 
early dark energy \citep{Xia09},  or Unified Dark Matter models \citep{Bertacca11}.

The ISW effect is also sensitive to any modifications of gravity, allowing  
sensitive constraints on gravity, and
can also be used to test models
for the evolution of the clustering and bias of radio
sources  \citep{Raccanelli08}, and to test models for the cosmological
evolution of radio sources  \citep{Massardi10}.
The predicted cross-correlation function for EMU
sources is shown in Fig.\ref{CCF}. EMU will provide sensitive constraints on any modifications of gravity.

\subsubsection{Cosmic Magnification}
 Large-scale structures along the line of sight to a distant source introduce distortions in the images of high-redshift sources as a result of gravitational lensing. The distortions by foreground (low-redshift) galaxies increase the apparent area occupied by background (high-redshift) galaxies, thus reducing the observed number density at a given flux density. On the other hand, the magnification increases the total flux density of unresolved high-redshift sources, thus increasing the observed number density at a given flux density. The size of the combined change in apparent number density  due to these two opposing effects is sensitive to the assumed cosmological geometry and parameters of GR.
 
 The combined effect, known as cosmic magnification, can be tested by cross-correlating number densities of low-redshift sources (e.g. selected from the optical surveys or from EMU star-forming galaxies) with number densities of high-redshift sources (selected from EMU). The effect was first unambiguously detected by  \cite{Scranton05} using SDSS foreground galaxies and quasars. The large database of EMU sources will develop this into a powerful technique for testing cosmological models.

\subsubsection{Observational tests with EMU}

Deviations from GR or dark energy physics will influence the auto-correlation, the ISW and cosmic magnification \citep{Zhao10}. By modelling the EMU source distribution and bias,  \cite{Raccanelli11} have shown that significant limits can be placed on cosmological parameters (such as $w$ and $w'$) that describe dark energy, and on parameters $\eta$ and $\mu$ that describe modifications to GR  \citep{Pogosian10}. GR predicts that $\eta =\,\mu = 1$. Figs.\ref{DE} and \ref{MG} show that, even without redshifts, EMU will place significant constraints on both Dark Energy and Modified GR.

Once shot noise is fully overcome, ISW measurements
are limited only by cosmic variance. Substantial improvements over current
measurements can therefore be achieved by using a bigger
survey area.
Combining the EMU survey
with the WODAN survey
will allow us to make a radio measurement of  the ISW effect over the
entire sky  for the first time, perhaps leading to
the highest significance ISW measurement yet.

\subsubsection{Tomographic Analysis}
The discussion of cosmological tests above makes the conservative assumption that no redshifts are available for EMU sources. If statistical redshifts (\S \ref{redshifts}) are available, the radio sources can be split into redshift bins for measuring the ISW effect or cosmic magnification, enabling a ``tomographic'' analysis of auto- and cross-correlation data, leading to even more significant constraints than those discussed above. For example, strong polarised sources are known to have a high median redshift, even when they are undetected in optical surveys.

If there are enough sources in each bin to achieve a statistically significant ISW detection, EMU will enable a measurement of the redshift evolution of the ISW effect and so better constrain cosmological models. The results at low redshifts will give an independent confirmation of detections achieved at optical wavelengths. However, the radio sources extend to far higher redshifts, and so test the ISW at epochs that cannot be reached at other wavelengths: a detection of the ISW at high redshift would be a clear signature of a non \LCDM+GR evolution  \citep{Raccanelli11}.

\subsection{Galactic Science}
\label{galaxy}
The EMU survey will include the Galactic Plane, thus creating a sensitive wide-field atlas of Galactic continuum emission, which can address several science goals including:
\begin{itemize}
\item A complete census of the early stages of massive star formation in the Southern Galactic Plane,
\item Understanding the complex structures of giant HII regions and the inter-relationship of dust, ionised gas and triggered star formation,
\item Detection of the youngest and most compact supernova remnants to the edge of the Galactic disk, some of which may have exploded within the past century,
\item Detection of supernova remnants, especially those detected by eRosita but which are undetected by previous radio surveys,
\item Detection of planetary nebulae, which are the most abundant compact Galactic sources in the NVSS, and
can be useful tools for measuring extinction, and estimating the
star-formation rate of stars too small to make SNe or HII regions \citep{Condon98b, Condon99},
\item Detection of radio stars and pulsars,
\item Serendipitous discoveries, such as the radio flares from ultra-cool dwarfs found by  \cite{Berger01}.
\end{itemize}
 
 In providing a sensitive, high-resolution continuum image of the Galactic Plane, EMU will complement the GASKAP HI Galactic HI survey \citep{Dickey10}.
 Existing interferometric radio continuum surveys of the Galactic Plane are either at high angular resolution but over a limited survey area, or cover a wide area at low angular resolution. For example, the MAGPIS  \citep{Helfand06} and CORNISH surveys  \citep{purcell2010} cover an area $\sim$100\sqdeg\  at an angular resolution of 1--6\arcsec, while the International Galactic Plane Survey consists of a number of studies over several hundred\sqdeg\  at a typical resolution of $\sim$1\,\arcmin\  \citep{naomi05, taylor2003,stil2006,haverkorn2006}. EMU, with its full sky coverage, high sensitivity, and $\sim$10\arcsec\ angular resolution, will bridge the gap between these two types of survey to reveal newly discovered populations of compact HII regions, planetary nebulae and young supernova remnants. When combined with the WODAN Northern Hemisphere survey (see \S \ref{others}), EMU will provide a complete census of centimetre-wave emission in the Galaxy.

At centimetre wavelengths the primary mechanisms for emission from Galactic objects are free-free emission from HII regions and planetary nebulae, synchrotron radiation emitted by supernova remnants, and diffuse synchrotron emission emitted by relativistic cosmic-ray electrons accelerated by SNRs. The known radio populations of each of these types of object are limited by a combination of issues including the limited area covered by existing surveys, frequency-dependent selection bias (in the case of optically thick HII regions), or biases against large scale structure introduced by limited \emph{uv} coverage snapshot surveys.

Thermal emission from HII regions may be separated from non-thermal synchrotron by using the correlation between thermal free-free and infrared emission  \citep[e.g.][]{Helfand06,thompson2006,conti2004}. The radio spectral index from EMU, and the polarisation data from POSSUM, will also be excellent discriminants between thermal and non-thermal emission.
The resolution of EMU is particularly well-matched to MIPSGAL 24 $\mu$m  \citep{carey2009} and Hi-GAL 70 $\mu$m  \citep{molinari2010} survey images, of which the latter is an effective tracer of the intensity of the exciting radiation field  \citep{compiegne2010}.

\subsubsection{Star Formation}
EMU's sensitivity allows access to all stages of the evolution of a compact HII region (hyper-compact, ultracompact and compact), even though the 1.4 GHz continuum emission will be optically thick, and EMU's 10 \arcsec\ resolution is insufficient to resolve the ultracompact HII regions.

Ultracompact HII  (UCHII) regions represent a young ($\sim$10$^{5}$ years old) phase in the development of an HII region   \citep{churchwell2002}. Most UCHII regions have been discovered by snapshot VLA surveys at 5 GHz  \citep[e.g.][]{woodchurchwell89,kcw94,walsh1998,purcell2010, urquhart09}. However, the snapshot surveys missed an entirely new class of optically thick HII region known as hypercompact HII (HCHII) regions   \citep{kurtz2005} and missed a diffuse component  \citep{kurtz1999,longmore2009} in which many UCHII regions are embedded.

\begin{figure}[h]
\begin{center}
\includegraphics[width=7cm, angle=0]{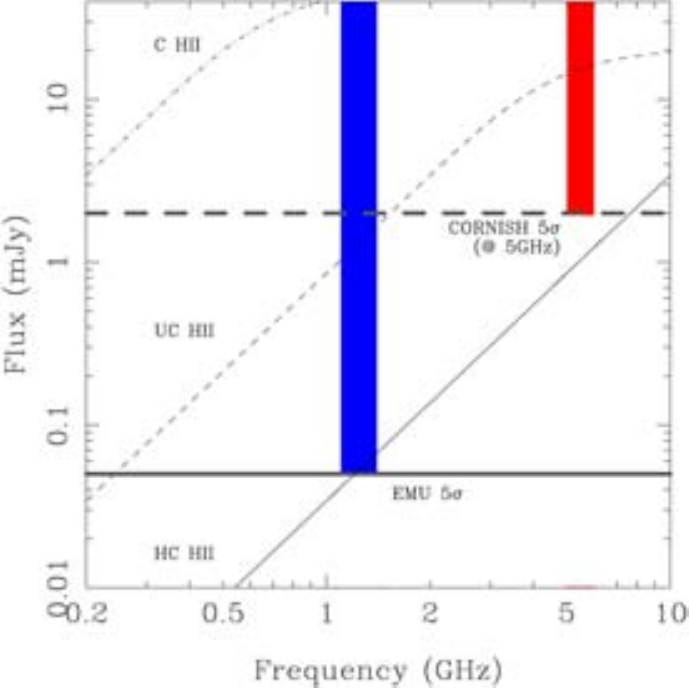}
\caption{Spectral energy distribution of HII regions over the frequency ranges of the EMU \& CORNISH surveys. The solid, dashed and dot-dashed lines represent the brightnesses of spherical
isothermal and homogeneous hypercompact (HC), ultracompact (UC) and compact (C) HII regions at an assumed distance of 18 kpc. The red band shows the frequency range of the 5 GHz CORNISH survey and the blue band the range of the 1.3 GHz EMU survey. $5\,\sigma$ detection limits are indicated by horizontal lines.}
\label{hii}
\end{center}
\end{figure}

To build a complete census of such objects requires a radio survey that is sensitive to both faint objects and also structure on scales of a few tens of arcsec. Fig.~\ref{hii} shows the typical spectral energy distribution expected from three types of HII region (compact, ultracompact and hypercompact) located at a distance of 18 kpc. In the optically thick regime the spectral index of free-free emission is proportional to $\nu{^2}$, falling to $\nu^{-0.1}$ in the optically thin regime  \citep{mezger1967}. For compact HII regions this turnover occurs at $\sim$1.5 GHz, whereas for the denser and more optically thick UCHII and HCHII regions the turnover frequency is shifted to higher values. This high-frequency turnover causes HCHII regions to be essentially undetectable with snapshot 5 GHz surveys (see Fig. \ref{hii}).
Even though HCHII regions are extremely optically thick at 1.4 GHz, the sensitivity of EMU is sufficient to detect them at distances up to 18 kpc, except in regions limited by confusing strong sources.
Consequently, EMU will be able to detect all types of HII region over most of the Galaxy, although higher resolution follow-up observations with other arrays may be needed to separate closely associated ultracompact and hypercompact HII regions. Many of these observations will be made as part of higher frequency surveys like CORNISH \citep{Purcell08}, and MeerGAL \citep{Thompson11}. Together with these surveys, EMU will be able to determine the turnover frequencies (and hence electron densities) of these regions.

\subsubsection{Supernova Remnants}
 
 Only $\sim$275 supernova remnants have so far been identified within the Milky Way  \citep{Green09}, out of an estimated total population of between 500 and 1000 remnants  \citep{Helfand06}.
 Although the \cite{Green09}
 catalogue represents the result of more than $\sim 50$ years of
 intensive searches with the world largest radio telescopes, it
 is considered to be incomplete and strongly biased by selection
 effects, such as the lack of compact (and hence young) and faint remnants, both of which should be addressed by the high sensitivity and angular resolution of EMU. There also exist strong synergies between EMU and the high-energy observatories HESS, {\it Fermi} \& {\it Chandra}, which may be used to confirm EMU non-thermal candidates as pulsar wind nebulae.

The increased surface-brightness sensitivity and higher angular resolution of EMU, combined with X-ray data from eROSITA  \citep{Predehl10}
 should allow the identification of more than 200 faint and diffuse X-ray sources which are currently classified as supernova remnant candidates  \citep{Becker09}.

\subsubsection{Stellar radio emission}

\begin{figure}[h]
\begin{center}
\includegraphics[width=7cm, angle=0]{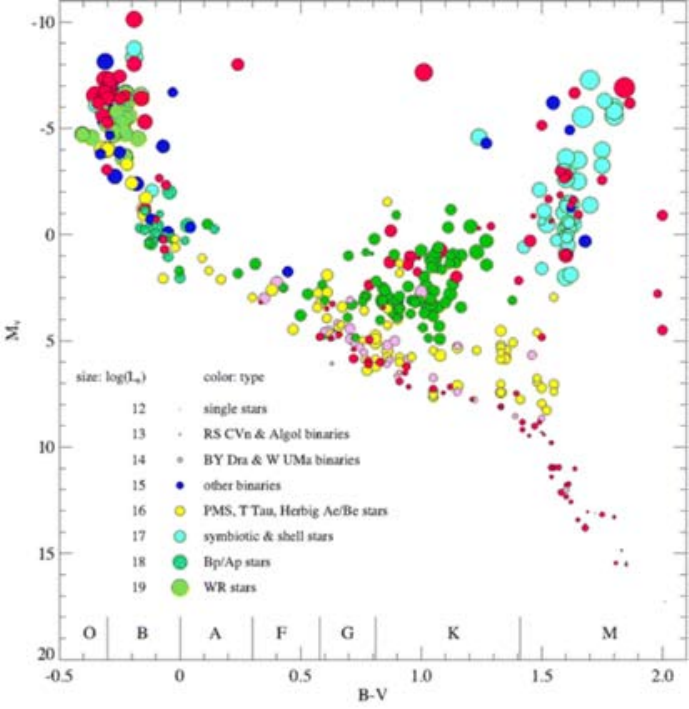}
\caption{H-R diagram of all known radio stars, taken from  \cite{Gudel02}. Radio stars can be found in almost every segment of the stellar H-R diagram.}
\label{stars}
\end{center}
\end{figure}

In the last few years, the improvement of observational capabilities has led to the
discovery of radio emission in a broad variety of stellar objects at diverse stages of their evolution, as shown in
Fig. \ref{stars}. In many cases, radio observations have revealed astrophysical phenomena not
detectable by other methods \citep{Gudel02}.

Radio stars have so far been detected only by targeted observations
directed at small samples of stars thought likely to be radio emitters.
Existing observations suffer
from limited sensitivity (e.g. no radio stars have been detected
at the radio luminosity of the quiescent Sun,
$L_{radio}\sim 10^{4} ~ \rm{W}\rm{Hz}^{-1}$),
and suffer from a strong selection bias in that observations
have been targeted to study
a particular aspect of stellar radio emission.
Consequently, it is difficult to forecast
the impact of EMU on the field
of stellar radio astronomy.

Much of our knowledge of radio stars comes from the study
of active stars and binary systems  \citep{Slee_1987, Drake_89, Umana_93,Umana_95, Seaquist93, Seaquist94, Umana_98, Budding_1999}.
The strongest radio emission appears to be associated with mass-loss and
magnetic phenomena and is often highly variable \citep{Gudel02, Dulk85, White_2004}.
Non-thermal radio emission can also originate from shocks of colliding winds in massive binaries
 \citep{Dougherty_2000}.

Transient events are also observed as
narrow-band, rapid, intense and highly polarised (up to 100~\%) radio bursts
 in stellar objects that have a
strong (and often variable) magnetic field and a source of energetic particles, including RS CVns and flare
stars  \citep{Slee_2008}, Brown dwarfs \citep{Hallinan_2008}, and
chemically peculiar stars  \citep{Trigilio_2000, Trigilio_2008, Ravi10}.
These radio flares have generally been
interpreted as a result of coherent emission mechanisms, including electron cyclotron maser emission and plasma emission.
This emission has been detected in only a
few tens of stars, partly because of the limited sensitivity of available instruments, but mainly due to the
absence of a deep all-sky radio survey.

These imperfect statistics suggest that EMU will provide an unbiased sample of several thousand stellar radio sources, enabling a detailed investigation into the physics of radio stars. This will include stellar physical parameters, magnetic activity, the fraction of active single stars and binary objects that show coherent emission, the time-scales of
its variability and their relationship to stellar parameters.
These are important  diagnostics for studies of stellar
magnetospheres, and can probe fundamental parameters
such as magnetic field intensity and topology, and electron energy distribution.

However, it will be essential to distinguish radio stars
from extragalactic sources using properties such as radio polarisation, SEDs, and  the proper motions available from  
Hipparcos, TYCHO-2 and UCAC3  \citep{Helfand_1999, Kimball09}.

\subsubsection{Pulsars}

The number of pulsars at high Galactic latitude is poorly known, and  \cite{Cameron11} have found that none of the IFRS detected in the ATLAS survey  \citep{Norris06, Middelberg08a} are pulsars, implying that the surface density of pulsars with a continuum flux level $> 150\mu$Jy at high Galactic latitude is $<$1 per 7\sqdeg. On the other hand, G. Hobbs (private communication) has estimated on the basis of other surveys that about one pulsar with a continuum flux level $> 50\mu$Jy should occur roughly every four\sqdeg\ , in which case EMU will detect about 8000 pulsars, which exceeds the total number of currently known pulsars. However, it will be difficult to distinguish these from other continuum sources. Diagnostics will include (a) lack of an optical/IR identification, (b) steep spectrum, (c) polarisation, especially circular polarisation (d) sidelobes caused by variability and scintillation during the observation. Candidates satisfying these criteria will will be searched for pulsar emission using the ASKAP COAST \citep{Ball09} project or conventional single-dish pulsar searches.

\subsection{Unexpected outcomes}
\label{WTF}

Experience has shown that whenever the sky is observed to a significantly greater sensitivity, or a significantly new volume of observational phase space is explored,  new discoveries are made.
 Even ATLAS, which expanded the phase space of wide-deep radio surveys by only a factor of a few, identified a previously unrecognised class of object (IFRS: see \S \ref{hizagn}). Because EMU will be much more sensitive than any previous large-scale radio survey, it is likely to discover new types of object, or new phenomena. Although it is impossible to predict their nature, we might reasonably expect new classes of galaxy, or perhaps even stumble across new Galactic populations. Furthermore, the large EMU dataset, covering large areas of sky, will be able to identify extremely rare objects.

Historically, discoveries of new classes of objects occur when an open-minded researcher, intimate with the telescope and with their science, recognises something odd in their data  \citep{Ekers10, Burnell10}. EMU is unlikely to be different: people will make unexpected discoveries while carefully using EMU data to test hypotheses. It is therefore arguable that this process needs no planning: any new class of object in the data will eventually be discovered anyway.
On the other hand, given the large volume of data, it is possible that a class of objects will lie undiscovered for decades because it didn't happen to fall within the selection criteria of any astronomer.
 
It is therefore important for EMU to plan to identify new classes or phenomena, rather than hoping to stumble across them. EMU is therefore taking the novel approach of developing data-mining techniques to identify those objects that don't fall into one of the known categories of astronomical object. A data-mining project (named Widefield ouTlier Finder, or WTF) is being established that will attempt to assign each object in the EMU catalogue to a known class of astronomical object, using the available cross-identifications to compare colours, luminosities, and any other available data. It will then identify those that are outliers or which depart systematically from known examples. Most such outliers will simply represent bad data, and are thus valuable in their own right for debugging ASKAP, while a few may be exciting new discoveries.

\subsection{Legacy value}
The largest existing radio survey, shown in Fig. 1, is the NRAO VLA Sky Survey (NVSS), whose release paper  \citep{Condon98} is one of the most-cited papers in astronomy. EMU will cover the same fraction ($\sim$ 75\%) of the sky as NVSS, but will be 45 times more sensitive, 4.5 times higher in resolution, with higher sensitivity to extended structures. As a result, EMU will detect $\sim$ 38 times as many sources as NVSS. More importantly, the greater sensitivity means that EMU breaks into a different regime. While most NVSS sources are radio-loud AGNs, EMU will provide estimates of star formation rate and radio-mode accretion activity in the galaxies currently being studied by mainstream astronomers at all other wavelengths.

The legacy value of large radio surveys depends on the ability to obtain a radio image of any object being studied at other wavelengths. Thus the legacy value depends critically on the area of sky covered: a survey covering the entire sky has more than simply twice the scientific value of a survey covering half the sky.
For example, the number of powerful (FRII-type) radio sources predicted at the epoch of reionisation is only $\sim$100 over the entire sky. 
The ability to identify radio counterparts (or upper limits) for any future observation (for example the highest redshift gamma-ray burst, supernova host galaxy, etc.) is essential. Serendipitous discoveries, too, are most likely to be maximised by surveying the greatest possible area. As a result,  EMU extends as far North as $+30\deg$ declination, and, together with WODAN (see \S \ref{others}), will cover the entire sky. The combined EMU and WODAN catalogues
are expected to become the primary radio source catalogue for all astronomers, and will not be superseded until after the SKA begins operation.

\section{EMU Design Study}
\subsection{The need for a Design Study }
In the two years prior to commencement of the EMU Survey,  the EMU Design Study is addressing the following challenges:
\begin{itemize}
\item Developing an optimum observing strategy, while taking into account the needs of potentially commensal surveys,
\item Working with the ASKAP computing group to ensure an optimum processing pipeline, and an optimum source extraction algorithm, including extraction of extended sources,
\item Addressing technical issues of dynamic range, confusion, etc.,
\item Developing an optical/IR identification pipeline, including a citizen science project,
\item Exploring the use of Statistical Redshifts (see \S \ref{redshifts}),
\item Developing the data access process, and identifying potential data issues,
\item Developing ``value-added'' projects to maximise the science return from EMU,
\item Refining the EMU Science Goals, to ensure that the EMU survey is optimised to address them, and planning early science papers.
\end{itemize}

The EMU Design Study  takes place through a number of working groups who each take responsibility for a number of goals, milestones, and deliverables.
The overall Design Study goal is to have a preliminary processing and analysis pipeline in place by November 2011, with a fully-functioning pipeline in place by November 2012.



\subsection{The role of ATLAS in the EMU Design Study }

An important component of the EMU  Design Study is the ATLAS project  \citep{Norris06, Middelberg08a, Hales11, Grant11} , which has a sensitivity, resolution and science goals similar to EMU, but over a much smaller area of sky (7\sqdeg\  surrounding the CDFS and ELAIS-S1 fields). The ATLAS survey, together with earlier observations of the HDFS  \citep{Norris05, Huynh05},  is being used as a test-bed for many of the techniques being developed for EMU, as well as guiding the development of the science goals and survey strategy.
 
 For example, the prototype EMU source extraction and identification pipeline will be used for the final ATLAS data release in late 2011. During the course of the ATLAS project, dynamic ranges approaching $\sim10^5$ have been reached, which are amongst the highest achieved in radio astronomy, and  imaging artefacts have been encountered which have been rectified in some specific cases, and a solution for the general case is being tackled. Such imaging problems must be solved not only for EMU, but for other SKA Pathfinder projects.

\subsection{Science Data Processing}

The data from the correlator will be reduced in an automated processing pipeline
 \citep{Cornwell11}, which includes the following steps for EMU:
\begin{itemize}
\item Flag for known radio frequency interference, working from a database of known RFI sources,
\item Identify unknown radio frequency interference, saving candidate identification, and identify and flag further bad data,
\item Solve for calibration parameters (i.e. frequency-dependent complex gains)  using least squares fits of the predicted visibility (from previously obtained model of the sky) to the observed visibility,
\item Apply calibration parameters, predicting forward from the last previous solution,
\item Average visibility data to required temporal and spectral resolution,
\item Construct an image by (a) gridding the data using convolutional resampling, (b) Fourier transforming to the image plane, and (c) Deconvolving the point spread function,
\item Find sources in the resulting image or cube,
\item Save science data products to the ASKAP Science Data Archive.
\end{itemize}

The measured flux densities from most radio surveys have a typical accuracy of 10\%, although the calibration accuracy should, in principle, enable flux densities to be measured to an accuracy of 
$\sim$ 1\%. This latter figure has been adopted as the target for EMU, and the EMU Design Study will investigate the reason that most
surveys fail to achieve this figure. A number of
instrumental effects and completeness corrections have been identified \citep{Hales11}, and may need to be
applied to the data, including the following.

\begin{itemize}
\item The primary beam model must be accurate to $\sim 1\%$ so that the data can be corrected for the primary beam response to this accuracy. In the case of ASKAP, this will vary across the PAF, and so needs to be determined for each of the 36 beams.

\item Position-dependent bandwidth smearing (chromatic aberration) over the mosaiced
 image due to the finite bandwidth of frequency channels  \citep{Condon98, Ibar09}.
 Surprisingly, mosaiced images suffer more from bandwidth smearing than pointed observations, since any location in the image, even at the centre of a pointing, may include many contributions from adjacent pointings in which bandwidth smearing is significant. Thus, even for tight angular spacing between pointings in a mosaic, bandwidth smearing will always be non-zero at any location over the final mosaiced image.  \cite{Hales11} found that bandwidth
smearing in central regions of the ATLAS mosaiced images, which used 8 MHz channels, typically caused a 10\% decrement in peak flux due to
this overlapping effect. Even with ASKAP's 1 MHz channels, this correction will be necessary if  1\% accuracy is to be achieved.
 
 \item Position-dependent time-average smearing over the mosaiced image
 due to finite integration time and Earth rotation  \citep{Bridle99}. This is thought not to be a significant issue for ASKAP, but is included here for completeness.

\item Clean bias in measured fluxes, which may be mitigated by suitable
 weighting schemes (to control beam sidelobe properties) and cleaning depth.
 
\end{itemize}

In addition, there are observational effects which may bias the fluxes of individual sources and/or the statistical properties of the sample as a whole, as follows.
\begin{itemize}

\item The probability of measuring a faint source located on a noise peak is higher than the probability of measuring a strong source located in a noise trough, because faint sources are more numerous \citep{Hogg98}. This results in a bias on all measured source fluxes, which decreases with increasing signal-to-noise. When considered in terms of source counts this is known as Eddington bias \citep{Eddington13, Simpson06}. \cite{Hales11} have shown that Eddington bias can also be a sensitive probe of number counts below the flux limit of a survey.

\item Resolution bias due to the lack of sensitivity to resolved sources, which can be manifested in two ways. First, the lack of short baselines can limit the maximum angular scale observable, although this is not likely to be a problem for EMU. Second, faint resolved sources may have integrated fluxes sufficient to be included in the final catalogue, but may be missed in the source extraction process because their peak fluxes fall below the signal-to-noise source detection threshold \citep[e.g.][]{Prandoni01}. This necessitates a resolution bias correction.

\item Sensitivity will generally vary across an image, so that the area surveyed to any limit is a function of that limit.   Since source counts must be normalised by the area surveyed, a completeness correction must be applied  to account for the position-dependent
 sensitivity to faint sources across the mosaiced image  \citep[e.g.][]{Bondi03},
 which must also take into account the position-dependent time-average and bandwidth-smearing effects \citep{Hales11}. For the same reason, large variations in sensitivity across the field are likely to lead to uncertainties in source counts, and so it is essential to measure the sensitivity across the field to high accuracy, and keep it as uniform as possible.

\end{itemize}

\subsection{Simulations and Imaging Pipeline}
\label{simulations}
The goals of the simulations and imaging pipeline working group are to:
\begin{itemize}
\item Develop a realistic extragalactic simulated sky,
\item Develop a realistic Galactic simulated sky,
\item Work with ASKAP engineers to choose optimum PAF configuration and weighting scheme,
\item Ensure ASKAP imaging processes are robust to Galactic-Plane observations,
\item Optimise imaging algorithms and parameters,
\item Ensure that EMU data will reach a dynamic range of $10^5$,
\item Ensure that EMU measured flux densities are accurate to 1\%.
\end{itemize}

\subsubsection{Simulation Analysis}
\begin{figure}[h]
\begin{center}
\includegraphics[width=7cm, angle=0]{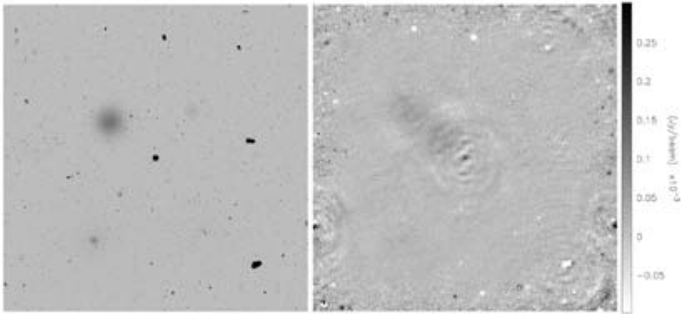}

\caption{Part of the simulated sky as observed by the simulated ASKAP telescope and processed using {\em ASKAPsoft}. (Left) The input model sky. (Right) The difference between the observed sky image and the input sky showing artefacts caused by the observing process and deconvolution errors. The intensity scale of both images is increased to highlight these errors.
}
\label{simulation}
\end{center}
\end{figure}

\begin{figure}[h]
\begin{center}
\includegraphics[scale=0.4, angle=0]{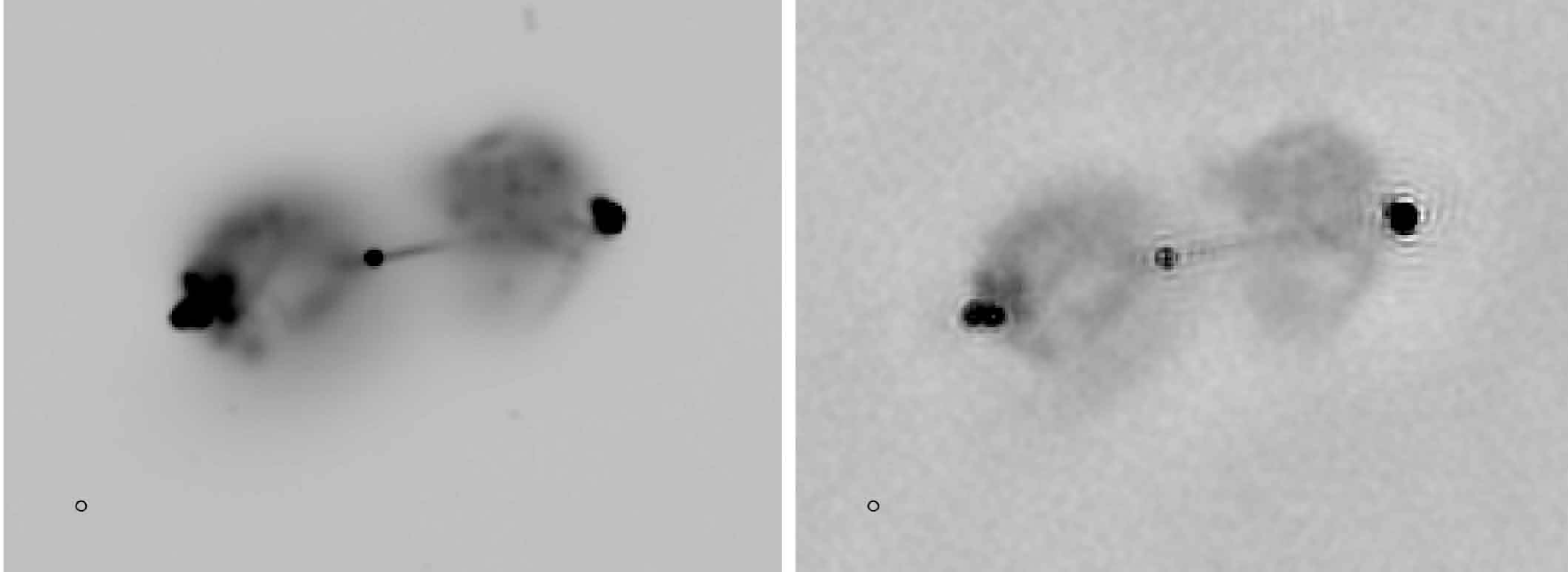}

\caption{ (Left) The simulated sky around Pictor-A as observed by the simulated ASKAP telescope and processed using
{\em ASKAPsoft}. The greyscale ranges from -0.01 to 0.03 Jy/beam, with a peak of 0.9 Jy/beam. (Right) The difference between the observed sky image and the input sky showing artefacts. This image is enhanced by a factor of ten to show up artefacts, so the greyscale ranges from -0.001 to 0.003 Jy/beam.
}
\label{pictor}
\end{center}
\end{figure}

Simulations have been performed using the SKA  Design Study (SKADS) simulated sky  \citep{Wilman08} as initial input. 
This sky is augmented by the addition of extended, diffuse, or complex sources, which are not fully represented in the SKADS sky. This sky is then used to generate \emph{uv} data similar to that which will be produced by ASKAP, using the antenna, beam and array characteristics. Thermal noise is added to the visibilities but calibration is assumed to be ideal. The data are then processed using the {\em ASKAPsoft} processing pipeline  \citep{Cornwell11} which is the prototype version of the ASKAP processing software. Current simulations use all 36 antennas, the full 300 MHz bandwidth, and 8 hours of on-source time with approximately uniform weighting. A typical observed simulated sky is shown in Fig. \ref{simulation}. Results so far suggest that the majority of fields will reach the required sensitivity and dynamic range, and the rest of this section considers those few remaining fields containing strong sources, such as that shown in Fig. \ref{pictor}, where this may not be the case.

The analysis of the simulations focusses on the  difference image, which is the difference between the dirty image and the input model convolved with the point spread function (PSF). In regions containing no strong sources, the rms noise in the difference image  is close to the value predicted from receiver performance (i.e. 10 $\mu$Jy for an EMU image). However, it is higher in regions containing bright or extended sources, because of the following effects.
\begin{itemize}
\item Insufficient sampling in the image domain, particularly for bright point sources that are not pixel-centred  \cite{Briggs92}. Simulations suggest that  4 pixels/beam will limit the dynamic range to $\sim$ 2000-3000 in these regions, and so $\sim$ 5-6 pixels across the beam are planned to achieve the required $10^5$ dynamic range..
\item Insufficient scales for the multi-scale CLEAN. The three scales used in earlier simulations were insufficient to deconvolve the more extended sources in the field, and so five scales (as used in the simulation results shown in Fig. \ref{simulation}) will be implemented in the final processing pipelines.
\item A large CLEAN threshold. A CLEAN threshold of 1 mJy/beam was used in the simulation, which will leave residual lobes of $\sim$ 20 \ujybm\  around sources slightly weaker than 1 mJy/beam. A lower value of CLEAN threshold will be used for ASKAP.
\end{itemize}

As described in \S \ref{askap}, ASKAP will generate a sky model of sources stronger than $\sim$ 1 mJy, which will be subtracted from the observed $uv$ data before imaging. As a result, many of the effects described above are not expected to be significant in ASKAP data, but simulations are continuing to test this hypothesis.


\subsubsection{Future Simulations}
A number of simulations are needed to ensure that the pipeline will deliver EMU science:
\begin{itemize}
\item Simulate low surface-brightness large-scale \\
structures (e.g. Fig. \ref{pictor}) and test the effectiveness of the pipeline in retaining these structures,
\item Ensure that preconditioning (see \S \ref{taper}) can recover the source structures required for EMU science,
\item Simulate faint (well below the rms) low surface-brightness large-scale structures to check that they can be recovered using correlation and stacking techniques,
\item Test deconvolution effectiveness on sources with curved radio spectra.
\end{itemize}

\subsubsection{Tapering}
\label{taper}
Radio synthesis imaging offers a trade-off between naturally weighted images (to optimise sensitivity) and uniformly weighted (to optimise resolution) images. The ASKAP processing software will use preconditioning \citep{Cornwell11}, rather than traditional weighting, and multi-scale clean  \citep{Cornwell08}, which reduces the impact of these effects. Nevertheless, in effect both naturally weighted and uniformly weighted data may be required, implying two processing pipelines for continuum data. Extended structures may require an even more heavily tapered weighting. Future simulations will explore potential solutions, including the use of Briggs robust weighting that may provide an optimal compromise between the two. They will also explore the possibility of off-line processing to generate more heavily tapered images than those produced in the real-time pipeline, and to determine the optimal on-line weighting to facilitate this with minimum loss of information.

\subsection{Source Extraction and Measurement}

Several radio source extraction tools are widely used, such as the {\em Miriad} /AIPS Gaussian fitting routines
IMSAD, SAD and VSAD,
Sextractor \citep{BA:96}, SFIND  \citep{Hop:02} and Duchamp  \citep{Whiting08}. There
are also numerous specific tools, such as
HAPPY (a modified version of SAD used in the
FIRST survey  \citep{Whi:97}, a machine-learning  back-end to VSAD,
used to construct the SUMSS catalogue  \citep{Mau:03},
BDSM (used for the LOFAR source-finding, N. Mohan, in prep), and the flood-fill algorithm being used in the ATLAS Data Release 2  \citep{Murphy07, Hales11}.

However, none of these is fully adequate for EMU, particularly  (a) in extended
sources, (b) in  the presence of artefacts from bright sources, and (c) in images
whose signal-to-noise ratio varies across the image. It appears that
an optimum strategy has not yet been constructed, particularly for extended sources, and there is still no truly general-purpose and robust
source-finding tool appropriate for the application to large-area
radio surveys. Moreover, there has not been a published
analysis that thoroughly and quantitatively compares the details of
source-finding (from defining thresholds, identifying pixels to
fit, fitting approaches, approaches to multiple-component and
extended sources, etc.) as implemented in different existing tools.

Since the size of the EMU survey precludes manual intervention in the source
extraction process, it is necessary to build efficient source extraction algorithms
that are robust to varying signal-to-noise across the image, sensitive to
source scales ranging from unresolved through complex and multiple-component
sources, to extended low-surface-brightness structures, and that require no
manual intervention. The Design Study is therefore building on existing expertise to identify the optimum
approach to source-finding, and produce a robust
continuum source finder for both point sources
and extended emission.

The primary goal is to develop an algorithm (or suite of algorithms) that
will be implemented both in the {\em ASKAPsoft} data reduction pipeline and also in
at least one of the standard radio-astronomical reduction packages, such as
 {\em casa\/}, as well as being published in a journal
paper. The software tool or tools will also be made available for use by all
members of the astronomical community for processing data from other
instruments.

Specific goals are to develop optimum source identification and measurement strategies and
algorithms, quantify the effects and limitations of confusion, and  identify mitigation
strategies for source-measurement of overlapping or multiple sources.

The problem is split into two
parts, one dealing with point sources (including marginally extended
sources still well-modelled by Gaussian fits), and the other with extended
and complex source structures. These initial approaches will be extended to include more sophisticated
investigations, contrasting wavelet or other decomposition approaches
with machine-learning and genetic algorithms, in particular for the extended
and complex source identification case, where information available at other
wavelengths may also be introduced.

An important part of this project is to build linkages with other survey groups
 through SPARCS (see \S \ref{sparcs}) so that the various
efforts underway worldwide can be coordinated, taking advantage of best practice where
established, and combining resources where practical.

\subsubsection{Compact Source extraction and measurement}

The point-source investigation is comparing different
background subtraction approaches (e.g., the polynomial mesh of SExtractor,
and the erosion/dilation algorithm of  \cite{Rudnick02}), different approaches to thresholding (e.g. simple ``canonical''
thresholds of $n \sigma$,  and False Discovery Rate (FDR) methods
 \citep{Hop:02}),
 and different approaches to Gaussian fitting.
 
This includes a comparison of existing source-finding and
measurement tools, to identify their advantages and disadvantages, and to determine whether improvements can be incorporated
into the existing ASKAP source-finder \citep[Duchamp: ][]{Whiting08}. For each tool, the completeness (fraction of artificial sources identified) and reliability
 (fraction of identified sources that are true model sources, as opposed to
false detections) are measured as a function of flux density.

While all source detection algorithms become incomplete and unreliable
at low signal to noise, detailed comparisons can identify
whether particular source finders are more
reliable at fainter levels  Furthermore, algorithms with the same completeness and
reliability can produce different source catalogues. Understanding how
these catalogues differ, and how they affect the science of
the ASKAP projects, is currently being studied \citep{Hancock11}.

The primary result so far has been to recognise the importance
of accurately measuring the background noise level, which may vary across the field because of artefacts, diffuse emission, and varying sensitivity, and subtracting any varying low-surface-brightness emission.
Given the size of ASKAP images, this background variation often appears on
scales larger than the filters coded into standard source-finders.
Errors on the measured flux densities are clearly correlated with the background structure, which can be removed  using a filter such as that shown in Fig. \ref{spatialfilter}. However, it may be challenging to determine an appropriate scale size for the filter that can be used automatically and robustly in EMU observations.

The next step is to understand why sources are identified by one
source-finder but not by others. This is often caused by the
way tools group pixels into objects, such as objects assigned as
extended by SExtractor, but measured as multiple separate sources by Gaussian
fitting tools such as VSAD and Sfind. This analysis will then be extended to
explore those objects that are poorly fitted by any of the source-finding
tools, to produce a list of recommendations for implementation in the ASKAP
source detection pipeline to achieve optimal source-extraction and
measurement.

\subsubsection{Extended and Complex Source Extraction and Measurement}

\begin{figure}[h]
\begin{center}
\includegraphics[scale=0.25, angle=0]{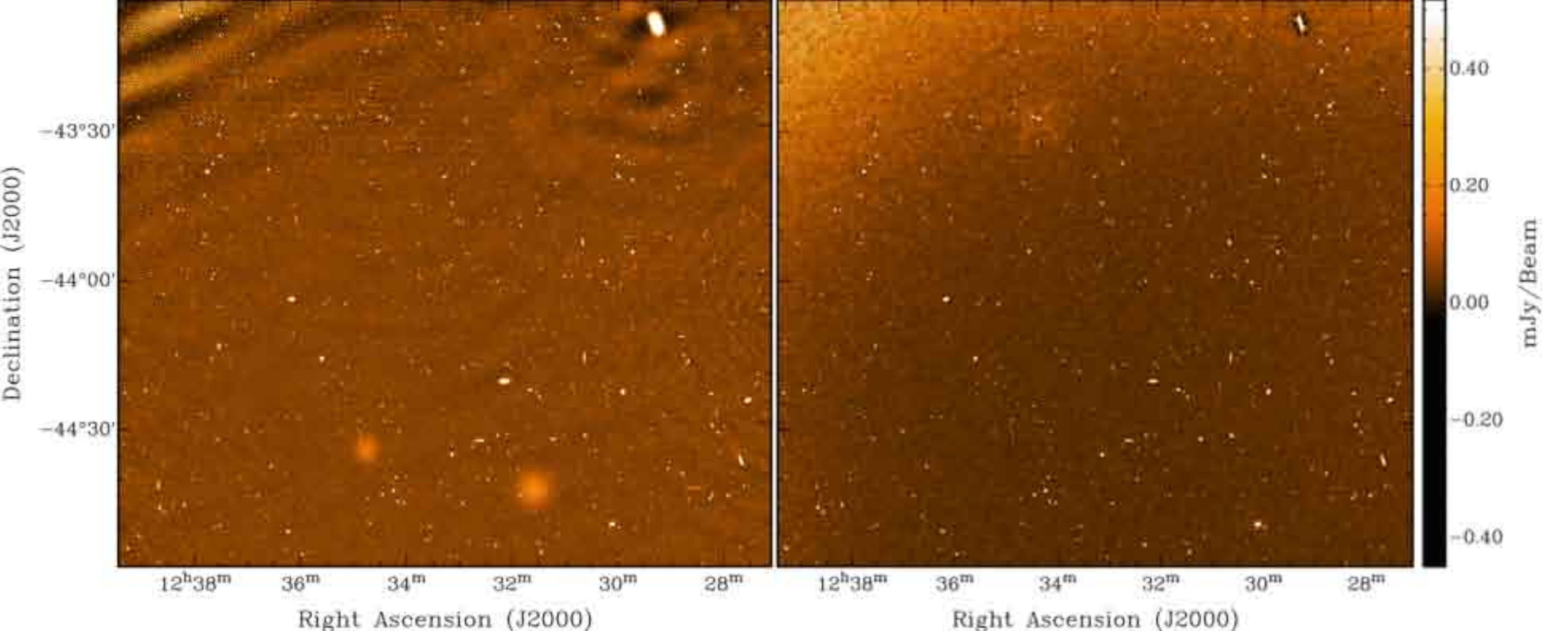}

\caption{Left: a portion of the original SST3 continuum simulation, including artefacts in the top left-hand corner caused by a strong source just outside the field; Right:
result of applying a spatial filter to remove large-scale structure. Note that
the diffuse Gaussians in the South are also completely removed. Both images have the same scaling, with an rms of $\sim 15$ \ujybm.
}
\label{spatialfilter}
\end{center}
\end{figure}

The investigation into optimising detectability of
extended structure is exploring a variety of filters that maximise
the detectability of extended structure, such as
 a
spatial scale filter \citep{Rudnick09} to identify extended
emission, as shown in Fig. \ref{spatialfilter}.

Other approaches include the use of the Hough transform \citep{Bal:81} to
measure circular or elliptical structures (such as SN remnants, or radio relic sources),  
and using
compressed sensing imaging techniques to radio interferometric
data \citep{Wiaux09}. The study  focusses on metrics to
compare techniques, and implementing suitable model
sources in the simulations.
Each  technique seems to hold promise for different kinds of structure
and measures, and the outputs are now being quantified and compared, to
identify a suitable implementation for the ASKAP pipeline.

\subsection{Survey Strategy}
\label{strategy}
One challenge in planning a large survey like EMU is to develop the optimal survey strategy to achieve the primary science goals, while taking into account the capabilities and limitations of ASKAP and the need for the most efficient use of telescope time. Prior to commissioning, several key issues must be considered.

\subsubsection{Source Density and Confusion}
\label{s_confusion}

\cite{Condon08} has shown that a beamsize of 10 \arcsec\ is near-optimum for ASKAP in terms of the trade-off between resolution and confusion, and that the rms sensitivity is increased by $\sim$ 10\% by confusion noise.
This is consistent with previous observations \citep{Hopkins03, Norris05, Huynh05} which have shown that the confusion level at the rms sensitivity (10 \ujybm) and resolution (10 \arcsec) of EMU  is $\sim$ 50 synthesised beams per source, so that only a small fraction of EMU sources will be affected by confusion.
Nevertheless, several science goals of EMU, in particular those focusing on diffuse emission and stacking experiments, rely on solid understanding and mitigation of confusion and noise variations. 

\begin{figure}[h]
\begin{center}
\includegraphics[width=6cm]{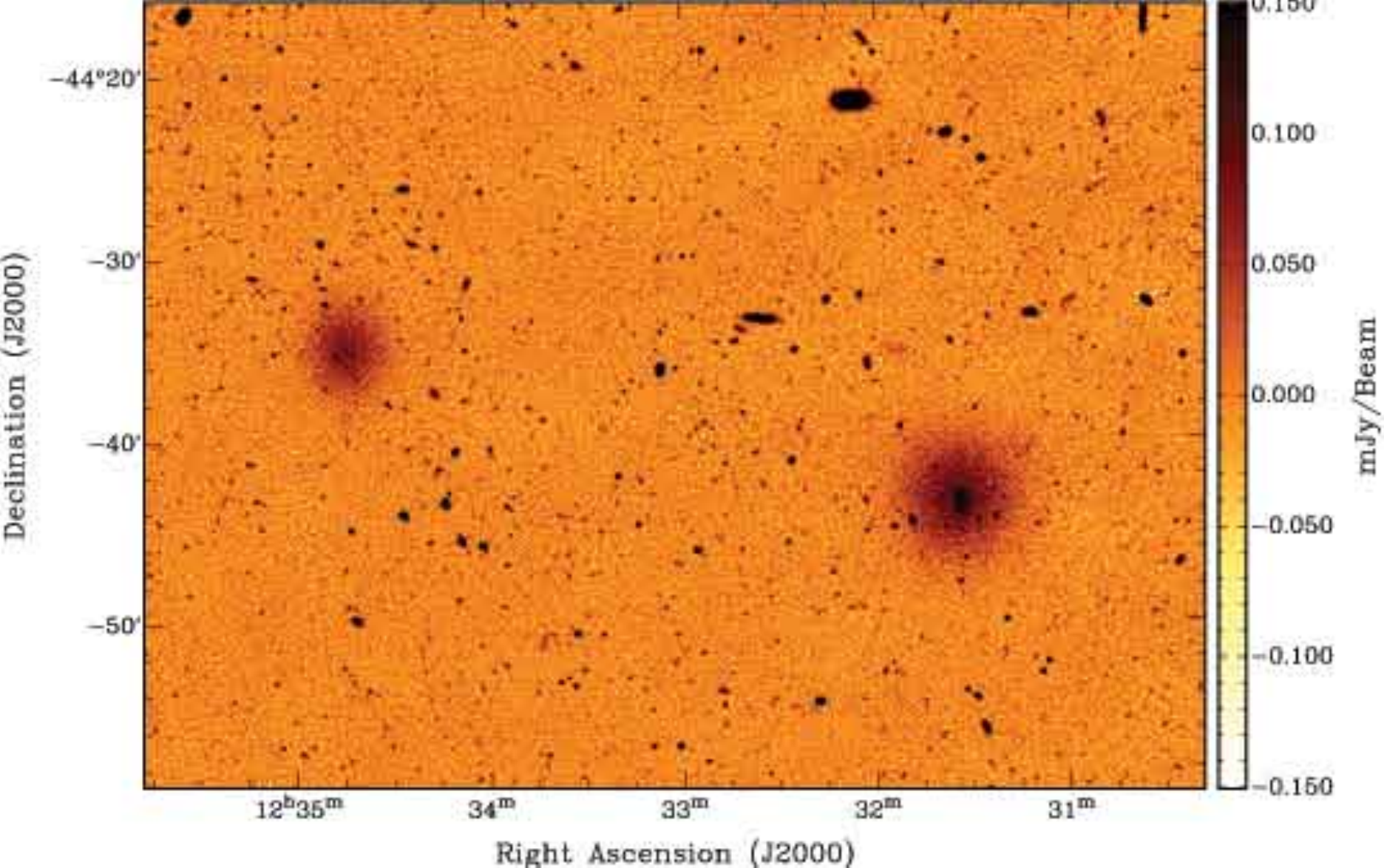}
\includegraphics[width=6cm]{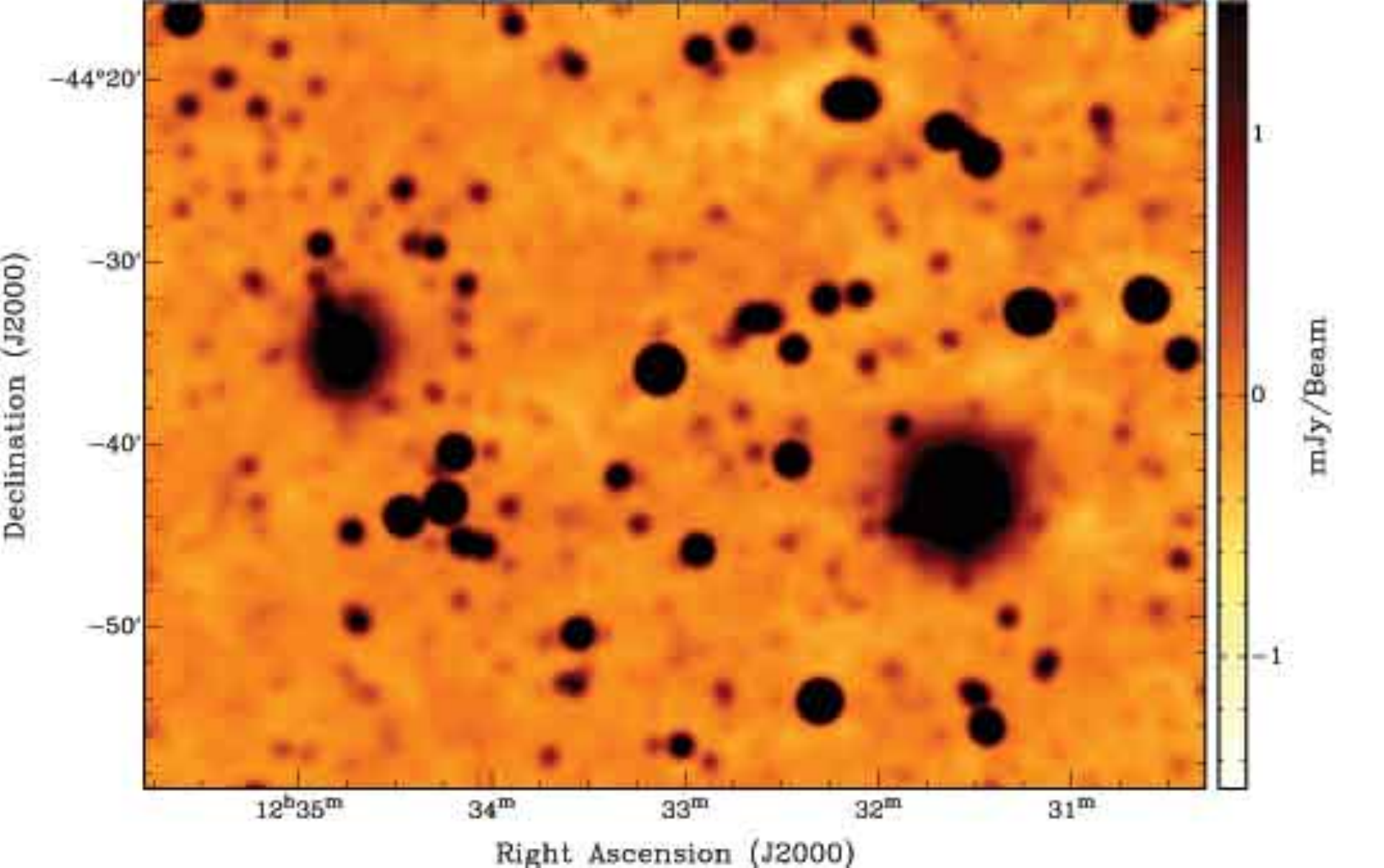}
\includegraphics[width=6cm]{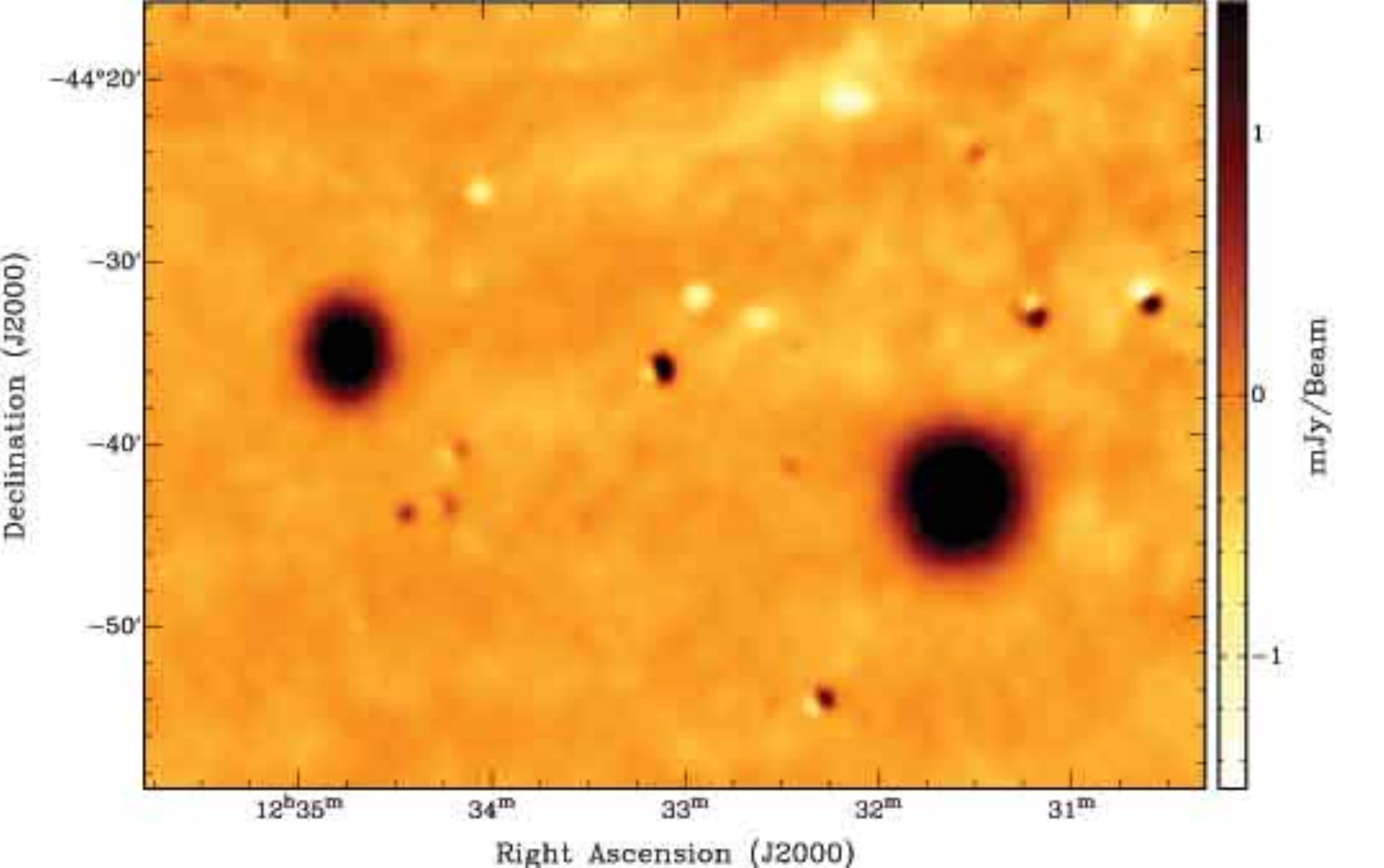}

\caption{Top: Original 10-\arcsec\ resolution ASKAP simulation with two diffuse radio halos injected; Centre: Same as above convolved to 60\arcsec\ resolution. The image is confusion limited at 40-100~\ujybm; Bottom: Same as above but with the SKADS source model subtracted (with errors). This image reaches thermal noise on small scales, with a typical rms of 20-50~\ujybm.
}
\label{confusion}
\end{center}
\end{figure}

Simulations conducted by the ASKAP computing team, supplemented by model diffuse sources, confirm that confusion is not the limiting factor at 10 \arcsec, but will be an issue at the 1 arcmin resolution ideal for the detection of large-scale diffuse emission. Compact source subtraction and filtering is being explored in order overcome this issue. Initial experiments (e.g., Fig. \ref{confusion}) have improved our understanding of these issues, and will inform the diagnostics needed in both the science and imaging simulations.

\subsubsection{Choice of Observing Frequency}
EMU will observe a single contiguous band at 1130-1430 MHz, which is a compromise between (a) the reduced PAF beam filling factor above 1400 MHz, (b) the poorer spatial resolution at lower frequencies, and (c) the desire to observe commensally with the WALLABY project, for which the frequency range is determined by redshifted hydrogen. On the other hand,  other factors (design constraints, choice of commensal observing programs) may necessitate revising this choice, and some continuum data may also be available as a by-product from other ASKAP projects (e.g. POSSUM, DINGO, FLASH) . Therefore the EMU Design Study simulations are including observing frequency as a free parameter.

\subsubsection{Observing Strategy}
To achieve the expected sensitivity in EMU it will be necessary either to
integrate at each pointing for $\sim$12h and then move over to the next pointing, or
take many short integrations at each pointing.
The first option allows the required depth to be reached without requiring any historical data, and is robust to variability (on timescales $>$12h), and changes in the dirty beam caused by instrumental failures. It will also allow an early identification of any instrumentation or systematic issues that may restrict our dynamic range. The second option is difficult to implement as it requires the processing pipeline to have access to historical data, but will detect variability, and will provide graceful degradation to array faults. For example, a hardware failure on a particular day may result in a reduction of sensitivity over several fields, rather than causing a total loss of one field.  Option 1 is maximally commensal with WALLABY while option 2 is maximally commensal with VAST. It is likely that a variant of Option 1 will be adopted, augmented by some dithering or interlacing to achieve the necessary sensitivity uniformity (see \S \ref{uniform}).

\subsubsection{Tiling Strategy}
Because of computational overheads, it will be inefficient for the ASKAP pipeline processor to combine data from overlapping pointings obtained on different dates. The loss of observing efficiency caused by overlapping pointings is minimised by the square feed array that has been adopted for ASKAP.
Initial tiling experimentation has begun with tiling stripes of constant declination, motivating a coordinated EMU/WALLABY move towards a unified formalism.

The trade-offs between tiling strategies and PAF weighting schemes are not yet fully understood. Weighting schemes may be optimised for (a) maximum field of view, (b) maximum sensitivity, (c) uniform sensitivity, or (d) best polarisation response, and that decision may influence the tiling strategy. A further factor is the need to combine (or ``stitch'') images and catalogues across adjacent fields, which may also have some impact on tiling strategy.


\subsubsection{Dynamic Range}
Dynamic range may be the biggest factor in limiting
the uniformity and completeness of EMU, and it drives many of the
ASKAP design requirements, such as the 3-axis telescopes, the strict pointing
requirements, and the good gain stability of the antenna front-end subsystems. It also drives much of the current EMU design study, such as the simulations (see \S \ref{simulations}).

Each 30\sqdeg\  field will contain about two sources stronger than 1\,Jy. The target rms noise is 10 \ujybm, and so in principle  a dynamic range of $\sim10^5$ across the 30\sqdeg\ FOV is required. Although the effect of a strong source on adjacent beams is currently unclear, the 1\,Jy sources will not be present in all the sub-beams within the 30\sqdeg. Therefore, even if dynamic range limits some sub-beams,  10 \ujybm\  rms should still be reached in the majority of sub-beams. Simulated ASKAP data is being used to test these effects and others, such as the effect of strong diffuse emission from the Galactic Plane. Other issues include the optimum weighting of the PAF feeds, the required uniformity across the beam, the optimum observing strategy, and choice of observing parameters such as weighting. So far, simulations (see \S \ref{simulations}) suggest that the dynamic range will not be limited by the data processing, but  an unambiguous answer must await the characterisation of the PAFs in the BETA observations.

\subsubsection{Sensitivity Uniformity}
\label{uniform}
Sensitivity can in principle vary significantly across the field of view of a single pointing, although this will be mitigated by a planned dithering pattern that should reduce non-uniformity to below 10\%. Nevertheless, it is important to estimate what the required uniformity is for EMU science goals before a final observing/dithering strategy is reached. We consider two cases.
\begin{itemize}

\item Galaxies, AGNs, \& Star Formation: Completeness is a critical component, and science goals permit some sensitivity variations provided the target 10~\ujybm\  rms is reached at the completeness cutoff. However, large sensitivity variations result in loss of observing time: a 20\% gain variation in one spot would require the completeness threshold to be set 20\% higher across the field, and since {\em sensitivity $\propto$ (observing time)$^2$}, this will result in $\sim$ 40\% of the data being discarded.

\item Diffuse emission: The scale size of any sensitivity variations must be much smaller or much larger than the diffuse emission of interest, otherwise the variations act as confusion noise. Simulations have shown that variations larger than $\sim$20-25\,\arcmin~can be effectively removed through spatial scale filtering. Noise variations on scales between 2 and 20\,\arcmin~are the most problematic, though  detailed knowledge and modelling of the variations will enable the data to be partially corrected for such variations before diffuse source extraction.

\end{itemize}

\subsection{Commissioning and Science Verification}
The EMU team will play an active role in the ASKAP science verification tests, which will include observations of a well-studied field with existing 20cm data with the same sensitivity and resolution as EMU. In addition,  that field will be observed regularly throughout the commissioning and operational life of ASKAP. Not only will such a process enable checks against images and results obtained with other instruments, but it will also provide a regular check on stability, and give warning of subtle problems that may arise during the lifetime of ASKAP.
 
These observations will also result in a set of radio data which can be co-added to provide an extremely deep field down to the confusion limit, and so the field should be chosen to maximise the availability of multi-wavelength data.  Probably the best studied field in the sky is  that surrounding ATLAS-CDFS, which offers:
\begin{itemize}
\item 30\sqdeg\  which has been mapped with the VLA at 20cm as part of NVSS to an rms of 0.45 mJy/beam,
\item 4\sqdeg\  which has been mapped with the ATCA at 20cm as part of ATLAS, to an rms of 10\ujybm, reproducing the rms, resolution, and dynamic range that are closely matched to the Design Study goals of ASKAP, including polarisation, spectral index, wide-field VLBI, and variability studies,
\item Probably the most intensely studied field in the sky at optical IR and X-ray wavelengths, resulting in important science from the repeated (and hence deep) ASKAP observations,
\item Areas of sky which have been mapped to an rms of 10\ujybm\  situated a few degrees away from sources with flux densities $>$ 1\,Jy, which will be invaluable for checking dynamic range.
\end{itemize}

The first observations will take place on BETA (see \S \ref{askap}). Although the instantaneous sensitivity and spatial resolution of BETA are modest, it will be unmatched for surveying large diffuse structures until the advent of the full ASKAP array. It is not yet clear whether BETA will be fully committed to engineering tests, or whether some science might be achievable with BETA. For the latter eventuality, science targets  have been chosen that can yield early science publications from ASKAP, including observations of clusters and superclusters including large diffuse galaxies, relics, and low-surface brightness shock structures.

Once the full ASKAP array is operational, the EMU survey will start with a number of stringent data quality checks before being released into the public domain, and this Quality Control process, described in \S \ref{quality}, is the responsibility of the EMU team.
In particular, it is planned to observe one or more fields that have been deeply surveyed by other telescopes, such as the ATLAS and COSMOS fields, to check for consistency of results and calibration. It is planned also to survey a large Northern field at a declination about $+25\deg$ which will also have been observed with the EVLA, and which will subsequently be observed with Northern surveys such as WODAN to ensure consistency between the surveys.

\subsection{Data Format and Access}
\subsubsection{EMU Data Requirements}
An EMU data requirements document has been continuously updated since the earliest days of EMU. The resulting list of primary data products (termed ``level 5 products'' in ASKAP documentation), is summarised in Table \ref{datatable}. However, new requirements continue to arise. For example, a recently recognised requirement is to optimise EMU data delivery for stacking. To correctly interpret stacking results requires detailed knowledge of the point-spread function at each point of the image, resulting in further data requirements such as (a) storing uncleaned residual images without cleaned components being added back in, (b) storing (or being able to reconstruct) an accurate representation of the dirty beam (which varies significantly across the field) at each point in the image.

In addition to the science-driven data requirements, discussed here, a number of statistical tests will be made on the data to check the data quality, as described  in \S \ref{quality}.

\begin{table}[h]
\begin{center}
\caption{EMU Primary Data Products}
\begin{tabular}{ll}
\hline
UVDATA\\
\hline
\emph{uv} data will be available in all four Stokes \\
parameters, and in all 300 frequency channels.\\
\hline
IMAGES (in IQUV)\\
\hline
An intensity image,\\
A coverage and sensitivity map \\
An uncleaned residual image$^a$,\\
Images of spectral index and spectral curvature,\\
Point-spread functions across the image$^b$,\\
Detailed metadata for all the above\\
\hline
CATALOGUE OF COMPONENTS$^c$\\
\hline
A unique IAU-registered component name,\\
Ra, Dec (J2000) of the peak,\\
A peak and integrated flux density,\\
The spectral index of the peak and integrated flux,\\
The parameters of a fitted Gaussian,\\
For failed Gaussian fits, an estimate of the extent \\
~~~~of the component,\\
A postage-stamp image,\\
A history of changes to the catalogue entry.\\
\hline
\end{tabular}
\label{datatable}
\medskip\\
\end{center}
$^a$The residual image after cleaning, in which cleaned components, and sources in the ASKAP sky model, have not been added back in. This image will be valuable for stacking and for detecting diffuse emission.\\
$^b$ The point spread function of the uncleaned emission, which will vary across the FOV. It is important to know this point spread function at any point to calculate the statistics of stacked images (see \S \ref{stacking}).\\
$^c$ All measured quantities will be accompanied by an estimate of their standard error.
\end{table}

\subsubsection{Value-Added Data}
EMU has an ambitious plan to provide source associations and identifications in near real-time, and place them in the public domain. These value-added data constitute the ``Secondary Data Products'', or, in ASKAP terminology, level 7 data. Their generation includes grouping radio components into sources (e.g. classical radio double and triple sources etc) and cross-identifying with optical and infrared surveys. EMU is also collaborating with the Galaxy Zoo group  to extend this process into the Citizen Science arena (see \S \ref{zoo}). To achieve this,
collaborations have been initiated with the appropriate surveys,  and discussions are taking place on how best to produce these associations and cross-identifications, using a probabilistic approach. In addition, NED (NASA Extragalactic Database) have agreed to accept a near-real-time automated query for this process, and have also agreed to host the resulting source catalogue and associations.

\subsubsection{Virtual Observatory Server}

Management of primary ASKAP data will be the responsibility of the ASKAP Science Data Archive Facility (ASDAF). However, existing Virtual Observatory (VO) standards have not yet been fully implemented on a large radio survey, and challenges will probably be encountered when we do so. To identify these,  an internal prototype VO server and associated MySQL Database is being developed for ATLAS data and is now in regular use by ATLAS team members. It is expected that further VO development may include modifications to VO protocols, or a requirement for additional VO Tools. The EMU Design Study  includes plans to place ATLAS cross-identifications and other data (e.g. spectroscopy) on the ATLAS VO server, as a step to identifying the EMU VO requirements.

\subsection{Source Characterisation and Cross-Identification}
\label{crossid}

The EMU component catalogue will be cross-matched in a near-real-time automated pipeline with available large-area optical/IR surveys shown in Table \ref{surveys}. For each survey,  the quoted sensitivity has been used to estimate the fraction of radio sources in EMU with $S_{20cm} > 50\,\mu\,$Jy that will be detected by that survey, based on the measured flux densities of sources in COSMOS  \citep{Scoville07, Schinnerer07}.

%
%
%

\begin{table*}[h]
\begin{center}
\caption{Key Multi-Wavelength surveys with which EMU data will be cross-identified (restricted to surveys larger than 1000 \sqdeg). All magnitudes in AB. The EMU detected column is the fraction of EMU sources that are detectable by the multi-wavelength survey to its $5-\sigma$ limit. The survey matched column is the fraction of EMU sources in the previous column which are in the area of sky covered by the multi-wavelength survey.  The sensitivity shown for the WISE survey is for the 3.5$\mu$m band.
}
\vspace{0.5cm}
\begin{tabular}{lllllll}
\hline
Survey  & Area & Wavelength & Limiting& EMU & Survey & Data\\
Name & (\sqdeg) & Bands & Mag. & Detected &  Matched & Release\\
&&&or flux$^{\rm a}$&(\%)&(\%)&Date\\
\hline
WISE$^{\rm 1}$ & 40000 & 3.4, 4.6, 12, 22 $\mu$m & $80\,\mu$Jy & 23 & 100 & 2012\\
Pan-Starrs$^{\rm 2}$ & 30000 & $g$, $r$, $i$, $z$, $y$ & $r<24.0$ & 54 & 50 & 2020\\
Wallaby$^{\rm 3,b}$ & 30000 & $20\,$cm (HI) & $1.6\,$mJy$^{\rm c}$ & 1 & 100 & 2013\\
LSST$^{\rm 4}$ & 20000 &$u$, $g$, $r$, $i$, $z$, $y$ & $r<27.5$ & 96 & 67 & 2020\\
Skymapper$^{\rm 5}$ & 20000 & $u$, $v$, $g$, $r$, $i$, $z$ & $r<22.6$& 31 & 66 & 2015\\
VHS$^{\rm 6}$ & 20000 & Y, J, H, K & K$<20.5$ & 49 & 66 &2012\\
SDSS$^{\rm 7}$ & 12000 & $u$, $g$, $r$, $i$, $z$ & $r<22.2$ & 28 & 22 & DR8 2011 \\
DES$^{\rm 8}$ & 5000 & $g$, $r$, $i$, $z$, $y$ & $r<25$ & 71 & 17 & 2017\\
VST-ATLAS$^{\rm 9}$ & 4500 & $u$, $g$, $r$, $i$, $z$ & $r<22.3$& 30 & 15 & 2012?\\
Viking$^{\rm 10}$ & 1500 & Y, J, H, K & K$<21.5$ & 68 & 5 & 2012\\
Pan-Starrs Deep$^{\rm 2}$ & 1200 & $0.5-0.8$, $g$, $r$, $i$, $z$, $y$ & $g<27.0$ & 57 & 4 & 2020\\
\hline
\end{tabular}
\medskip\\
\label{surveys}
\end{center}
$^{\rm a}$Denotes $5\sigma$ point source detection.  However, in many cases, \emph{a priori} positional information will enable  $3\sigma$ data to be used, resulting in a higher detection rate.\\
$^{\rm b}$Being an HI survey, WALLABY will measure redshifts for all detected galaxies out to $z=0.26$.\\
$^{\rm c}$per $4\,$km s$^{-1}$ channel achieved in $8\,$hours integration.\\
$^{\rm 1}$\citet{Wright2010}\\
$^{\rm 2}$\citet{Kaiser2010}\\
$^{\rm 3}$\citet{Koribalski11}\\
$^{\rm 4}$\citet{Ivezic2008}\\
$^{\rm 5}$\citet{Keller07}\\
$^{\rm 6}$http://www.ast.cam.ac.uk/research/instrumentation.surveys.and.projects/vista\\
$^{\rm 7}$\citet{Abazajian09}\\
$^{\rm 8}$\citet{DES2005}\\
$^{\rm 9}$\citet{Shanks05}\\
$^{\rm 10}$http://www.eso.org/sci/observing/policies/PublicSurveys/sciencePublicSurveys.html\\
\end{table*}

\twocolumn

Experience with surveys such as ATLAS and COSMOS indicates that $\sim$ 5\% of EMU radio sources will consist of multiple components such as radio double or triple sources with a core-lobe morphology. A key goal of the EMU Design Study is to develop an automated cross-identification pipeline, significantly more sophisticated than a simple nearest-neighbour algorithm, to group the \ngal million radio components detected by EMU into ``sources'', and cross-identify them with optical and infrared sources found by other surveys. The algorithm will be run (in an automated pipeline) on the primary components catalogue described in Table \ref{datatable} to produce a secondary data product, the ``source'' catalogue, in which components are grouped together to form sources, and are associated with optical/IR identifications. In cases where several cross-identifications are possible, each will be given in the source catalogue along with an estimated probability.

This Design Study  builds on the experience developed through a number of international and multi-wavelength surveys in which EMU team members are named investigators, and some of whom are developing similar tools for {\it Herschel}, {\it Spitzer} and SCUBA/SCUBA-2 cross-identifications
  \cite[e.g.][]{Roseboom09, Roseboom10, Smith10, Chapin10}.

Given the expected source density, and redshift distribution, of both radio sources and potential matching catalogues (see Table \ref{surveys}), a likelihood ratio matching technique \citep{Sutherland92, Smith10}
 or similar (i.e. Budavari \& Szalay 2009, Roseboom et al. 2009) should be adequate in the majority of cases, especially if source SED/colour information is incorporated (Roseboom et al.  2009; Chapin et al.  2010). The $\sim5$\% of radio sources \citep[see e.g.][]{Norris06} consisting of multiple components represent a considerable challenge to automated cross-identification algorithms, and our proposed algorithm will need to account for these. Some potential solutions (i.e. using geometric information in matching, simultaneous radio self matching and ancillary data matching) are currently being developed, and will be tested as part of the  Design Study.

We will use Monte-Carlo simulations to derive probable misidentification rates for various classes of source, and will publish the algorithm and the results of the simulations in a journal paper, as well as provide code to the astronomical community. During the early part of the  Design Study, ATLAS data will be used to investigate possible algorithms. The later part of the  Design Study will include setting up collaborations, ensuring access to appropriate surveys, constructing software for the pipeline processor, and running simulations to measure the misidentification rates.
 
Applying the limits shown in Table \ref{surveys} to the sources detected in ATLAS and COSMOS, it is expected that, of the \ngal million sources detected and catalogued by EMU, $\sim40$\% will initially have identified counterparts in wide-field optical and IR surveys (e.g. SkyMapper, VHS, WISE, LSST, etc.) and a further few percent will be identified by smaller but deeper surveys such as HERMES \citep{Seymour11}. This fraction will rise to $\sim$70\% in 2020 with the release of the LSST data. 

Of those 30 million sources with potential identifications initially, we estimate that $\sim$ 25 million will yield reliable cross-identifications using a simple likelihood ratio algorithm as discussed above, and a further 3 million will be handled satisfactorily by a more sophisticated cross-identification algorithm which will take account of the spectral energy distribution at other wavelengths. For example, three radio components in a line could either be three low-redshift star-forming galaxies, or a high-redshift FRII galaxy. Three IR galaxies with a star-forming SED coincident with the three radio components would indicate the first solution, whereas a bright elliptical component at the centre, and no identification with the outer components, would indicate the latter solution.

The remaining 2 million are likely to be extended, confused, or complex sources for which automated algorithms are unlikely to be able to make a confident identification. Ironically, these are the sources most likely to yield the most interesting science, but a human brain will be required for an optimum cross-identification, and there are simply not enough human brains within the EMU team (or within the global community of professional astronomers) to look at each galaxy.

\label{zoo}
In cases where the automated pipeline fails, EMU will use a ``Citizen Science'' approach to enlist the help of thousands of enthusiastic amateurs, in collaboration with the Galaxy Zoo group  \citep{Lintott08}. Galaxy Zoo have used this technique very successfully for classification of SDSS galaxies, and have agreed to collaborate with EMU on creating ``Radio Zoo''. An initial prototype RadioZoo project will be established in mid 2011 using ATLAS data. If resources permit, machine-learning cross-identification will also be explored, following the approach of  \cite{Djorgovski10} who have extended the ``Citizen Science'' approach by using it to train neural nets and other algorithms.

\subsection{Stacking and Data-Intensive Research}
EMU has an additional requirement for facilities equipped for data exploration and mining, and data-intensive research, such as the technique of stacking.
\label{stacking}

The process of stacking involves identifying a class of objects which are generally below the detection limit of a survey, but then combining (typically by taking a censored mean or median) together the data at the position where such objects are expected in the survey. The noise tends to cancel, while any low level of flux density in the sources adds, resulting in a detection threshold very much lower than that of the unstacked survey. Stacking at radio wavelengths has been used very successfully  \citep[e.g.][]{White07, Ivison07, Ivison10, Dunne09, Messias10, Bourne11} on high-resolution data for the purpose of studying faint populations which are below the detection threshold of the radio image, and has proven to be a powerful tool for studying star formation rates, AGN activity, and measuring the fraction of the extragalactic background contributed by various source populations. For example, the radio-infrared correlation can be probed down to levels well below the threshold of any radio survey by averaging the radio flux density at the position of all galaxies with an infrared flux density within a particular bin  \citep{Boyle07}.

The unprecedented area-depth product of EMU makes it remarkably suitable for the stacking process. For example, stacking a sample of a million optically selected galaxies in the EMU data will result in a noise level of $\sim$10 nJy. Because of the wide area of EMU, even rare classes of source can be stacked, and it should also be possible to create stacked images of extended sources, such as clusters. However, the extent to which such deep stacking will be successful in EMU will depend on the extent to which imaging artefacts are cancelled by stacking, and the effectiveness of stacking EMU data will be simulated, and stacking science projects conducted using ATLAS, NVSS, and FIRST, to see whether particular observing strategies or reduction algorithms are necessary.

In particular, it has been established that stacking can be biased by a number of effects such as failure to account for the point-spread function, particularly in the presence of confusion and in highly clustered fields  \citep{Bourne11, Greve10,
Penner11, Chary10, Beelen10} and so stacking experiments and simulations will be conducted during the Design Study to (a) determine what requirements these place upon the EMU data (e.g. storing the point-spread function at all positions across the primary beam) and (b) to establish stacking algorithms and software which can be used robustly on EMU data.

Other examples of data-intensive research include:
\begin{itemize}
\item Identification of sources which do not fit into known categories of radio source, and so are likely to be artefacts or exciting new classes of source, as discussed in \S \ref{WTF}. 
\item Extraction of low surface brightness emission cross-correlated with millions of galaxies selected from other surveys. For example, this can be used to detect the synchrotron emission from cosmic filaments or sheets.
\item Cross-correlation of low-redshift galaxies with high-redshift galaxies, or the CMB, to test cosmology and fundamental physics as discussed in \S \ref{cosmology}.
\end{itemize}
 
The computational facilities required for such data-intensive research are currently unfunded, but it is hoped that the computational facilities associated with the ASDAF will become available.
 
\subsection{Measuring Redshifts}
\label{redshifts}

To interpret data from EMU, redshifts are invaluable. Most of the science goals discussed in \S 2 assume that redshifts will not be available for most EMU sources, although much of the EMU science can be greatly enhanced if redshifts \emph{are} available. However, no existing or planned spectroscopic redshift survey can cover more than a tiny fraction of EMU's \ngal million sources.

For nearby galaxies, HI redshifts will be available from WALLABY, which will provide $\sim 5 \times 10^5$ redshifts, and smaller numbers will be provided by other redshift surveys such as SDSS \citep{York00} and GAMA \citep{Driver09}. The remaining $\sim$ 99\% of EMU galaxies will not have spectroscopic redshifts.

Photometric redshifts, in which SEDs of various templates are fitted to the measured multi-band photometry of target galaxies, are widely used as a surrogate for spectroscopic redshifts, with considerable success.
As discussed below, $\sim$ 30\% of EMU sources will have multi-wavelength optical/IR photometric data at the time of data release, increasing to $\sim$ 70\% in 2020. The SED of star-forming galaxies has a characteristic shape which provides good fits to SED modelling, and so it is expected that the majority of EMU star-forming galaxies (whose host galaxies will mainly be low-redshift and hence optically bright) will have reliable photometric redshifts. 

Similarly, low-power AGN (such as Seyfert galaxies) and AGNs hosted by elliptical galaxies at z $<$ 1 are also likely to yield reliable photometric redshifts.

High-redshift radio-loud AGNS present more of a challenge. Photometric redshifts are notoriously bad at fitting the power-law SEDs of AGNs, and hence strong radio-selected sources. For example, \citet{Rowan08} found a 99\% success rate in photometric redshifts of galaxies in the SWIRE survey, while  \citet{Norris06} found only a 50\% success rate in Rowan-Robinson's fits for radio sources within that sample, and a success rate indistinguishable from chance for radio sources at z$>$1. On the other hand,  \cite{Salvato10} have achieved reliable photometric redshifts of AGNs in the COSMOS field by using high-quality photometry in a large number of bands, and taking account of effects such as source variability.

While the relatively sparse photometry available for most EMU sources (listed in Table \ref{surveys}) can not generate such accurate photometric redshifts for high-redshift sources, it can provide an estimate of the probability of a given source being at a particular redshift, for about 30\% of EMU galaxies. Such estimates are valuable for statistical purposes such as  cosmological tests (see \S \ref{cosmology}), and we term such redshifts ``statistical redshifts''.  

 Even a non-detection can carry useful information, and radio data themselves can add significantly to the choice of SED template, and hence to a probabilistic estimate of redshift.  For example, an AGN template may be indicated by the radio morphology, the detection of polarisation, a radio spectral index that is significantly different from $-0.7$, or a low far-infrared-radio ratio. Alternatively, high-redshift radio galaxies can be identified from their strong radio emission coupled with a K-band non-detection \citep{Willott03}. Even the radio data alone can provide information about the probable redshift range. For example, a steep radio spectral index increases the probability of a high redshift \citep{Breuck02}, while the angular size of some classes of  galaxy can be correlated with redshift \citep[e.g. ][]{Wardle74}.

Radio polarisation data  can also provide statistical redshift information \citep{Norris11c}.  The POSSUM  project \citep{Gaensler10}  will run commensally with EMU to generate a catalogue of polarised fluxes and Faraday rotation measures (or upper limits) for all sources detected by EMU. Sources that are strongly polarised are nearly always AGNs \citep{Hales11}, and so have a mean $z \sim 1.88$, while unpolarised sources are mainly star-forming galaxies with a mean $z \sim 1.08$ (see Fig. \ref{nz}). Consequently, cosmological tests may be made by treating unpolarised sources as a low-redshift screen in front of background high-redshift polarised sources.

A significant fraction of statistical redshifts are likely to be incorrect. A correction for this can be applied  in a statistical study of a sufficient number of objects, provided that reliability and completeness are carefully calibrated in a small well-studied area with accurate spectroscopic redshifts.
The ATLAS survey  \citep{Norris06,Middelberg08a} is expected to yield 16000 radio sources at its completion in 2011 \citep{Grant11}. Spectroscopy on these sources is already underway  \citep{Mao11b} and other large spectroscopy proposals have also been submitted to measure redshifts for the ATLAS radio sources. This spectroscopy should enable the training of algorithms to yield, and correct, statistical redshifts for $\sim$ 30\% of EMU sources.

A further initiative is a proposed project
to make use of the wide-field spectroscopic capabilities
of the UK Schmidt Telescope at Siding Spring.
The aim of this project, named TAIPAN (Transforming Astronomical Imaging-surveys through Polychromatic Analysis of Nebulae),
is to extend the depth of the 6dFGS  \citep{Jones09} to achieve a complete spectroscopic
survey over the full southern hemisphere, potentially approaching
the depths of the SDSS spectroscopic survey. This would
provide invaluable spectroscopic data to complement all
wide-field ASKAP surveys, and especially EMU.

\subsection{Imaging the Galactic Plane}
Although EMU focuses primarily on extragalactic science, Galactic science is also an important secondary goal, as discussed in \S \ref{galaxy} above. Imaging the Galactic Plane, or even imaging extragalactic sky near the Galactic Plane,  introduces additional issues which must be tackled in the Design Study, such as:
\begin{itemize}
\item What are the limitations in the depth of EMU due to the bright extended sources in the Galactic Plane?
\item How can the sensitivity to larger-scale structures such as HII regions be maximised?
\item How close to the Galactic Plane can  the full sensitivity and dynamic range be achieved for the main extragalactic EMU survey?
\end{itemize}
The Design Study  attempts to answer these by conducting simulations of the standard SKADS simulated sky to which Galactic continuum images have been added.

\subsection{SPARCS: SKA Pathfinders Radio Continuum Survey Working Group}
\label{sparcs}
Several next-generation radio telescopes and upgrades are already being built around the world, including LOFAR (The Netherlands), ASKAP (Australia), Meerkat (South Africa), EVLA (USA), eMERLIN (UK), the Allen Telescope Array (USA), and Apertif (The Netherlands). Most of these projects have multiple science goals, but they have one goal in common, which is to survey the radio continuum emission from galaxies, in order to understand the formation and evolution of galaxies over cosmic time, and the cosmological parameters and large-scale structures which drive it. In pursuit of this goal, the different teams are developing techniques such as multi-scale deconvolution, source extraction and classification, and multi-wavelength cross-identification. Furthermore, these projects share specific science goals, which in most cases require further definition before a well-planned survey can be executed. Finally, each of these new instruments has different strengths, and coordination of surveys between them can help maximise the science from each of them.

The SKA PAthfinder Radio Continuum Survey (SPARCS) Working Group has therefore been established, under the auspices of the SKA project, with the following goals:
\begin{itemize}
\item To coordinate developments of techniques, to avoid duplication of effort and ensure that each project has access to best practice,
\item To facilitate cross-project discussions of the specific science goals, to ensure cross-fertilisation of ideas and optimum survey strategies,
\item To coordinate the surveys in their choice of area, depth, location on the sky, and other survey parameters, to maximise the science return from the surveys.
\end{itemize}

SPARCS consists of a core group of the leaders of the radio continuum surveys such as EMU, and is also open to the estimated 200 other astronomers who are engaged in the continuum survey projects on SKA Pathfinders. The first meeting of SPARCS took place in February 2011 at the Lorentz Center in Leiden, The Netherlands, and was attended by about 60 astronomers. The Proceedings will be published as a multi-author paper  \citep{Norris11b} in a refereed journal. SPARCS will continue to hold such meetings annually, as well as facilitating other interactions between members of the various survey groups.

\section{EMU Observing and Operations}

It is expected that the observations will take place commensally with other ASKAP surveys such as WALLABY and POSSUM, implying that observing strategies and schedules will need to be agreed on by the different teams. The actual observing is expected to be conducted autonomously, with observing schedules delivered to the telescope in advance, and telescope performance monitored throughout the observations. Little or no interaction is expected to take place between the astronomers and the telescope until the data products are delivered from the pipeline for quality control.

Observations are expected to take place on one field at a time, rather than revisiting each field many times, to minimise the effect of source variability and to minimise the large processing overheads associated with combining several epochs of data. Typically, ASKAP will survey one field of 30\sqdeg\  for 12 hours before moving to the next, although this may be complicated by the need for dithering to achieve a uniform sensitivity across the field.

\subsection{Data Quality Control}
\label{quality}
Data produced by the {\em ASKAPsoft} ~processing pipeline will be subject to a number of Data Quality checks before being released into the public domain, and this Quality Control process is the responsibility of the EMU team. In real-time, the statistics of each field will be compared with those obtained from other radio surveys, previous ASKAP pointings, and the expected performance of ASKAP, including:
\begin{itemize}
\item rms noise level,
\item source counts of detected sources (i.e. do they fit published logN/logS distributions?),
\item spectral indices,
\item the circularly polarised source components (expected to be close to zero),
\item rotation measures of linearly polarised source components,
\item measured positions of cross-identified components, compared with their positions from multi-wavelength surveys.
\end{itemize}

In the early days of the survey, these automated tests will be followed by a visual inspection of each image for artefacts or other problems. However, it is expected that as experience and confidence grows with time, the visual inspections will be replaced by automated tests.

A further level of testing will be provided on a longer timescale by the cross-identification program, which will, for example, show up any systematic positional errors. Additional quality control data will come from the WTF program (section \ref{WTF}) which will not only identify astronomically interesting objects which do not fall into known classes, but will also identify errors in the EMU data which may masquerade as a new class of objects.

The final quality control step is from external users. As all data is released into the public domain in near-real time, feedback will be encouraged from external users of EMU data which will enable the identification of rare or subtle problems in the data.

\subsection{Data Release}

The primary data products consist of the \emph{uv} data, image data, and a catalogue of radio components, as listed in Table \ref{datatable}. These data products are delivered to the ASKAP Science Data Archive Facility (ASDAF) by the real-time processing pipeline, and are then subjected to the quality control processes described above. After passing quality control, the data will then be released into the public domain, with no proprietary period. Once routine observing is established, the time taken for the quality control step is expected to be no more than about a day, although in the early days of the survey the data quality control step is likely to take longer.

Because the time to process the primary data is comparable to the observing time, routine reprocessing of data is not an option. It is therefore important that the primary data products
are as final as possible, and it is likely that there will be only one data release for each field.

The secondary data products consist of a catalogue of sources, each of which may consist of one or more components in the primary catalogue, together with cross-identifications with optical/IR data. This will also be produced by an automated pipeline which will run asynchronously from the primary data processing pipeline. The secondary data products will also be placed in the public domain as soon as possible after processing, although the need to protect proprietary optical/IR data with privileged access is also recognised.
 
It is likely that as additional survey data become available over time, and as expertise increases, the cross-identification process will also be refined. Several releases of the secondary catalogue are therefore expected, each with a release number embedded in the source identifiers.

Both the primary and secondary data will be available from the ASDAF using Virtual Observatory (VO) tools such as Simbad and TopCat, and through simple cone-search and cut-out servers. It is expected that they will also be available from the NASA Extragalactic Database (NED) and other astronomical databases.

\subsection{Outreach and Communications}

EMU provides an excellent opportunity to inform and engage the public about new developments in radio astronomy and ASKAP. An EMU website aimed at the public is currently under development. Initially it will be a fairly traditional ``static'' site but there is great potential for it to grow and incorporate new technologies as data become available, once ASKAP is operational. Linkage with the Radio Zoo site (see \S \ref{zoo} ) should expose it to a large and diverse audience.

The advent of a new ``Australian Curriculum'' for both Science (K-10) and Physics (Years 11-12) (http://www.australiancurriculum.edu.au) over the next few years is an opportunity to develop new, targeted formal education resources utilising not just EMU data for the ``Science Understanding'' and ``Science Inquiry Skills'' strands but examples of scientists involved in EMU and ASKAP for the ``Science as a Human Endeavour'' strand for students across Australia.

Communication within the EMU team is primarily via a closed Wiki, augmented by an electronic newsletter produced three times per year.

\section{Conclusion}

The most obvious benefit from EMU is the legacy survey. Its greater depth, greater resolution, and program of cross-identifications will impact all areas of astronomy. EMU's sensitivity means that the majority of objects in EMU will be star-forming galaxies rather than radio-loud AGNs. Thus EMU plays a key role in the continuing emergence of extragalactic radio astronomy from the niche that it once occupied, focussing largely on radio-loud AGNs, to one that impacts on all areas of astronomy. As a legacy survey, EMU will provide a resource to all astronomers, leveraging astronomical observations at all wavelengths both through the VO server and also through data centres like NED and CDS. It is likely to find rare, interesting targets (such as the $\sim$100 EoR radio galaxies) that can then be followed up by other observatories such as ALMA, ELT, JWST, and Meerkat, which are better suited to deep pointed observations.

In addition to the legacy program,   this paper has described a number of science areas in which EMU is likely to have a major impact. For example, it will distinguish between alternative models of cosmology and General Relativity, trace the evolution of star formation and AGNs, and significantly increase the number of known clusters of galaxies. 

It is even possible that the greatest impact of EMU is on science that hasn't yet been thought of. EMU will occupy a new region of observational phase space, and we plan to exploit this by mining the data for unexpected types of astronomical object.
 
EMU is novel in that cross-identifications with all available data at optical, infrared, and radio wavelengths will be produced by the EMU cross-identification pipeline, and will be placed in the public domain, as a service to the community, as soon as the data quality has been checked. They will be made available using the VO server (and thus immediately via services such as Aladin and TopCat), through the EMU web page, and also through the ASKAP Science Data Archive Facility services.

EMU is an open collaboration and we welcome interactions and collaborations with other surveys and other instruments, to increase our common scientific productivity. We also encourage other individual scientists to join the EMU collaboration, to help generate even better science, by contacting Ray Norris. Further information on EMU can be found on http://askap.org/emu

\section*{Acknowledgments}
The EMU team consists of over 180 members from 15 countries, all of whom are listed on \\
\small{http://askap.pbworks.com/TeamMembers}. We thank them all for their significant contributions  to the various stages of the EMU project. Of course, EMU will not be possible without ASKAP, and so we especially thank all the ASKAP staff, too numerous to name individually, who are actually designing and building the instrument on our behalf.  We particularly thank the architects of the ASKAP science processing document: Tim Cornwell, Ben Humphreys, Emil Lenc, Max Voronkov, and Matt Whiting, for permission to use sections of that document. We also thank Lakshmi Saripalli for providing data on diffuse sources in ATLBS prior to publication, and Paul Nulsen for comments on a draft of this paper.
The Centre for All-sky Astrophysics (CAASTRO) is an Australian Research Council Centre of Excellence, funded by grant CE11E0090. ASKAP is sited on the Murchison Radio-astronomy Observatory, which is jointly funded by the Commonwealth Government of Australia and State Government of Western Australia. We acknowledge the Wajarri Yamatji people as the traditional owners of the Observatory site.

\section*{Authors Affiliations}
{\small \affil{1}\,CSIRO Astronomy \& Space Science, PO Box 76, Epping, NSW 1710, Australia}\\
{\small \affil{2}\,Australian Astronomical Observatory, PO Box 296, Epping, NSW 1710, Australia}\\
{\small \affil{3}\,Centro de Astronomia e Astrof\'{\i}sica da Universidade de Lisboa, Observat\'{o}rio Astron\'{o}mico de Lisboa, Tapada da Ajuda, 1349-018 Lisboa, Portugal}\\
{\small \affil{4}\,National Radio Astronomy Observatory, 520 Edgemont Road, Charlottesville, VA 22903, USA}\\
{\small \affil{5}\,School of Physics \& Astronomy, University of Nottingham, University Park, Nottingham, NG7 2RD, UK}\\
{\small \affil{6}\, Centre for Astrophysics Research, Science \& Technology Research Institute, University of Hertfordshire, Hatfield, Herts., UK}\\
{\small \affil{7}\,School of chemical \& Physical Sciences, Victoria University of Wellington, PO Box 600, Wellington 6140, New Zealand}\\
{\small \affil{8}\,Astronomisches Institut, Ruhr-Universit\"at Bochum, Universit\"atsstr. 150, 44801 Bochum, Germany}\\
{\small \affil{9}\,European Southern Observatory, Karl-Schwarzschild-Str. 2, D-85748 Garching bei M\"unchen, Germany}\\
{\small \affil{10}\,INAF-IRA, Via P. Gobetti 101, 40129 Bologna, Italy}\\
{\small \affil{11}\,Department of Astronomy, University of Minnesota, 116 Church St. SE, Minneapolis, MN 55455, USA}\\
{\small \affil{12}\,University College London, Department of Space \& Climate Physics, Mullard Space Science Laboratory, Holmbury St. Mary, Dorking, Surrey RH5 6NT, UK.}\\
{\small \affil{13}\,INAF-Catania Astrophysical Observatory, Via S. Sofia 78, 95123 Catania, ITALY}\\
{\small \affil{14}\,Depto.\ de Astronom\'{\i}a, Universidad de Guanajuato, Guanajuato, C.P.\ 36000, GTO, Mexico}\\
{\small \affil{15}\,NASA Herschel Science Center, Caltech, 1200E. California Blvd, Pasadena, CA 91125, USA}\\
{\small \affil{16}\,Institute of Cosmology and Gravitation, University of Portsmouth, Dennis Sciama Building, Burnaby Road, Portsmouth, PO1 3FX, UK}\\
{\small \affil{17}\,Max-Planck Institut fŸr extraterr. Physik, Giessenbachstrasse 1, PO BOX 1312, 85741 Garching, Germany}\\
{\small \affil{18}\,School of Physics, Monash University, Clayton, VIC 3800, Australia}\\
{\small \affil{19}\,INAF - OABO, Via Ranzani 1, 40127 Bologna, Italy.}\\
{\small \affil{20}\,School of Physics and Astronomy, Cardiff University, The Parade,Cardiff CF24 3AA,UK}\\
{\small \affil{21}\,Department of Physics,Durham University,South Road,Durham, DH1 3LE, UK}\\
{\small \affil{22}\,Sydney Institute for Astronomy, School of Physics, The University of Sydney, NSW 2006, Australia}\\
{\small \affil{23}\,International Centre for Radio Astronomy Research, University of Western Australia, M468, 35 Stirling Hwy, Crawley WA 6009, Australia}\\
{\small \affil{24}\,UK Astronomy Technology Centre, Royal Observatory, Blackford Hill, Edinburgh EH9 3HJ, UK}\\
{\small \affil{25}\,Institute for Astronomy, University of Edinburgh, Blackford Hill, Edinburgh EH9 3HJ, UK}\\
{\small \affil{26}\,Institute of Astronomy, University of Cambridge, Madingley Road, Cambridge, CB3 0HA, UK}\\
{\small \affil{27}\,Space Telescope Science Institute, 3700 San Martin Dr., Baltimore MD 21218, USA}\\
{\small \affil{28}\,School of Mathematics \& Physics, University of Tasmania, Private Bag 37, Hobart, 7001, Australia}\\
{\small \affil{29}\,Centro de Astronomia e Astrof\'{\i}sica da Universidade de Lisboa, Observat\'{o}rio Astron\'{o}mico de Lisboa, Tapada da Ajuda, 1349-018 Lisboa, Portugal}\\
{\small \affil{30}\, Argelander Institute for Astronomy, Bonn University, Auf dem H{\"u}gel 71, 53121 Bonn, Germany}\\
{\small \affil{31}\,Dept of Physics and Astronomy, University of Sussex, Falmer, East Sussex, BN1 9RH, U.K.}\\
{\small \affil{32}\,Sterrewacht, University of Leiden, The Netherlands}\\
{\small \affil{33}\,National Centre for Radio Astrophysics, TIFR, Pune 411 007, India}\\
{\small \affil{34}\,Mount Stromlo Observatory, Canberra, Australia}\\
{\small \affil{35}\,Dept. of Physics and Astronomy, University of British Columbia, Canada}\\
{\small \affil{36}\,ARC Centre of Excellence for All-sky Astrophysics (CAASTRO)}\\
{\small \affil{37}\,Department of Physics and Astronomy, Macquarie University, NSW 2109,Australia}\\
{\small \affil{38}\,Physics Department, University of the Western Cape, Cape Town, 7535, South Africa}\\
{\small \affil{}\,Email: Ray.Norris@csiro.au}


\onecolumn

%
%
%


\begin{thebibliography}{}
\bibitem[Abazajian et al.(2009)]{Abazajian09}Abazajian, K.N., et al. 2009, \apjs, 182, 543
\bibitem[Afonso et al.(2005)]{Afonso05} Afonso, J., Georgakakis, A., Almeida, C., Hopkins, A.~M., Cram, L.~E., Mobasher, B., \& Sullivan, M.\ 2005, \apj, 624, 135 
\bibitem[Afonso et al.(2006)]{Afonso06} Afonso, J., Mobasher, 
B., Koekemoer, A., Norris, R.~P., \& Cram, L.\ 2006, \aj, 131, 1216 
\bibitem[Allen et al.(2008)]{Allen08} Allen, S.~W., Rapetti, D.~A., Schmidt, R.~W., Ebeling, H., Morris, R.~G., \& Fabian, A.~C.\ 2008, \mnras, 383, 879 
\bibitem[Bagchi et al.(2002)]{Bagchi02}Bagchi, J., En{\ss}lin, T.~A., Miniati, F., Stalin, C.~S., Singh, M., Raychaudhury, S., \& Humeshkar, N.~B.\ 2002, \na, 7, 249 
\bibitem[Ball et al.(2009)]{Ball09}Ball, L., et al., 2009, ASKAP memo 26, 
{\tiny http://www.atnf.csiro.au/SKA/ASKAP\_SSP\_Sep09\_final.pdf} 
\bibitem[Ballard (1981)]{Bal:81}Ballard, D. H., 1981, Pattern Recognition, 13.2, 111
\bibitem[Banfield et al.(2011)]{Grant11}Banfield, J.~K., et al., 2011, in preparation 
\bibitem[Barrena et al.(2009)]{Barrena09}Barrena, R., Girardi, M., Boschin, W., \& Das{\'{\i}}, M.\ 2009, \aap, 503, 357
\bibitem[Battaglia et al.(2009)]{Battaglia09}Battaglia, N., Pfrommer, C., Sievers, J.~L., Bond, J.~R., \& En{\ss}lin, T.~A.\ 2009, \mnras, 393, 1073
\bibitem[Becker et al.(1995)]{Becker95}Becker, R. H., White, R. L., \& Helfand, D. J. 1995, \apj, 450, 559
\bibitem[Becker et al.(2009)]{Becker09} Becker, W., 2009, Astrophysics and Space Science Library, 357, 91
\bibitem[Bell (2003)]{Bell03} Bell, E.~F.\ 2003, \apj, 586, 794
\bibitem[Bell-Burnell (2009)]{Burnell10}Bell-Burnell, J., 2009, in ÒAccelerating The Rate Of Astronomical DiscoveryÓ, edited by Ray P. Norris \& Clive L. N. Ruggles, 2009, PoS (SpS5) 
\bibitem[Berger et al.(2001)]{Berger01} Berger, E., et al.  2001, \nat, 410, 338
\bibitem[Bertacca et al.(2011)]{Bertacca11} Bertacca, D., Raccanelli, A., Piattella, O.~F., Pietrobon, D., Bartolo, N., Matarrese, S., \& Giannantonio, T.\ 2011, \jcap, 3, 39 
\bibitem[Bertin \& Arnouts (1996)]{BA:96}Bertin, E., Arnouts, S. 1996, A\&AS, 117, 393
\bibitem[Best et al.(2003)]{Best03} Best, P.~N., Arts, J.~N., R{\"o}ttgering, H.~J.~A., Rengelink, R., Brookes, M.~H., \& Wall, J.\ 2003, \, \mnras, 346, 627
\bibitem[Best et al.(2006)]{Best06} Best, P.~N., Kaiser, C.~R., Heckman, T.~M., \& Kauffmann, G.\ 2006, \, \mnras, 368, L67
\bibitem[B{\'e}thermin et al.(2010)]{Beelen10} B{\'e}thermin, M., Dole, H., Beelen, A., \& Aussel, H.\ 2010, \aap, 512, A78 
\bibitem[Biggs \& Ivison (2006)]{Biggs06} Biggs, A.~D., \& Ivison, R.~J., 2006, \mnras, 371, 963
\bibitem[Biggs \& Ivison (2008)]{Biggs08} Biggs, A.~D., \& Ivison, R.~J., 2008, \mnras, 385, 893
\bibitem[Biggs, Younger \& Ivison (2010)]{Biggs10} Biggs, A.~D., Younger J.~D., \& Ivison, R.~J., 2010, \mnras, 408, 342
\bibitem[Blake 
\& Wall(2002)]{Blake02} Blake, C., \& Wall, J.\ 2002, \mnras, 329, L37 
\bibitem[Blanton et al.(2003)]{2003AJ....125.1635B} Blanton, E.~L., et al., \ 2003, \aj, 125, 1635 
\bibitem[Bondi et al.(2003)]{Bondi03} Bondi, M., et al.  2003, A\&A, 403, 857
\bibitem[Bongiorno et al.(2007)]{Bongiorno07}Bongiorno, A., et al. 2007, A \&A, 472, 443
\bibitem[Boughn \& Crittenden (2004)]{Boughn04} Boughn, S., \& Crittenden, R.\ 2004, \nat, 427, 45
\bibitem[Bourne et al.(2011)]{Bourne11} Bourne, N., Dunne, L., Ivison, R.~J., Maddox, S.~J., Dickinson, M., \& Frayer, D.~T.\ 2011, \mnras, 410, 1155 
\bibitem[Bower et al.(2006)]{Bower06} Bower, R.~G., Benson, A.~J., Malbon, R., Helly, J.~C., Frenk, C.~S., Baugh, C.~M., Cole, S., \& Lacey, C.~G.\ 2006, \mnras, 370, 645
\bibitem[Boyle et al.(2007)]{Boyle07}Boyle, B.J., et al., 2007, \mnras,376, pp1182-1188
\bibitem[Brax \& van de Bruck (2003)]{Brax03}Brax, P., \& van de Bruck, C., 2003, Class.Quant.Grav., 20, R201 (also arXiv:hep-th/0303095)
\bibitem[Bridle \& Schwab (1999)]{Bridle99} Bridle, A.~H., \& Schwab, F.~R.\ 1999,
Synthesis Imaging in Radio Astronomy II, 180, 371
\bibitem[Briggs \& Cornwell (1992)]{Briggs92}Briggs, D., \& Cornwell, T.J., 1992, EVLA Memo 114
\bibitem[Brown \& Rudnick (2009)]{Brown09} Brown, S., \& Rudnick, L.\ 2009, \aj, 137, 3158
\bibitem[Brown et al.(2008)]{Brown08} Brown, M.~J.~I., et al.  2008, \apj, 682, 937
\bibitem[Brown et al.(2010)]{Brown10} Brown, S., Farnsworth, D., \& Rudnick, L.\ 2010, \mnras, 402, 2\\
\bibitem[Brown et al.(2011a)]{Brown11a} Brown, S., Duesterhoeft, J., \& Rudnick, L.\ 2011, \apjl, 727, L25 
\bibitem[Brown et al.(2011b)]{Brown11b} Brown, S., et al.  2011, in preparation 
\bibitem[Brunetti et al.(2009)]{Brunetti09} Brunetti, G., Cassano, R., Dolag, K. \& Setti, G.\ 2009, \aap, 507, 661
\bibitem[Budding et al.(1999)]{Budding_1999} Budding, E., Jones, K. L., Slee, O. B., Watson, L., 1999, \mnras 305, 966.
\bibitem[Bunton \& Hay (2010)]{Bunton10}Bunton, J. D., \& Hay, S. G., 2010, in International Conference on Electromagnetics in Advanced Applications (ICEAA), 
{\tiny http://ieeexplore.ieee.org/xpls/abs\_all.jsp?arnumber=5651120}
\bibitem[Cameron et al.(2011)]{Cameron11}Cameron, A., et al., 2011, \mnras, in press
\bibitem[Carey et al.(2009)]{carey2009} Carey, S.~J., et al.  2009, \pasp, 121, 76
\bibitem[Carilli \& Rawlings (2004)]{Carilli04}Carilli, C., \& Rawlings, S., 2004, New Astronomy Reviews, Vol. 48, Elsevier.
\bibitem[Carilli et al.(2002)]{Carilli02} Carilli, C.~L., Gnedin, N.~Y., \& Owen, F.\ 2002, \apj, 577, 22
\bibitem[Cassano et al.(2010)]{Cassano10} Cassano, R., Ettori, S., Giacintucci, S., Brunetti, G., Markevitch, M., Venturi, T. \& Gitti, M.\ 2010, \apjl, 721, L82
\bibitem[Cen \& Ostriker (1999)]{Cen99}Cen, R., \& Ostriker, J. P. (1999), \apj, 514, 1
\bibitem[Chapman et al.(2003)]{Chapman03}Chapman, S.C., Blain, A.W., Ivison, R.J., \& Smail, I.R. 2003, Nature, 422, 695
\bibitem[Chapin et al.(2011)]{Chapin10} Chapin, E.~L., et al.\ 2011, \mnras, 411, 505  
\bibitem[Chary et al.(2008)]{Chary10} Chary, R.-R., Cooray, A., \& Sullivan, I.\ 2008, \apj, 681, 53 
\bibitem[Chatterjee et al. (2010)]{Vast10} Chatterjee, S., Murphy, T., \& VAST Collaboration 2010, Bulletin of the American Astronomical Society, 42, \#470.12
\bibitem[Churchwell (2002)]{churchwell2002} Churchwell, E.\ 2002, \araa, 40, 27
\bibitem[Cole et al.(2005)]{Cole05} Cole, S., et al.  2005, \mnras, 362, 505
\bibitem[Compi{\`e}gne et al.(2010)]{compiegne2010} Compi{\`e}gne, M., Flagey, N., Noriega-Crespo, A., Martin, P.~G., Bernard, J.-P., Paladini, R., \& Molinari, S.\ 2010, \apjl, 724, L44
\bibitem[Condon (1974)]{Condon74}Condon, J.~J.\ 1974, \apj, 188, 279
\bibitem[Condon (1988)]{Condon88}Condon, J.~J.\ 1988, in "Galactic and Extragalactic Radio Astronomy", 2nd edition, eds. G.L. Verschuur and K.I. Kellermann. Also in 
{\tiny http://nedwww.ipac.caltech.edu/level5/Sept04/Condon/Condon\_contents.html}
\bibitem[Condon (1992)]{Condon92}Condon, J. J. (1992), ARA \&A, 30, 575
\bibitem[Condon (2008)]{Condon08}Condon, J. J. (2008), ASKAP memo 015, 
{\tiny http://www.atnf.csiro.au/SKA/newdocs/condon\_memo.pdf}
\bibitem[Condon et al.(1998)]{Condon98} Condon, J.~J., Cotton, W.~D., Greisen, E.~W.,
Yin, Q.~F., Perley, R.~A., Taylor, G.~B., \& Broderick, J.~J.\ 1998, AJ, 115, 1693 
\bibitem[Condon \& Kaplan(1998)]{Condon98b} Condon, J.~J., \& Kaplan, D.~L.\ 1998, \apjs, 117, 361 
\bibitem[Condon et al. (1999)]{Condon99} Condon, J.~J., Kaplan, D.~L., \& Terzian, Y.\ 1999, \apjs, 123, 219 
\bibitem[Conti \& Crowther (2004)]{conti2004} Conti, P.~S., \& Crowther, P.~A.\ 2004, \mnras, 355, 899
\bibitem[Cornwell (2008)]{Cornwell08} Cornwell, T.~J.\ 2008, IEEE Journal of Selected Topics in Signal Processing, 2, 793 
\bibitem[Cornwell (2010)]{Cornwell10}Cornwell, T.J., 2010, pers. communication.
\bibitem[Cornwell et al.(2011)]{Cornwell11}Cornwell, T.J., et al. 2011, ASKAP Memo ASKAP-SW-0020
\bibitem[Cotton et al.(2003)]{Cotton03}Cotton, W. D., et al. 2003, \pasa, 20, 12
\bibitem[Cowie et al.(1996)]{Cowie96}Cowie, L. L., et al.J. G. (1996), \aj, 112, 839
\bibitem[Cram et al.(1998)]{Cram98} Cram, L., Hopkins, A., Mobasher, B., \& Rowan-Robinson, M.\ 1998, \apj, 507, 155
\bibitem[Crittenden \& Turok (1996)]{Crittenden96}Crittenden, R. \& Turok, N. (1996), Phys. Rev. Lett. 76, 575
\bibitem[Croom et al.(2004)]{Croom04}Croom, S. M., et al. 2004, \mnras, 349, 1397
\bibitem[Croton et al.(2005)]{Croton05}Croton, D. J., et al. 2005, \mnras, 356, 1155 
\bibitem[Croton et al.(2006)]{Croton06}Croton, D. J., et al. 2006, \mnras, 365, 11 
\bibitem[Cowie \& Binney(1977)]{Cowie77} Cowie, L.~L., \& Binney, J.\ 1977, \apj, 215, 723 
\bibitem[Cruz et al.(2006)]{Cruz06}Cruz, M. J., et al. 2006, \mnras, 373, 1531
\bibitem[Cruz et al.(2007)]{Cruz07}Cruz, M. J., et al. 2007, \mnras, 375, 1349
\bibitem[Daddi et al.(2007)]{Daddi07}Daddi, E., et al. 2007, \apj, 670, 173
\bibitem[Dalal et al. (2008)]{Dalal08} Dalal, N., Dor{\'e}, O., Huterer, D., \& Shirokov, A.\ 2008, Phys. Rev D., 77, 123514
\bibitem[The Dark Energy Survey Collaboration(2005)]{DES2005} The Dark Energy Survey Collaboration 2005, arXiv:astro-ph/0510346
\bibitem[Deboer et al.(2009)]{Deboer09} Deboer, D.~R., et al.  2009, IEEE Proceedings, 97, 1507
\bibitem[De Breuck et al.(2001)]{Breuck01}De Breuck, C., et al. 2001, \aj, 121, 1241
\bibitem[De Breuck et al.(2002)]{Breuck02}De Breuck C., et al., 2002, \aj, 123, 637
\bibitem[Deghan et al.(2011)]{Deghan11} Deghan, S., et al., 2011, J. Astrophys. Astr. in press 
\bibitem[Dewdney et al.(2009)]{Dewdney09}Dewdney, P., et al., 2009, Proc. IEEE, 97, 1482
\bibitem[Dickey et al. (2010)]{Dickey10} Dickey, J.~M., Gibson, S.~J., Gomez, J.~F., Imai, H., Jones, P.~A., McClure-Griffiths, N.~M., Stanimirovic, S., \& van Loon, J.~T.\ 2010, arXiv:1008.4640 
\bibitem[Djorgovski et al.(2010)]{Djorgovski10} Djorgovski, S.~G., Mahabal, A., Donalek, C., Graham, M., Moghaddam, B., Drake, A., \& Williams, R.\ 2010, Bulletin of the American Astronomical Society, 42, \#231.05 
\bibitem[Dolag et al.(2008)]{Dolag08}Dolag, K., et al. 2008, Space Science Rev., 134, 311
\bibitem[Dougherty \& Williams, 2000]{Dougherty_2000} Dougherty, S. M., Williams, P. M., 2000, \mnras, 319, 143.
\bibitem[Drake et al.(1989)]{Drake_89}Drake, S. A., Simon, T., Linsky, J. L., 1989, \apjs, 71, 90.
\bibitem[Driver et al.(2009)]{Driver09} Driver, S.~P., et al.  2009, {\it Astr. \& Geophys.}, 50,12 
\bibitem[Dulk (1985)]{Dulk85}Dulk, G. A., 1985 \araa, 23, 169
\bibitem[Dunne et al.(2009)]{Dunne09}Dunne, L., et al. 2009, \mnras, 394, 3
\bibitem[Eddington (1913)]{Eddington13} Eddington, A.~S.\ 1913, \mnras, 73, 359
\bibitem[Edge et al. (1959)]{Edge59} Edge, D.~O., Shakeshaft, J.~R., McAdam, W.~B., Baldwin, J.~E., \& Archer, S.\ 1959, \memras, 68, 37
\bibitem[Eisenstein et al.(2005)]{Eisenstein05} Eisenstein, D.~J., et al.  2005, \apj, 633, 560
\bibitem[Ekers (2010)]{Ekers10}Ekers, R. D., 2010 in ÒAccelerating The Rate Of Astronomical DiscoveryÓ, edited by Ray P. Norris \& Clive L. N. Ruggles, 2010, PoS (SpS5) 
\bibitem[Elbaz et al.(2007)]{Elbaz07}Elbaz, D., et al. 2007, A\&A, 468, 33
\bibitem[Elbaz et al.(2009)]{Elbaz09} Elbaz, D., Jahnke, K., Pantin, E., Le Borgne, D., \& Letawe, G.\ 2009, \aap, 507, 1359
\bibitem[Evans et al.(2006)]{Evans06} Evans, D.~A., Lee, J.~C., Kamenetska, M., Gallagher, S.~C., Kraft, R.~P., Hardcastle, M.~J., \& Weaver, K.~A.\ 2006, \apj, 653, 1121 
\bibitem[Fabian \& Nulsen(1977)]{Fabian77}Fabian, A.~C., \& Nulsen, P.~E.~J.\ 1977, \mnras, 180, 479 
\bibitem[Fanaroff \& Riley (1974)]{FRI} Fanaroff, B.~L., \& Riley, J.~M. 1974, \mnras, 167, 31
\bibitem[Fanti (2009a)]{cfanti09a} Fanti, C. 2009a, Astron. Nachr., 330, 120
\bibitem[Fanti (2009b)]{rfanti09b} Fanti, R. 2009b, Astron. Nachr., 330, 303
\bibitem[Feretti(2000)]{Feretti00} Feretti, L.\ 2000, Invited review at IAU 199, arXiv:astro-ph/0006379 
\bibitem[Ferrarese \& Merritt (2000)]{Ferrarese00} Ferrarese, L., \& Merritt, D. 2000, \apj, 539, L9
\bibitem[Feulner et al.(2005)]{Feulner05} Feulner, G., et al. 2005, \apj, 633, L9
\bibitem[Fraser-McKelvie et al.(2011)]{Fraser11}Fraser-McKelvie, A., Pimbblet, K.~A., \& Lazendic, J.~S.\ , \mnras, in press (arXiv:1104.0711)
\bibitem[Fujita et al.(2007)]{Fujita07} Fujita, Y., Kohri, K., Yamazaki, R., \& Kino, M.\ 2007, \apjl, 663, L61 
\bibitem[Gaensler et al.(2010)]{Gaensler10} Gaensler, B.~M., Landecker, T.~L., Taylor, A.~R. \& POSSUM Collaboration, 2010, BAAS, 42, 515
\bibitem[Garn \& Alexander (2008)]{Garn08}{Garn}, T., \& {Alexander}, P. 2008, \mnras, 391, 1000 
\bibitem[Garn et al.(2009)]{Garn09} Garn, T., et al. 2009, \mnras, (arXiv:0905.1218)
\bibitem[Gebhardt et al.(2000)]{Gebhardt00}Gebhardt, K., et al. 2000, \apj, 539, L13
\bibitem[Giannantonio et al.(2008a)]{Giannantonio08a}Giannantonio, T., Scranton, R., Crittenden, R.~G., Nichol, R.~C., Boughn, S.~P., Myers, A.~D., \& Richards, G.~T.\ 2008a, 2008PhRvD..77l3520G (arXiv:0801.4380v2)
\bibitem[Gil de Paz et al. (2007)]{Galex07} Gil de Paz, A., et al.  2007, \apjs, 173, 185 
\bibitem[Giovannini \& Feretti (2004)]{Giovannini04}Giovannini, G., \& Feretti, L. 2004, J. Korean Ast. Soc., 37, 323
\bibitem[Giovannini et al.(1999)]{Giovannini99}Giovannini, G., Tordi, M., \& Feretti, L.\ 1999, \na, 4, 141 
\bibitem[Giovannini et al.(2009)]{Giovannini09}Giovannini, G. et al. 2009, \aa, 507, 1257
\bibitem[Giovannini et al.(2010)]{Giovannini10}Giovannini, G. et al. 2010 \aa, 511, L5
\bibitem[Giovannini et al.(2011)]{Giovannini11}Giovannini, G. et al. 2011, J. Astrophys. Astr. in press 
\bibitem[Gladders \& Yee (2005)]{Gladders05} Gladders, M.~D., \& Yee, H.~K.~C.\ 2005, \apjs, 157, 1
\bibitem[Gobat et al.(2011)]{Gobat11} Gobat, R., et al.  2011, \aap, 526, A133
\bibitem[Gold et al.(2011)]{Gold11} Gold, B., et al.  2011, \apjs, 192, 15
\bibitem[Gomez et al.(2003)]{Gomez03} Gomez, P. L., et al. 2003, \apj, 584, 210
\bibitem[Gopal-Krishna, \& Wiita (2001)]{Gopal01}Gopal-Krishna, \& Wiita, P. J. 2001, \apj, 560, L115
\bibitem[Govoni et al.(2001)]{Govoni01} Govoni, F., Feretti, L., Giovannini, G., B{\"o}hringer, H., Reiprich, T.~H., \& Murgia, M.\ 2001, \aap, 376, 803 
\bibitem[Granett et al.(2008)]{Granett08}Granett, B.~R., Neyrinck, M.~C., \& Szapudi, I.\ 2008, \apjl, 683, L99 
\bibitem[Green (2009)]{Green09} Green, D.~A.\ 2009, Bulletin of the Astronomical Society of India, 37, 45
\bibitem[Greve et al.(2010)]{Greve10}Greve, T.~R., et al.\ 2010, \apj, 719, 483 
\bibitem[G\"udel (2002)]{Gudel02} G\"udel., M., $ARA\&A$ 40, 217 
\bibitem[Gupta et al.(2006)]{Gupta06}Gupta, N., Salter, C. J., Saikia, D. J., Ghosh, T.
   \& Jeyakumar, S. 2006, \mnras, 373, 972
\bibitem[Gupta et al.(2008)]{Gupta08}Gupta, N., et al., 2008, ``The Initial Array Configuration for ASKAP'', ASKAP Internal Report.
\bibitem[Haarsma et al.(2000)]{Haarsma00}Haarsma, D. B., Partridge, R. B., Windhorst, R. A., Richards, E. A., 2000, \apj, 544, 641
\bibitem[Hales et al.(2011)]{Hales11} Hales, C.~A.\ et al., 2011, in prep.. 
\bibitem[Hallinan et al., 2008]{Hallinan_2008} Hallinan, G., Antonova, A., Doyle, J. G., Bourke, S., Lane, C., Golden, A., 2008, \apj, 684, 644.
\bibitem[Hancock et al.(2011)]{Hancock11} Hancock, P., Murphy, T., \& Hopkins, A., 2011, in prep.
\bibitem[Hardcastle et al.(2006)]{Hardcastle06} Hardcastle, M.~J., Evans, D.~A., \& Croston, J.~H.\ 2006, \mnras, 370, 1893 
\bibitem[Hardcastle et al.(2007)]{Hardcastle07}Hardcastle, M. J., et al. 2007, \mnras, 376, 1849
\bibitem[Hasinger et al.(2005)]{Hasinger05}Hasinger, G., et al. 2005, A \&A, 441, 417
\bibitem[Haverkorn et al.(2006)]{haverkorn2006} Haverkorn, M., Gaensler, B.~M., McClure-Griffiths, N.~M., Dickey, J.~M., \& Green, A.~J.\ 2006, \apjs, 167, 230
\bibitem[Helfand et al.(1989)]{Helfand89} Helfand, D.J., Velusamy, T., Becker, R.H., Lockman, F.J., 1989, \apj, 341, 151
\bibitem[Helfand et al.(1999)]{Helfand_1999} Helfand, D. J., Schnee, S., Becker, R. H., White, R. L., McMahon, R. G. \aj, 117, 1568.
\bibitem[Helfand et al.(2006)]{Helfand06} Helfand, D.~J., Becker, R.~H., White, R.~L., Fallon, A., \& Tuttle, S.\ 2006, \aj, 131, 2525
\bibitem[Herbert et al.(2010)]{Herbert10}Herbert et al. 2010, \mnras, 406, 1841
\bibitem[Hine \& Longair(1979)]{Hine79} Hine, R.~G., \& Longair, M.~S.\ 1979, \mnras, 188, 111 
\bibitem[Hoeft et al.(2011)]{Hoeft11}Hoeft, M., et al. 2011, J. Astrophys. Astr. in press 
\bibitem[Hogg \& Turner (1998)]{Hogg98} Hogg, D.~W., \& Turner, E.~L.\ 1998, PASP, 110, 727
\bibitem[Hopkins \& Beacom (2006)]{Hopkins06}Hopkins, A. M., \& Beacom, J. F. 2006, \apj, 651, 142
\bibitem[Hopkins et al.(2002)]{Hop:02}Hopkins, A. M., et al., 2002, AJ, 123, 1086
\bibitem[Hopkins et al.(2003)]{Hopkins03}Hopkins, A. M. et al. 2003, AJ, 125, 465
\bibitem[Hopkins (2004)]{Hopkins04}Hopkins, A. M. 2004, \apj, 615, 209
\bibitem[Huynh et al.(2005)]{Huynh05}Huynh, M., et al. 2005, AJ, 130, 1373
\bibitem[{Huynh} et~al. (2010)]{Huynh10}{Huynh}, M.~T., {Norris}, R.~P., \& {Middelberg}, M., 2010, \apj, 710, 698
\bibitem[Ibar et al.(2009)]{Ibar09} Ibar, E., Ivison, R.~J., Biggs, A.~D., Lal, D.~V., Best, P.~N., \& Green, D.~A.\ 2009, \mnras, 397, 281
\bibitem[Ibar et al.(2010)]{Ibar10} Ibar, E., Ivison, R.~J., Best, P.~N., Coppin, K., Pope, A., Smail, I., \& Dunlop, J.~S.\ 2010, \mnras, 401, L53

\bibitem[Ivezic et al. (2008)]{Ivezic2008} Ivezic, Z., Tyson, J.~A., Allsman, R., Andrew, J., Angel, R., \& for the LSST Collaboration 2008, arXiv:0805.2366
\bibitem[Ivison et al.(2007)]{Ivison07}Ivison, R.~J., et al. 2007, \apj, 660, L77 
\bibitem[Ivison et al.(2010)]{Ivison10}Ivison, R.~J., et al. 2010, \mnras, 402, 245 
\bibitem[Jackson \& Rawlings(1997)]{Jackson97} Jackson, N., \& Rawlings, S.\ 1997, \mnras, 286, 241 
\bibitem[Jackson (2005)]{Jackson05}Jackson, C., 2005, PASA, 22, 36
\bibitem[Jarvis et al.(2001a)]{Jarvis01a}Jarvis, M. J., et al. 2001a, \mnras, 326, 1563
\bibitem[Jarvis et al.(2001b)]{Jarvis01b}Jarvis, M. J., et al. 2001b, \mnras, 326, 1585
\bibitem[Jarvis et al.(2001c)]{Jarvis01c}Jarvis, M. J., et al. 2001c, \mnras, 327, 907
\bibitem[Jarvis et al.(2004)]{Jarvis04}Jarvis, M. J.,  \& Rawlings, S., 2004, NewAR, 48, 1173
\bibitem[Jarvis et al.(2009)]{Jarvis09} Jarvis, M.~J., Teimourian, H., Simpson, C., Smith, D.~J.~B., Rawlings, S., \& Bonfield, D. 2009, \, \mnras, 398, L83
\bibitem[Johnston et al.(2007)]{Johnston07}Johnston, S., et al.  2007, \pasa, 24, 174
\bibitem[Johnston et al.(2008)]{Johnston08}Johnston, S., et al. 2008, Experimental Astronomy, 22, 151
\bibitem[Johnston-Hollitt, M., et al.(2009)]{Johnston-Hollitt09}Johnston-Hollitt, M., et al. 2009, \mnras, (submitted)
\bibitem[Jonas(2009)]{Jonas10} Jonas, J.~L.\ 2009, IEEE Proceedings, 97, 1522
\bibitem[Jones et al.(2009)]{Jones09}Jones, D.~H., et al.\ 2009, \mnras, 399, 683  
\bibitem[Juneau et al.(2005)]{Juneau05} Juneau, S., et al. 2005, \apj, 619, L135
\bibitem[Kaiser et al. (2010)]{Kaiser2010} Kaiser, N., et al.  2010, \procspie 7733, 0E
\bibitem[Keller et al.(2007)]{Keller07}Keller, S. C., et al. 2007, \pasa, 24, 1
\bibitem[Kempner et al.(2004)]{Kempner04} Kempner, J.~C., Blanton, E.~L., Clarke, T.~E., En{\ss}lin, T.~A., Johnston-Hollitt, M., \& Rudnick, L.\ 2004, The Riddle of Cooling Flows in Galaxies and Clusters of galaxies, 335, http://www.astro.virginia.edu/coolflow
\bibitem[Keshet et al.(2004)]{Keshet04} Keshet, U., Waxman, E., \& Loeb, A.\ 2004, \apj, 617, 281\\
\bibitem[Kim et al.(1989)]{Kim89}Kim, K.-T., Kronberg, P.~P., Giovannini, G., \& Venturi, T.\ 1989, \nat, 341, 720 
\bibitem[Kimball et al. (2009)]{Kimball09} Kimball, A.~E., Knapp, G.~R., Ivezi{\'c}, {\v Z}., West, A.~A., Bochanski, J.~J., Plotkin, R.~M., \& Gordon, M.~S.\ 2009, \apj, 701, 535
\bibitem[Klamer et al.(2004)]{Klamer04}Klamer I. J., Ekers R. D., Sadler E. M., Hunstead R. W., 2004, \apj, 612, L97
\bibitem[Kodama et al.(2007)]{Kodama07} Kodama, T., et al.,\ 2007, \mnras, 377, 1717
\bibitem[Koribalski et al.(2008)]{Koribalski08}Koribalski, B.S., et al.,  2008, in ``Galaxies in the Local Volume'', Sydney, July 2007, Springer, p. 41 
\bibitem[Koribalski et al.(2011)]{Koribalski11}Koribalski, B. S., et al., 2011, \pasa, in preparation 
\bibitem[Kronberg et al.(2007)]{Kronberg07} Kronberg, P.~P., Kothes, R., Salter, C.~J., \& Perillat, P.\ 2007, \apj, 659, 267
\bibitem[Kurtz et al.(1994)]{kcw94} Kurtz, S., Churchwell, E., \& Wood, D.~O.~S.\ 1994, \apjs, 91, 659
\bibitem[Kurtz et al.(1999)]{kurtz1999} Kurtz, S.~E., Watson, A.~M., Hofner, P., \& Otte, B.\ 1999, \apj, 514, 232
\bibitem[Kurtz (2005)]{kurtz2005} Kurtz, S.\ 2005,  in ``Massive Star Birth: A Crossroads of Astrophysics'', IAU Symp. 227, p.\ 111, Cambridge: Cambridge University Press
\bibitem[Labiano et al.(2008)]{labiano08} Labiano, A., O'Dea, C.~P., Barthel, P.~D., de Vries, W.~H., \& Baum, S.~A. 2008, \aa, 477, 491
\bibitem[Laing et al.(1994)]{Laing94} Laing, R.~A., Jenkins, C.~R., Wall, J.~V., \& Unger, S.~W.\ 1994, The Physics of Active Galaxies, 54, 201 
\bibitem[Large et al.(1981)]{Large81} Large, M.~I., Mills, B.~Y., Little, A.~G., Crawford, D.~F., \& Sutton, J.~M.\ 1981, \, \mnras, 194, 693
\bibitem[Lewis et al.(2002)]{Lewis02} Lewis, I., et al. 2002, \mnras, 334, 673
\bibitem[Lilly et al.(1996)]{Lilly96} Lilly, S. J., et al.(1996), \apj, 460, L1
\bibitem[Lintott et al.(2008)]{Lintott08} Lintott, C. J., et al. 2008, \mnras, 389,1179
\bibitem[Liu et al.(2008)]{Liu08}Liu, F., et al. 2008,Proceedings of the SPIE, Vol. 7017, 70170M
\bibitem[Longmore et al.(2009)]{longmore2009} Longmore, S.~N., Burton, M.~G., Keto, E., Kurtz, S., \& Walsh, A.~J.\ 2009, \, \mnras, 399, 861
\bibitem[Madau et al.(1996)]{Madau96} Madau, P., et al.(1996), \mnras, 283, 1388
\bibitem[Magorrian et al.(1998)]{Magorrian98}Magorrian, J., et al.(1998), \aj, 115, 2285
\bibitem[Mantz et al.(2010)]{Mantz10}Mantz, A., Allen, S.~W., Rapetti, D., \& Ebeling, H.\ 2010, \mnras, 406, 1759
\bibitem[Mao et al.(2009)]{Mao09}Mao, M., et al. 2009, \mnras, 392, 1070 
\bibitem[Mao et al.(2010)]{Mao10}Mao, M., et al. 2010, \mnras, 406, 2578 
\bibitem[Mao et al.(2011a)]{Mao11a}Mao, M.~Y., Huynh, M.~T., Norris, R.~P., Dickinson, M., Frayer, D., Helou, G., \& Monkiewicz, J.~A.\ 2011, \apj, 731, 79  
\bibitem[Mao et al.(2011b)]{Mao11b}Mao, M., et al., 2011b, in preparation 
\bibitem[Mao et al.(2011c)]{Mao11c}Mao, M., et al., 2011c, J. Astrophys. Astr. in press 
\bibitem[Markevitch (2011)]{Markevitch10}Markevitch, M.,\ 2011, arXiv:1010.3660
\bibitem[Marriage et al.(2010)]{Marriage10}Marriage, T.~A., et al.  2010, arXiv:1010.1065
\bibitem[Massardi et al.(2010)]{Massardi10}Massardi M. et al., 2010, \mnras, 404, 532
\bibitem[Mauch et al.(2003)]{Mau:03}Mauch, T., et al., 2003, \mnras, 342, 1117
\bibitem[Mauch \& Sadler (2007)]{Mauch07}Mauch, T., \& Sadler, E. M. 2007, \mnras, 375, 931
\bibitem[McClure-Griffiths et al.(2005)]{naomi05} McClure-Griffiths, N.~M., Dickey, J.~M., Gaensler, B.~M., Green, A.~J., Haverkorn, M., \& Strasser, S.\ 2005, \apjs, 158, 178
\bibitem[McLure et al.(2006)]{Mclure06} McLure, R.~J., Jarvis, M.~J., Targett, T.~A., Dunlop, J.~S., \& Best, P.~N.\ 2006, \, \mnras, 368, 1395
\bibitem[McNamara \& Nulsen(2007)]{McNamara07} McNamara, B.~R., \& Nulsen, P.~E.~J.\ 2007, \araa, 45, 117 
\bibitem[Messias et al.(2010)]{Messias10} Messias, H., Afonso, J., Hopkins, A., Mobasher, B., Dominici, T., \& Alexander, D.~M.\ 2010, \apj, 719, 790 
\bibitem[Mezger \& Henderson (1967)]{mezger1967} Mezger, P.~G., \& Henderson, A.~P.\ 1967, \apj, 147, 471
\bibitem[Middelberg et al.(2008a)]{Middelberg08a}Middelberg, E., et al., 2008a, \aj, 135, 1276 
\bibitem[Middelberg et~al (2008b)]{Middelberg08b}Middelberg, E., Norris, R.~P., Tingay, S., Mao, M.~Y., Phillips, C.~J., \& Hotan, A.~W. 2008b, \aap, 491, 435 
 \bibitem[Middelberg et~al. (2011)]{Middelberg11}Middelberg, E., Norris, R.~P., Seymour, N., Huynh, M., Johnston-Hollitt, M., Mao, M.~Y., \& Hales, C.~A., \& {Hotan}, 2010, \aa in press
\bibitem[Mignano et al.(2008)]{Mignano08}Mignano, A., et al. 2008, A\&A, 477 459
\bibitem[Miley et al.(2006)]{Miley06}Miley, G. K., et al. 2006, \apj, 650, L29
\bibitem[Miller et al.(2008)]{Miller08} Miller, N.~A., Fomalont, E.~B., Kellermann, K.~I., Mainieri, V., Norman, C., Padovani, P., Rosati, P., \& Tozzi, P.\ 2008, \apjs, 179, 114
\bibitem[Miniati et al.(2001)]{Miniati01}Miniati, F., et al. 2001, \apj, 562, 233
\bibitem[Mitchell \& Condon. (1985)]{Mitchell85}Mitchell, K. J., \& Condon, J. J. 1985, \aj, 90, 1957
\bibitem[Mittal et al.(2009)]{Mittal09} Mittal, R., Hudson, D.~S., Reiprich, T.~H., \& Clarke, T.\ 2009, \aap, 501, 835 
\bibitem[Mobasher et al.(2009)]{Mobasher09} Mobasher, B., et al. 2009, \apj, 690, 1074
\bibitem[Molinari et al.(2010)]{molinari2010} Molinari, S., et al.  2010, \pasp, 122, 314
\bibitem[Morganti (2008)]{morganti08} Morganti, R. 2008, in "Extragalactic Jets: Theory and Observation from Radio to Gamma Ray", edited by Rector, T.~A., \& De Young, D.~S., 210
\bibitem[Morganti et al.(2009)]{morganti09} Morganti, R., Emonts, B., Holt, J., Tadhunter, C., Oosterloo, T., \& Struve, C. 2009, Astron. Nachr., 330, 789

\bibitem[Morrison et al.(2010)]{Morrison2010} Morrison, G.~E., et al., 2010, \apjs, 188, 178
\bibitem[Murgia et al.(1999)]{Murgia99}Murgia, M., Fanti, C., Fanti, R., Gregorini, L., Klein, U., Mack, K.-H., \& Vigotti, M. 1999, \aap, 345, 769
\bibitem[Murphy et al. (2007)]{Murphy07} Murphy, T., Mauch, T., Green, A., Hunstead, R.~W., Piestrzynska, B., Kels, A.~P., \& Sztajer, P.\ 2007, \mnras, 382, 382
\bibitem[Norris et al.(2005)]{Norris05} Norris, R.~P., et al.  2005, \aj, 130, 1358 
\bibitem[Norris et al.(2006)]{Norris06}Norris, R.P., et al. 2006, \aj, 132, 2409 
\bibitem[Norris et~al. (2007)]{Norris07}{Norris}, R.~P., {Tingay}, S., {Phillips}, C., {Middelberg}, E., {Deller}, A., \& {Appleton}, P.~N. 2007, \mnras, 378, 1434 
\bibitem[Norris et al.(2009)]{Norris09}Norris R.P., et al., 2009 in "Proceedings of Panoramic Radio Astronomy". Ed George Heald. PoS (PRA2009)033 
\bibitem[Norris, R. P. (2010)]{Norris10}Norris, R. P., 2010, in ÒAccelerating The Rate Of Astronomical DiscoveryÓ, edited by Ray P. Norris \& Clive L. N. Ruggles, 2010, PoS (SpS5) 
\bibitem[Norris et al.(2011a)]{Norris11a}Norris, R.P., et al., 2011a, \apj, in press. 2011arXiv1105.0960N 
\bibitem[Norris et al.(2011b)]{Norris11b}Norris, R.P., et al., 2011b, \pasa, in preparation. 
\bibitem[Norris et al.(2011c)]{Norris11c}Norris, R.P.,  et al., 2011c, J. Astrophys. Astr., in press. 
\bibitem[O'Dea (1998)]{Odea98} O'Dea, C. 1998, \pasp, 110, 493
\bibitem[Ogle et al.(2006)]{Ogle06} Ogle, P., Whysong, D., \& Antonucci, R.\ 2006, \apj, 647, 161 
\bibitem[Ogle et al. (2010)]{Ogle10} Ogle, P., Boulanger, F.,  Guillard, P.,  Evans, D. A., Antonucci, R., Appleton, P. N.., Nesvadba, N., \& Leipski, 2010, \apj, 724, 1193
\bibitem[Oliver et al.(2010)]{Oliver10} Oliver, S.~J., et al.\ 2010, \aap, 518, L21 
\bibitem[Oosterloo et al. (2009)]{Oosterloo10} Oosterloo, T., Verheijen, M., van Cappellen, W., Bakker, L., Heald, G., \& Ivashina, M.\ 2009, arXiv:0912.0093
\bibitem[Overzier et al.(2003)]{Overzier2003} Overzier, R.~A., R{\"o}ttgering, H.~J.~A., Rengelink, R.~B., \& Wilman, R.~J.\ 2003, \aap, 405, 53 
\bibitem[Owen \& Morison (2008)]{Owen08}Owen, F.N, Morison, G.E., 2008, \aj, 136, 1889
\bibitem[Owsianik et al.(1999)]{Owsianik99}Owsianik, I., Conway, J. E. \& Polatidis, A. G. 1999, NewAR, 43, 669
\bibitem[Padovani et al.(2009)]{Padovani09}Padovani, P., et al., 2009, \apj, 694, 235
\bibitem[Papovich et al.(2006)]{Papovich06} Papovich, C., et al. 2006, \apj, 640, 92
\bibitem[Peacock \& Smith (2000)]{Peacock00}Peacock, J. A., \& Smith, R. E. 2000, \mnras, 318, 1144
\bibitem[Pen et al.(2009)]{Pen09} Pen, U.-L., Staveley-Smith, L., Peterson, J.~B., \& Chang, T.-C.\ 2009, \mnras, 394, L6
\bibitem[Penner et al.(2011)]{Penner11} Penner, K., et al.\ 2011, \mnras, 410, 2749 
\bibitem[Perlmutter et al.(1999)]{Perlmutter99}Perlmutter, S. et al.(1999) Astrophys. J. 517, 565
\bibitem[Peroux et al.(2005)]{Peroux05}Peroux, C., et al. 2005, \mnras, 363, 479
\bibitem[Peterson et al.(2001)]{Peterson01} Peterson, J.~R., et al.\ 2001, \aap, 365, L104 
\bibitem[Peterson et al.(2003)]{Peterson03} Peterson, J.~R., Kahn, S.~M., Paerels, F.~B.~S., Kaastra, J.~S., Tamura, T., Bleeker, J.~A.~M., Ferrigno, C., \& Jernigan, J.~G.\ 2003, \apj, 590, 207 
\bibitem[Peterson \& Fabian(2006)]{Peterson06} Peterson, J.~R., \& Fabian, A.~C.\ 2006, Phys. Rep., 427, 1 
\bibitem[Pfrommer et al.(2007)]{Pfrommer07}Pfrommer, C., et al. 2007, \mnras, 378, 385
\bibitem[Pierre et al.(2004)]{Pierre04} Pierre, M., et al.  2004, J. Cosm. \& Astroparticle Phys., 9, 11
\bibitem[Pietrobon et al.(2006)]{Pietrobon06}Pietrobon, D., et al. 2006, Phys. Rev. D, 74, 043524
\bibitem[Pillepich et al.(2011)]{Pillepich11}Pillepich, A., Porciani, C., \& Reiprich. T.H., 2011, in preparation
\bibitem[Planck Collaboration (2011)]{Planck11} Planck Collaboration \ 2011, arXiv:1101.2024
\bibitem[Pogosian et al.(2010)]{Pogosian10}Pogosian, L., et al, 2010, Phys. Rev. D, 81, 104023
\bibitem[Prandoni et al.(2001)]{Prandoni01} Prandoni, I., Gregorini, L., Parma, P.,
de Ruiter, H.~R., Vettolani, G., Wieringa, M.~H., \& Ekers, R.~D.\ 2001, \aa, 365, 392
\bibitem[Predehl et al.(2010)]{Predehl10}Predehl, P., et al.  2010, \procspie, 7732,
\bibitem[Purcell et al. (2008)]{Purcell08} Purcell, C.~R., Hoare, M.~G., \& Diamond, P.\ 2008, Massive Star Formation: Observations Confront Theory, 387, 389 
\bibitem[Purcell \& Hoare (2010)]{purcell2010} Purcell, C.~R., \& Hoare, M.~G.\ 2010, Highlights of Astronomy, 15, 781
\bibitem[Raccanelli et al.(2008)]{Raccanelli08} Raccanelli, A., Bonaldi, A., Negrello, M., Matarrese, S., Tormen, G., \& de Zotti, G.\ 2008, \mnras, 386, 2161 
\bibitem[Raccanelli et al.(2011)]{Raccanelli11}Raccanelli, A., et al. 2011, in preparation
\bibitem[Randall et al.(2011)]{Randall11}Randall, K. E.,  et al. 2011,  \mnras, in press.
\bibitem[Ravi et al. (2010)]{Ravi10} Ravi, V., et al.  2010, \mnras, 408, L99
\bibitem[Rawlings \& Jarvis (2004)]{Rawlings04}Rawlings, S., \& Jarvis, M. J. 2004, \mnras, 355, L9
\bibitem[Reich et al.(1992)]{Reich92} Reich, W., F\"urst, E. Arnal, E.M., 1992, \aa, 256, 214
\bibitem[Riess et al.(1998)]{Riess98}Riess, A. G. et al., 1998, \aj, 116, 1009
\bibitem[Romer et al.(2001)]{Romer01} Romer, A.~K., et al., \ 2001, \apj, 547, 594
\bibitem[Rosati et al.(1998)]{Rosati98} Rosati, P., et al., \ 1998, \apjl, 492, L21
 \bibitem[Roseboom et al.(2009)]{Roseboom09} Roseboom, I.~G., Oliver, S., Parkinson, D., \& Vaccari, M.\ 2009, \mnras, 400, 1062 
\bibitem[Roseboom et al.(2010)]{Roseboom10} Roseboom, I.~G., et al.\ 2010, \mnras, 409, 48 
\bibitem[R\"ottgering et al.(1997)]{Rottgering97} R\"ottgering, H.~J.~A., Wieringa, M.~H., Hunstead, R.~W., \& Ekers, R.~D.\ 1997, \mnras, 290, 577 
\bibitem[R\"ottgering et al.(2010a)]{Rottgering10a}R\"ottgering, H., et al., 2010a, PoS (ISKAF2010)050 
\bibitem[R\"ottgering et al.(2010b)]{Rottgering10b}R\"ottgering, H., et al., 2010b, http://www.astron.nl/radio-observatory/apertif-eoi-abstracts-and-contact-information 
\bibitem[Rowan-Robinson et al.(2008)]{Rowan08}Rowan-Robinson, M., et al.(2008), \mnras, 386, 697
\bibitem[Rudnick (2002)]{Rudnick02}Rudnick, L., 2002, PASP, 114, 427
\bibitem[Rudnick \& Lemmerman (2009)]{Rudnick09} Rudnick, L., \& Lemmerman, J.~A.\ 2009, \apj, 697, 1341
\bibitem[Ryu et al.(2008)]{Ryu08}Ryu, D., et al. 2008, Science, 320, 909
\bibitem[Saikia et al.(2003)]{Saikia03}Saikia, D. J., Jeyakumar, S., Mantovani, F., Salter, C. J.,
 \& Spencer, R. E., Thomasson, P. \& Wiita, P. J. 2003, PASA, 20, 50
\bibitem[Saikia et al.(2009)]{Saikia09}Saikia, D. J. \& Jamrozy, M. 2009, BASI, 37, 63
\bibitem[Salvato et al.(2010)]{Salvato10}Salvato, M., et al., 2010, 
\bibitem[Santos et al.(2010)]{Santos10} Santos, J.~S., Tozzi, P., Rosati, P., B\"ohringer, H.\ 2010, \aap, 521, A64 
\bibitem[Saripalli et al. (2011)]{Saripalli11}Saripalli, L., et al., 2011, in preparation.
\bibitem[Schmidt (1968)]{Schmidt68}Schmidt, M. 1968, \apj, 151, 393
\bibitem[Schinnerer et al. (2007)]{Schinnerer07} Schinnerer, E., et al.  2007, \apjs, 172, 46
\bibitem[Schuecker et al.(2001)]{Schuecker01} Schuecker, P., B{\"o}hringer, H., Reiprich, T.~H., \& Feretti, L.\ 2001, \aap, 378, 408
\bibitem[Scoville et al.(2007)]{Scoville07}Scoville, N., 2007, \apjs, 172, 1
\bibitem[Scranton et al.(2005)]{Scranton05}Scranton, R., et al. 2005, \apj, 633, 589
\bibitem[Seaquist et al.(1993)]{Seaquist93} Seaquist, E. R., Krogulec, M., \& Taylor, A.~R., 1993, ApJ, 410, 260

\bibitem[Seaquist \& Ivison (1994)]{Seaquist94} Seaquist, E. R., \& Ivison, R.~J., 1994, \mnras, 269, 512

\bibitem[Seymour et al.(2008)]{Seymour08}Seymour, N., et al. 2008, \mnras, 386, 1695
\bibitem[Seymour et al.(2011)]{Seymour11} Seymour, N., et al.\ 2011, \mnras, 280 
\bibitem[Shanks (2005)]{Shanks05}Shanks, T., 2005,VST-ATLAS proposal, http://www.astro.dur.ac.uk/Cosmology/vstatlas/
\bibitem[Shaver et al.(1996)]{Shaver96} Shaver, P.~A., Wall, J.~V., Kellermann, K.~I., Jackson, C.~A., \& Hawkins, M.~R.~S.\ 1996, \nat, 384, 439
\bibitem[Simpson et al.(2006)]{Simpson06}Simpson, C., et al. 2006, \mnras, 372, 741
\bibitem[Skillman et al.(2008)]{Skillman08}Skillman, S. W., et al. 2008, \apj, 689, 1063
\bibitem[Skillman et al.(2010)]{Skillman10}Skillman, S.~W., Hallman, E.~J., O'Shea, B.~W., Burns, J.~O., Smith, B.~D., \& Turk, M.~J.\ 2010, arXiv:1006.3559
\bibitem[Slee et al.(2008)]{Slee_2008}Slee, O. B., Wilson, W., Ramsay, G., 2008, PASA 25, 94
\bibitem[Slee et al.(1987)]{Slee_1987}Slee, O. B., Nelson, G. J., Stewart, R. T., Wright, A. E., Innis, J. L., Ryan, S. G., Vaughan, A. E., 1987, \mnras 229, 659.
\bibitem[Smith et al.(2010)]{Smith10}Smith, D.J.B., Dunne, Maddox, 2010, arxiv:1007.5260 
\bibitem[Smolcic et al.(2008)]{Smolcic08}Smolcic, V., et al. 2008, \apjs, 177, 14
\bibitem[Snellen et al.(1998)]{snellen98} Snellen, I.~A.~G., Schilizzi, R.~T., de Bruyn, A.~G., Miley, G.~K., Rengelink, R.~B., R{\"o}ttgering H.~J.~A., \& Bremer, M.~N., 1998, \aas, 131, 435
\bibitem[Snellen et al.(1999)]{snellen99} Snellen, I.~A.~G., Schilizzi, R.~T., Miley, G.~K., Bremer, M.~N., R{\"o}ttgering H.~J.~A., \& van Langevelde, H.~J. 1999, NewAR, 43, 675
\bibitem[Snellen et al.(2009)]{Snellen09} Snellen, I.~A.~G., et al.\ 2009, Astronomische Nachrichten, 330, 297 
\bibitem[Springel et al.(2005)]{Springel05}Springel, V., et al. 2005, Nature, 435, 629
\bibitem[Stil et al.(2006)]{stil2006} Stil, J.~M., et al.  2006, \aj, 132, 1158
\bibitem[Subrahmanyan et al.(2010)]{Subrahmanyan10}Subrahmanyan, R., Ekers, R.~D., Saripalli, L., \& Sadler, E.~M.\ 2010, \mnras, 402, 2792
\bibitem[Sutherland \& Saunders(1992)]{Sutherland92} Sutherland, W., \& Saunders, W.\ 1992, \mnras, 259, 413 
\bibitem[Sutherland (2009)]{Sutherland09}Sutherland, W. 2009, in "Science with the VLT in the ELT Era", Astrophysics and Space Science Proceedings, 171
\bibitem[Taylor et al.(2003)]{taylor2003} Taylor, A.~R., et al.  2003, \aj, 125, 3145
\bibitem[Tegmark (2004)]{Tegmark04} Tegmark, M. et al.(2004), Phys. Rev. D 69, 103501
\bibitem[Thompson et al.(2006)]{thompson2006} Thompson, M.~A., Hatchell, J., Walsh, A.~J., MacDonald, G.~H., \& Millar, T.~J.\ 2006, \aap, 453, 1003
\bibitem[Thompson et al.(2011)]{Thompson11} Thompson, M.~A., et al., in preparation 
\bibitem[Trigilio et al., 2008]{Trigilio_2008} Trigilio, C., Leto, P., Umana, G., Buemi, C. S., Leone, F. 2008, \mnras 384, 1437.
\bibitem[Trigilio et al., 2000]{Trigilio_2000} Trigilio, C., Leto, P., Leone, F., Umana, G., Buemi, C. 2000, \aap, 362, 281.
\bibitem[Tschager et al.(2003)]{tschager03} Tschager, W., Schilizzi, R.~T., R{\"o}ttgering H.~J.~A., Snellen, I.~A.~G., Miley, G.~K., \& Perley, R. 2003, \pasa, 20, 75
\bibitem[Umana et al.(1998)]{Umana_98} Umana, G., Trigilio, C., Catalano, S., 1998, \aap, 329, 1010.
\bibitem[Umana et al.(1995)]{Umana_95} Umana, G., Trigilio, C., Tumino, M., Catalano, S., Rodon\'{o} M. 1995 \aap, 298, 143.
\bibitem[Umana et al.(1993)]{Umana_93} Umana, G., Trigilio, C., Hjellming, R. M.;, Catalano, S., Rodon\'{o}, M. 1993, \aap, 267, 126.
\bibitem[Urquhart et al.(2009)]{urquhart09} Urquhart, J.~S., et al.  2009, \aap, 501, 539
\bibitem[Uzan (2003)]{Uzan03}Uzan, J. P., 2003, Rev. Mod. Phys. 75, 403
\bibitem[van Breugel et al.(1999)]{Breugel99}van Breugel, W., et al.(1999), \apj, 518, L61
\bibitem[van der Heyden \& Jarvis (2010)]{Heyden10}Van der Heyden, K., \& Jarvis, M. J., 2010, MIGHTEE proposal to Meerkat
\bibitem[van Weeren et al.(2010)]{Weeren10}van Weeren, R. J., R\"ottgering, H. J. A., Br\"uggen, M., Hoeft, M., 2010, Science 310, 347
\bibitem[Venturi et al.(2008)]{Venturi08}Venturi, T., et al. 2008, A\&A, 484, 327
\bibitem[Vikhlinin et al.(2009)]{Vikhlinin09}Vikhlinin, A., et al.\ 2009, \apj, 692, 1060 
\bibitem[Wall et al.(2005)]{Wall05}Wall, J. V., et al. 2005, \aa , 434, 133
\bibitem[Walsh et al.(1998)]{walsh1998} Walsh, A.~J., Burton, M.~G., Hyland, A.~R., \& Robinson, G.\ 1998, \mnras, 301, 640
\bibitem[Wardle \& Miley (1974)]{Wardle74} Wardle, J. F. C., Miley, G.K., 1974, \aap, 30, 305
\bibitem[White et al.(1997)]{Whi:97}White, R. L., et al., 1997, \apj, 475, 479
\bibitem[White et al.(2005)]{white2005} White, R.~L., Becker, R.~H., \& Helfand, D.~J.\ 2005, \aj, 130, 586
\bibitem[White, 2004]{White_2004} White, S. M., 2004, New Astronomy Reviews, 48, 1319.
\bibitem[White, 2007]{White07} White, R. L., Helfand, D. J., Becker, R. H.,  et al., 2007,  \apj, 654, 99
\bibitem[Whiting (2008)]{Whiting08}Whiting, M. T., 2008; ``Galaxies in the Local Volume'', ed. B.S.Koribalski \& H.Jerjen, Astrophysics and Space Science Reviews, Springer, 343
\bibitem[Whysong \& Antonucci(2004)]{Whysong04} Whysong, D., \& Antonucci, R.\ 2004, \apj, 602, 116 
\bibitem[Wiaux et al.(2009)]{Wiaux09}Wiaux, Y., Jacques, L., Puy, G., Scaife, A.~M.~M., \& Vandergheynst, P.\ 2009, \mnras, 395, 1733 
\bibitem[Wilman et al.(2008)]{Wilman08} Wilman, R.~J., et al.  2008, \mnras, 388, 1335
\bibitem[Wilman et al.(2010)]{Wilman10} Wilman, R.~J., Jarvis, M.~J., Mauch, T., Rawlings, S., \& Hickey, S.\ 2010, \mnras, 405, 447 
\bibitem[Williamson et al.(2011)]{Williamson11} Williamson, R., et al.  2011, arXiv:1101.1290
\bibitem[Willott et al.(2003)]{Willott03} Willott, C.~J., et al., \ 2003, \mnras, 339, 173 
\bibitem[Wilson et al.(2008)]{Wilson08} Wilson, G., et al.  2008, Infrared Diagnostics of Galaxy Evolution, 381, 210
\bibitem[Wing \& Blanton(2011)]{Wing11} Wing, J.~D., \& Blanton, E.~L.\ 2011, \aj, 141, 88
\bibitem[Wood \& Churchwell (1989)]{woodchurchwell89} Wood, D.~O.~S., \& Churchwell, E.\ 1989, \apjs, 69, 831
\bibitem[Wright et al. (2010)]{Wright2010} Wright, E.~L., et al.  2010, \aj 140, 1868 
\bibitem[Wyithe \& Loeb (2009)]{Wyithe09} Wyithe, J.~S.~B., \& Loeb, A.\ 2009, \mnras, 397, 1926
\bibitem[Xia et al.(2009)]{Xia09}Xia J.-Q., Viel M., Baccigalupi C., Matarrese S., 2009, JCAP, 09, 003
\bibitem[York et al.(2000)]{York00}York, D. G., et al. 2000, \aj, 120, 1579 
\bibitem[Zhao et al.(2010)]{Zhao10}Zhao, G.-B., Giannantonio, T., Pogosian, L., Silvestri, A., Bacon, D.~J., Koyama, K., Nichol, R.~C., \& Song, Y.-S.\ 2010, \prd, 81, 103510 
\bibitem[Zheng et al.(2007)]{Zheng07}Zheng, X. Z., et al. 2007, \apj, 661, L41
\bibitem[Ziatev et al.(1999)]{Ziatev99}Zlatev, I., Wang, L., \& Steinhardt, P. J., 1999, Phys. Rev. Lett, 82, 896
\end{thebibliography}
\end{document}